\documentclass[useAMS,usenatbib]{mnras}
\usepackage[fleqn]{amsmath}

\usepackage{newtxtext,newtxmath}

\usepackage[T1]{fontenc}
\usepackage{ae,aecompl}

\usepackage{amsfonts,bm}
\usepackage{graphicx}
\usepackage[utf8]{inputenc}
\usepackage{adjustbox}   %Centering issues
\usepackage{subfig}
\usepackage{marvosym}
\usepackage{siunitx}
\usepackage{longtable, booktabs}
\usepackage[justification=centering]{caption}
\usepackage{rotating, lscape}
\newcommand{\R}{\textcolor{red}}

\newcommand  *{\diff}   {\mathop{}\!\mathrm{d}}
\renewcommand*{\vec}[1] {\boldsymbol{#1}}
\newcommand  *{\uvec}[1]{\hat{\vec{#1}}}
\newcommand  *{\s}[1]   {\mathsf{#1}}
\newcommand  *{\mat}[1] {\vec{\s{#1}}}
\newcommand  *{\Ups}    {\Upsilon}
\newcommand  *{\The}    {\Theta}
\newcommand  *{\I}      {\mathrm{i}}

\newcommand  *{\Exp}[1] {\mathrm{e}^{\textstyle #1}}
\newcommand  *{\phn}    {\phantom{-}}
\newcommand  *{\pfrac}[2] {\left(\frac{#1}{#2}\right)}
\newcommand   {\mbhm}{\text{M}_{\text{BH}}}
\newcommand   {\mbh}{$\text{M}_{\text{BH}}$}

\newcommand{\va}{$\left(\theta, \phi, \psi\right)$}

\title[Triaxial Schwarzschild Models of BCGs]{Triaxial Schwarzschild Models of Brightest Cluster Galaxies with Long-Slit LBT Data}

\author[S.~de Nicola et al.]{%
Stefano de Nicola,$^{\!\!1,2}$\thanks{E-mail: denicola@mpe.mpg.de}
Roberto P. Saglia,$^{\!\!2,1}$
Jens Thomas,$^{\!\!2,1}$
Jan Snigula,$^{\!\!2}$
Matthias Kluge,$^{\!\!2}$
\newauthor
and Ralf Bender $^{\!\!1,2}$
\\ \\
$^{1}$ Universit{\"a}ts-Sternwarte Muenchen, Scheinerstrasse 1, D-81679, Munich, Germany\\
$^{2}$ Max-Planck Institute for Extraterrestrial Physics, Giessenbachstrasse 1, D-85748, Garching (Germany) \\
}

\date{Accepted --- Received --- in original form ---}

\pubyear{2023}

\begin{document}
\label{firstpage}
\pagerange{\pageref{firstpage}--\pageref{lastpage}}
\maketitle

\begin{abstract}
We present new long-slit stellar kinematics for a sample of 21 Brightest Cluster Galaxies (BCGs) and triaxial Schwarzschild models for 16 of these objects using our orbit modelling code SMART. The new kinematics obtained with the Large Binocular Telescope (LBT) is complemented with high-resolution photometry from HST or new AO-assisted ground-based observations also obtained at LBT and combined with wide-field imaging from the Wendelstein Observatory. These data enable robust modeling from the innermost regions - where the Supermassive Black Hole dominates the potential - to larger radii, where stars and dark matter (DM) are the primary mass contributors. As already discussed in a companion paper, we discovered 8 Ultramassive Black Holes (UMBHs, with mass $> 10^{10}$ M$_\odot$) in this BCG sample, more than doubling the number of galaxies with dynamically detected UMBHs. We show that the DM halos display a wide variety of geometries. Purely kinematical results include low central velocity dispersion with increasing profiles towards the outskirts, and the discovery of one Kinematically Decoupled Core. 

\end{abstract}


%%%%%%%%%%%%%%%%%%%%%%%%%%%%%%%%%%%%%%%%%%%%%%%%%%

%%%%%%%%%%%%%%%%% BODY OF PAPER %%%%%%%%%%%%%%%%%%
%Two phase merging scenario: 1) infalling BHs form the core 2) bound system forms the tangential anisotropy - cite Rantala+18,19; Frigo+21

%Dichotomy ellipticals formation: references in Rantala+18, my thesis
%Boxy-disky: KB96

%In-situ z > 2 (old stellar population), halo accretion z < 1 (dwarfs disruption, tidal stripping from intermediate mass glxs)
%ICL is bluer

%GCs specific frequency higher at larger radii -> trace ICL

\section{Introduction}
\label{Sec.introduction}
Brightest Cluster Galaxies (BCGs) belong to the class of the most massive early-type galaxies (ETGs) found in the Universe. These objects are typically hosted at the cluster centre. Their peculiar evolutionary history, locked with that of their host cluster, makes them tracers of the properties of the cluster itself: they are aligned with the DM halo \citep{Matthias21}, have similar three-dimensional shapes as the halo \citep{dN22BCGs}, and are embedded in the intra-cluster light (ICL), whose origin is believed to lie in ex-situ accreted stellar material by the BCG. The rich (major) merger history of these objects is such that they exhibit different scaling relations with respect to ordinary ETGs—for example, following a different \citet{Faber76} relation \citep{Matthias23BCGs}. \\
Lying at the centre of the cluster, and thus of the potential well, they are in an ideal position to accrete material and feed their Supermassive Black Holes (SMBHs). Their mass can be estimated using the Schwarzschild orbit superposition technique \citep{Schwarzschild79, Schwarzschild93}, where a superposition of stellar orbits under a given gravitational potential is constructed. A typical application of this is the fit to kinematic data, so that the best-fit potential - and thus the mass components - can be recovered. To this extent, it is important to include, along with the stellar component and a central BH, also a DM halo; otherwise, \mbh\,might end up being underestimated \citep{Rusli13DM}. %Determining the stellar mass distribution and, hence, the mass-to-light ratio allows for a comparison with estimates derived directly from spectra by means of Single Stellar Population (SSP) fitting \citep{Thomas03, Maraston05}. It is still a matter of debate whether a \citet{Kroupa01} IMF can be universally assumed to hold for all galaxies, or if bottom-heavy (e.g. \citealt{Salpeter55}) IMFs are indeed observed in massive ETGs. \\
%, yielding a dynamically based determination of the Initial Mass Function (IMF), which overcomes the major limitation of the SSP models—namely, that in elliptical galaxies we only observe spectra coming from old stellar populations
One important assumption concerns the galaxy geometry. Since the pioneering work of \citet{Schwarzschild79}, several codes and/or their applications have appeared in the literature for spherical \citep{Richstone85, Rix97}, axisymmetric \citep{vdM98, Gebhardt00, Jens04, Valluri04}, and triaxial \citep{VDB08, Vasiliev20, Bianca21} geometries. This last case appears to be the best assumption when modeling BCGs, as both the photometry (see \citealt{Matthias20}) and the kinematics exhibit features (e.g. isophotal twists, kinematic misalignment) that would not be observed if these objects were axisymmetric. \citet{dN22BCGs} directly demonstrated this by recovering the intrinsic shape of 57 BCGs. \\
In order to apply triaxial Schwarzschild models to derive a robust black hole mass \mbh\,estimate and to constrain the mass profile, one needs photometric and kinematic data that sample the BH Sphere of Influence (SOI) but also extend to the region where DM dominates. Such photometric data are provided by deep, g$'$-band Wendelstein observations \citep{Hopp10, LangBardl16,Matthias20, Matthias21}, which deliver reliable photometry beyond 100 kpc, as well as by complementary high-resolution HST or AO-aided observations performed at the Large Binocular Telescope Observatory (LBTO). Similarly, we require a fair number of spectra ($>$ 10) inside the BH SOI.  \\
The extension of the BH SOI can be inferred measuring the sizes of cores \citep{Lauer07_BHs, Jens16}. Massive ETGs often show a light-deficient central region - the core - whose properties (amount of missing light, core size, surface brightness of the core itself) correlate well with \mbh\,\citep{Kormendy09, Jens16, Kianusch19}, while canonical scaling relations such as \mbh-$\sigma$ break down at the high-mass end (\citealt{Rob16}, de Nicola et al. submitted). By photometrically identifying candidates with large cores, we selected a sample of 22 galaxies from \citet{Matthias20} whose core sizes suggest $\mbhm > 10^{10}$ M$_\odot$, a population that is almost entirely unstudied \citep{Liepold24}. \\
Having obtained long-slit kinematics at LBT, along with photometric data satisfying the requirements described above, the main goal of this paper is to publish the measured kinematicss and present the results of triaxial Schwarzschild modeling for the BCG sample. To this extent, we use our triaxial Schwarzschild code SMART \citep{Bianca21} which can handle non-parametric LOSVDs and uses an information based generalised model selection framework to obtain robust mass models \citep{Mathias21,Jens22}. Having already applied this setup to the dynamical modeling of NGC~708, the BCG of A262 \citep{dN24}, which led to the discovery of a 10$^{10}$\,M$_\odot$\,BH, we take that study as a reference for our analysis of the other 21 BCGs. Other galaxies analysed with information-based triaxial models include NGC5419 \citep{Bianca23N5419} and NGC~1272 \citep{Rob24}. A detailed discussion about BHs and their correlations with host galaxies is presented in a companion paper (de Nicola et al. submitted), while deeper analyses of the stellar component (and, hence, the comparison between dynamical mass-to-light ratios and those derived from stellar population analysis) and the properties of the DM halos will be the subject of future works. The paper is structured as follows. Sections 2 and 3 focus on the photometric and kinematic data, respectively. Section 4 presents the results of the dynamical models, whereas Section 5 includes comments on the individual objects. Finally, Section 6 contains our summary and conclusions. Throughout the paper, we adopt the convention of identifying the BCG with the host cluster name. Moreover, we use arcseconds to measure projected quantities on the plane of the sky and kpc for intrinsic ones.

%In this work, however, we follow \citet{Matthias20} in identifying the BCG with the kinematical centre of the cluster, which can differ from the geometrical one (e.g. the Virgo Cluster).  \\

\begin{figure*}

\subfloat[\label{Fig.ell_hist}]{\includegraphics[width=.35\linewidth]{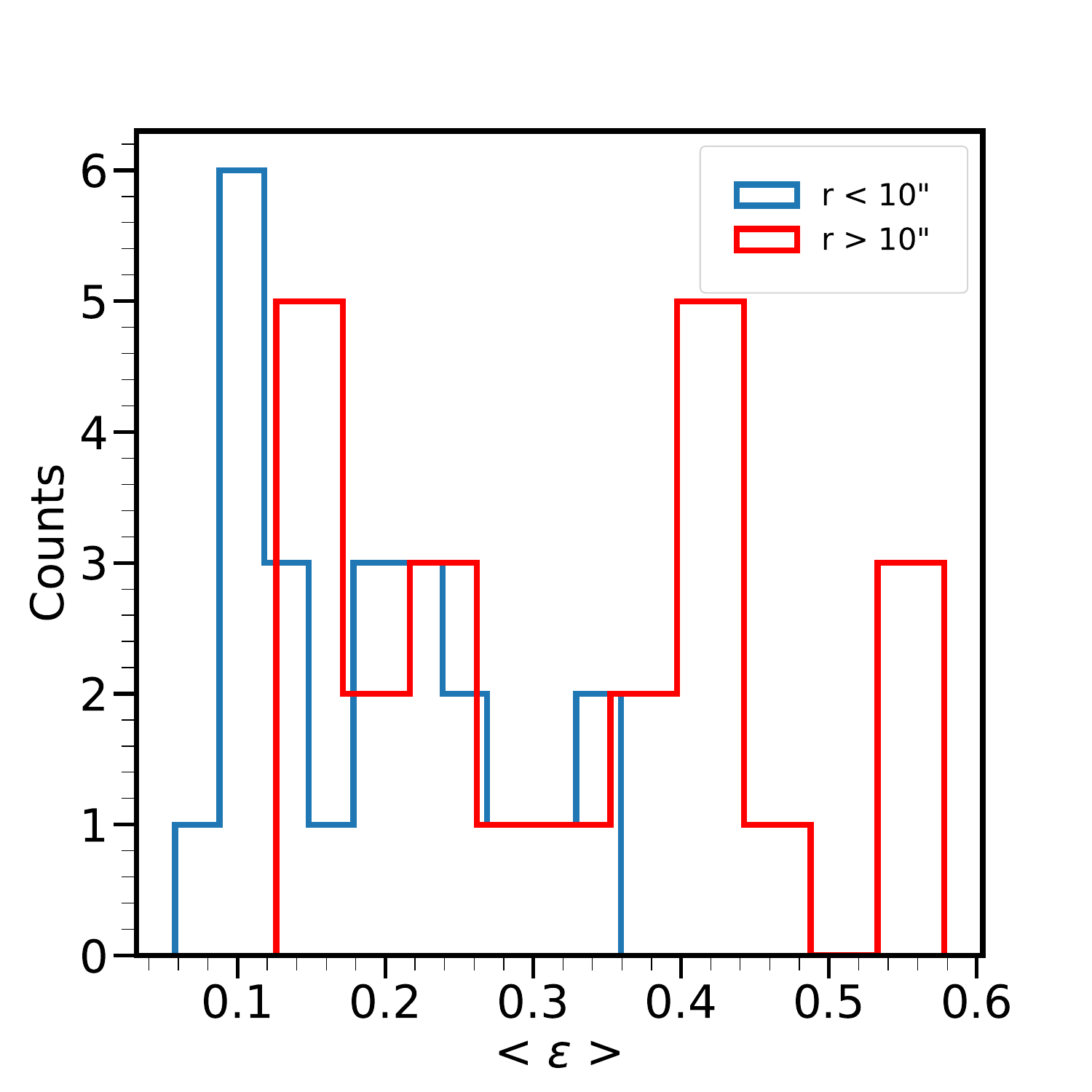}}
\subfloat[\label{Fig.PA_hist}]{\includegraphics[width=.35\linewidth]{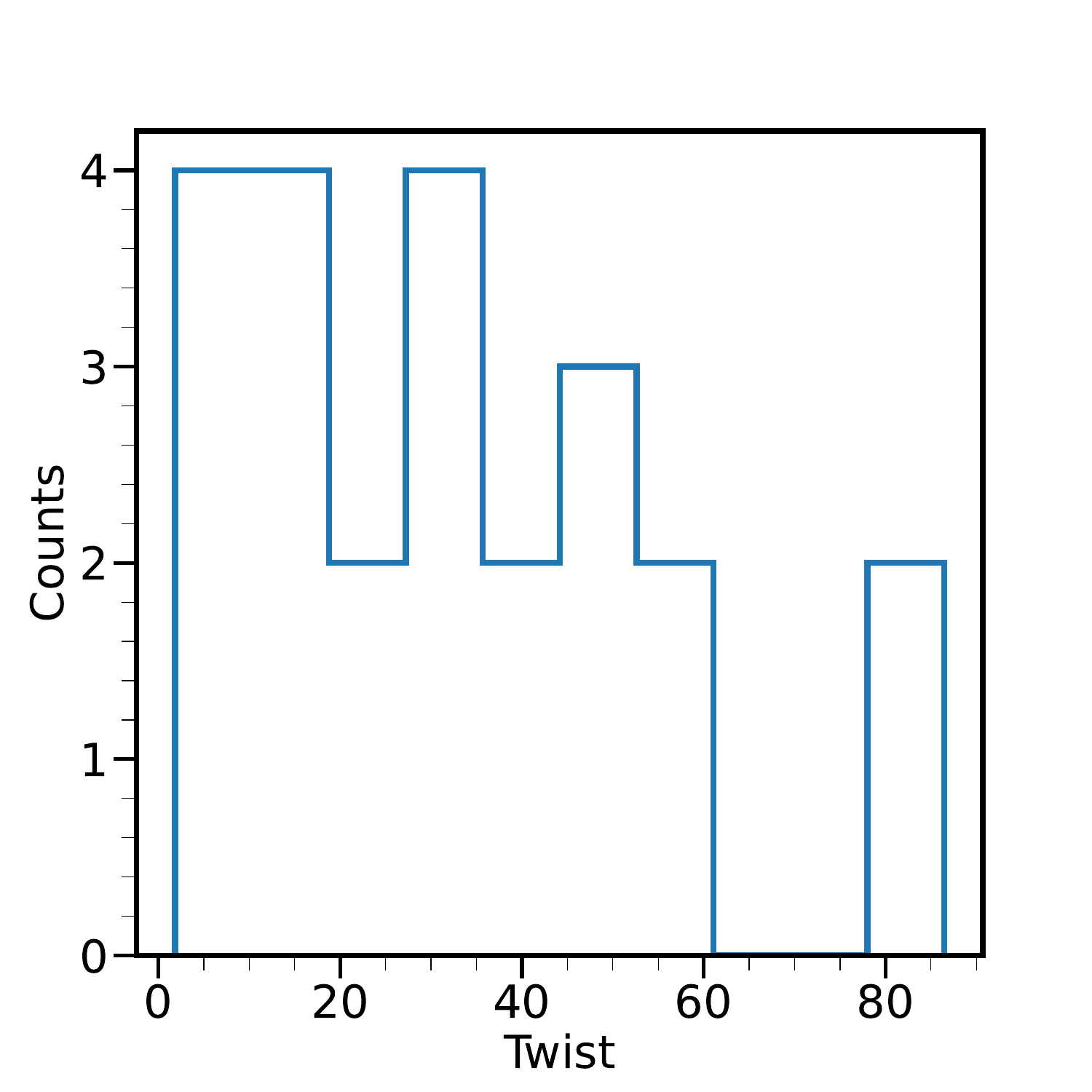}}

    \caption{Left: Histogram of the average ellipticity within and beyond 10", showing that our BCGs become flatter at large radii, although some objects remain round or flat at all radii. Right: Histogram of the twist, computed as max(PA) $-$ min(PA), showing that, in contrast to ordinary ETGs, BCGs often exhibit large twists.}
    \label{Fig.ell_PA}
\end{figure*}

%Check: MAC scripts + N708 paper
\section{Photometry}  \label{Sec.photometry}

In this section, we describe the photometric data used in this work. It largely builds on \citet{dN20, dN24} and addresses general aspects of the data and the deprojection pipeline. Details on individual objects can be found in Sec.~\ref{Sec.comments}.

%https://scienceops.lbto.org/luci/instrument-characteristics/filters/
\subsection{Data} \label{Ssec.phot_data}
The photometric observations are extensively described in \citet{dN24} (see also \citealt{Matthias20} for technical details). In brief, the $g'$-band wide-field photometry was obtained using images collected at the Wendelstein Observatory with the 2.2m Fraunhofer Telescope \citep{Hopp10, LangBardl16}. The total field of view (FOV) of $27.6 \times 28.9$ arcminutes allows for reliable photometry beyond 100 kpc. For A262, high-resolution photometry was obtained from WFPC2 HST observations, using the F555W, F814W, and F110W filters. The same applies to the BCGs in A160, A634, A1314, A1775, and A2147, observed with the F814W filter. The typical resolution is $\sim$0.1", and the camera features a grid of $800 \times 800$ pixels with a pixel size of 0.0455", yielding a FOV of $36.4 \times 36.4$". \\
All other BCGs - with the exception of A292 and A1982, for which we rely on PSF-deconvolved Wendelstein data only - were observed at the LBT using Adaptive Optics (AO), via the LUCI-I instrument\footnote{The LUCI-I and LUCI-II instruments are nearly identical near-infrared (NIR). The AO is only available on LUCI-I.}. We observed the galaxies first using the H-band ($\lambda_\mathrm{c} = 1.653,\mu$m, FWHM = 0.301 $\mu$m) and then the Ks-band\footnote{With the exception of A1185 and A2388, we always use this band for the deprojection.} ($\lambda_\mathrm{c} = 2.163,\mu$m, FWHM = 0.270 $\mu$m). The pixel scale is 0.0153". For each band we acquire 10 images with 60 second exposures, along with 5 sky exposures with the same exposure time, for a total of 30 minutes exposure time for almost every galaxy. The AO system requires a guide star\footnote{A compact elliptical galaxy, often found near a BCG, can also be used.} brighter than $m_\text{R} = 16.0$ mag and within 1' of the galaxy. If such a star is not available, we rely on the Enhanced Seeing Mode (ESM), which does not provide diffraction-limited correction but still improves the PSF \citep{Rothberg19} and permits guide star selection over a much larger area. In this case, the pixel scale is 0.1216". Given the $4 \times 4$ arcminute FOV, we do not acquire sky exposures and simply dither, for a total of 10 minutes exposure time per each band. Finally, to estimate the PSF for the AO observations we need to observe a star for which the distance to the AO guide star is roughly the same as that between the galaxy and the AO guide star. If such star is not available, we adopt a typical PSF of 0.3". Instead, for ESM this is not an issue, in that the large FOV always allows to find a star. \\
The modeling of HST/LBT and WWFI data is performed by identifying a suitable radial interval [$\mathrm{r}_\mathrm{min}$, $\mathrm{r}_\mathrm{max}$] where the high-resolution observations overlap with the Wendelstein data and interpolating the high-resolution photometry over to Wendelstein radii $r_W$. We minimize

\begin{equation}
\mathrm{I}_\mathrm{lr} (r_W) = \mathrm{I}_\mathrm{hr} (r_W) \times C + K_{sky}
\label{eq.high_low_phot}
\end{equation}

\noindent where $C$ is a scaling factor, $K_{sky}$ is an additive constant depending on the background sky and $\mathrm{I}_\mathrm{lr}, \mathrm{I}_\mathrm{hr}$ are the low- (i.e. WWFI) and high-resolution (i.e. HST or LBT) surface intensities. We then use the resulting coefficients to convert the HST or LBT data to the $g'$-band. After the conversion, we adopt high-resolution photometry for $\mathrm{r} < \mathrm{r}_\mathrm{match}$ and WWFI data for $r > \mathrm{r}_\mathrm{match}$, providing what is needed for robust deprojection and, subsequently, for dynamical modeling. The exact $\mathrm{r}_\mathrm{match}$, $\mathrm{r}_\mathrm{min}$, $\mathrm{r}_\mathrm{max}$ and PSF values are reported in Tab.~\ref{Tab.photometry}. To assess the contribution of intracluster light (ICL) both in the region used for deprojection and in the area covered by our kinematical data - the latter being particularly relevant for our dynamical models - we follow \citet{Matthias21} (see App.~\ref{App.ICL}). Our analysis shows that ICL contamination is mild in the photometry and negligible for the kinematics.\\
Finally, in Fig.~\ref{Fig.ell_hist} we show that while BCGs are typically round in the central regions and become flatter at large radii, some are already flat in the center, while others remain very round, whereas in Fig.~\ref{Fig.PA_hist} it can be seen that BCGs exhibit (very) large twists, in contrast to ordinary ETGs, where twists are typically mild \citep{Bender88, Bender92, Kormendy96}. 

\begin{table}
  \centering
  \begin{tabular}{cccc}
    \hline
    Galaxy & Photometry & PSF FWMH (arcsec) & ($\mathrm{r}_\mathrm{min}$, $\mathrm{r}_\mathrm{max}$, $\mathrm{r}_\mathrm{match}$) (arcsec) \\ \hline
    A150 & AO & 0.3 & (3,5,10) \\
    A160 & HST & 0.15 & (5,15,25) \\
    A240 & AO & 0.3 & (3,5,10) \\
    A292 & WWFI & 1.2 & - \\
    A399 & AO & 0.3 & (3,5,10) \\
    A592 & AO & 0.17 & (3,7,10) \\
    A634 & HST & 0.15 & (5,15,25) \\
    A688 & AO & 0.3 & (3,5,10) \\
    A1185 & AO & 0.3 & (3,5,10) \\
    A1314 & HST & 0.15 & (5,15,25) \\
    A1749 & AO & 0.3 & (3,5,10) \\
    A1775 & HST & 0.15 & (5,15,25) \\
    A1982 & WWFI & 1.2 & - \\
    A2107 & ESM & 0.4 & (3,6,10) \\
    A2147 & HST & 0.15 & (5,15,25) \\
    A2255 & ESM & 0.45 & (3,6,12) \\
    A2256 & AO & 0.3 & (3,7,10) \\
    A2319 & AO & 0.3 & (3,5,12) \\
    A2388 & ESM & 0.67 & (3,6,6) \\
    A2506 & ESM & 0.85 & (3,6,6) \\
    A2665 & AO & 0.3 & (3,5,10) \\

    \hline
  \end{tabular}
  \caption{Note. \textit{Col. 1}: BCG name; \textit{Col. 2}: Whether the high-resolution photometry comes from HST, LBT (AO or ESM) or WWFI. \textit{Col. 3}: FWHM of the PSF. The WWFI images of A292 and A1982 have been PSF-deconvolved before using them. \textit{Col. 4}: The three values $\mathrm{r}_\mathrm{min}$, $\mathrm{r}_\mathrm{max}$ and $\mathrm{r}_\mathrm{match}$, where $\mathrm{r}_\mathrm{min}$, $\mathrm{r}_\mathrm{max}$ enclose the interval where high- and low-resolution data are matched and $\mathrm{r}_\mathrm{match}$ is the radius from which we start using low-resolution data.}
  \label{Tab.photometry}
\end{table}

\subsection{Core radii} \label{Ssec.core_radii}
As stated in the introduction, the core size is a robust estimator of the BH mass: it tightly correlates with \mbh\,\citep{Jens16} and can be used at the high-mass end of the BH-hosts scaling relations where the canonical \mbh-$\sigma$ relation breaks down (\citealt{Rob16}, de Nicola et al. submitted). The core size, quantified in terms of the core radius r$_\mathrm{c}$, can be measured by fitting a Nuker \citep{Lauer95} or a Core-S{\'e}rsic (CS, \citealt{Graham03, Trujillo04}) profile to the observed Surface Brightness (SB). In this case, the core radius corresponds to the break radius r$_\mathrm{b}$ of the profile. Another way of estimating the core size is the cusp radius r$_\gamma$ \citep{Carollo97}, defined as dSB/dlogr (r$_\gamma$) = -1/2. In this work, we employ PSF-convolved CS fits for all galaxies, where the adopted PSF depends on the individual object (see Tab.~\ref{Tab.photometry}). We describe the procedure in App.~\ref{App.core_size_PSF}. \\
The results are shown in Fig.~\ref{Fig.cores} and reported in Tab.~\ref{Tab.core_fit}. For nearly all galaxies the central slope $\gamma$ is smaller than 0.5, indicating that the SB profile does indeed become flatter with respect to the outer parts. The only galaxy for which this is not the case is A2107: here, the central slope $\gamma_\text{CS}$ is too large. We thus classify the galaxy as power-law elliptical. Instead, for two galaxies (A292 and A1185) we could not obtain a reliable CS fit and adopt r$_\gamma$ as core size estimate. \\
The large number of parameters of the CS profile can lead to degeneracies and/or to implausible values/error bars. In these cases, we try different fits varying the outermost radius, finding that neither the core radius r$_\mathrm{b}$ nor the central slope $\gamma_\mathrm{CS}$ are significantly affected by this choice. Moreover, at large radii BCGs exhibit almost power-law SB profiles \citep{Matthias23BCGs} which can cause large $n$, r$_\mathrm{e}$ and $\mu_\mathrm{e}$ values. \\
For 7 BCGs in our sample (A160, A1314, A2107, A2147, A2319, A2388, and A2665), \citet{Matthias20} found that adding an additional S{\'e}rsic component provides a better reproduction of the data. In \citet{Matthias21} (see also \citealt{dN25}) it is suggested that this second component typically traces the ICL. Since we fit the entire SB profile, thus extending into the regions where the ICL contribution becomes significant, we fit a Core-S{\'e}rsic + S{\'e}rsic model to these 6 galaxies (A2107 does not have a core), presenting the results in App.~\ref{App.CSS_fit}. With the exception of A2388, for which we could not obtain a converging CSS fit, the break radii are always consistent within 2$\sigma$.\\
Finally, for the BCGs A399, A592, A688, A1982, A2255, A2319, and A2506 we find large values of $\alpha_\text{CS}$, signaling a sharp transition between the core and the S{\'e}rsic-dominated regions. Although Fig.~\ref{Fig.cores} does not show any obvious discontinuity in the profiles, we produce CS fits for these 7 cases by fixing $\alpha_\text{CS} = 5$. The results, presented in App.~\ref{App.alphaCS_fit}, show no significant differences. The only exception is A688, which indeed appears to require a large $\alpha_\text{CS}$ value.

%Errors with 2 sign. digits
\begin{table*}
  \centering
  \begin{tabular}{ccccccc}
    \hline
    Galaxy & r$_\mathrm{e}$ (arcsec) & $\mu_\mathrm{e}$ (mag/arcsec$^2$) & $n$ & r$_\mathrm{b}$ (arcsec) & $\gamma_\mathrm{CS}$ & $\alpha_\mathrm{CS}$ \\ \hline
    A150 & $85.1 \pm 5.8$ & $26.33 \pm 0.14$ & $6.47 \pm 0.62$ & $0.53 \pm 0.11$ & $0.00^{+0.12}_{-0.00}$ & $1.20 \pm 0.33$ \\
    A160 & $ 271\pm 87 $ & $27.97 \pm 0.60$ & $ 8.73 \pm 0.99 $ & $1.014 \pm 0.12$ & $0.068 \pm 0.052$ & $4.5 \pm 3.1$ \\
    A240 & $50.3 \pm 1.5$ & $25.62 \pm 0.062$ & $5.10 \pm 0.53$ & $0.559 \pm 0.044$ & $0.170 \pm 0.031$ & $2.14 \pm 0.36$ \\
    A399 & $49.8 \pm 1.2$ & $25.20 \pm 0.051$ & $2.910 \pm 0.064$ & $1.10 \pm 0.15$ & $0.137 \pm 0.021$ & $63 \pm 14$ \\
    A592 & $186 \pm 27$ & $27.06 \pm 0.28$ & $4.71 \pm 0.29$ & $0.45 \pm 0.19$ & $0.14^{+0.16}_{-0.14}$ & $13 \pm 12$ \\
    A634 & $38.1 \pm 5.3$ & $23.92 \pm 0.34$ & $6.52 \pm 0.98$ & $0.45 \pm 0.16$ & $0.00^{+0.27}_{-0.00}$ & $0.99 \pm 0.44$ \\
    A688 & $15.48 \pm 0.28$ & $24.654 \pm 0.036$ & $3.364 \pm 0.034$ & $0.163 \pm 0.017$ & $0.221 \pm 0.091$ & $32 \pm 24$ \\
    A1314 & $4861 \pm 3896$ & $33.0 \pm 1.6$ & $20.0 \pm 2.6$ & $1.118 \pm 0.029$ & $0.326 \pm 0.020$ & $7.4 \pm 1.7$ \\
    A1749 & $131 \pm 20$ & $27.13 \pm 0.32$ & $14.3 \pm 1.2$ & $0.772 \pm 0.096$ & $0.12 \pm 0.11$ & $2.72 \pm 0.71$ \\
    A1775 & $116.6 \pm 8.7$ & $26.82 \pm 0.15$ & $7.37 \pm 0.35$ & $1.480 \pm 0.061$ & $0.118 \pm 0.013$ & $4.7 \pm 1.2$ \\
    A1982 & $32.9 \pm 1.6$ & $24.23 \pm 0.10$ & $4.40 \pm 0.23$ & $1.098 \pm 0.082$ & $0.127 \pm 0.079$ & $130 \pm 66$ \\
    %A2107 & $66.0 \pm 3.7$ & $24.93 \pm 0.079$ & $4.497 \pm 0.090$ & $0.48 \pm 0.43$ & $0.61 \pm 0.34$ & $104 \pm 27$ \\
    A2147 & $152 \pm 17$ & $25.98 \pm 0.22$ & $5.44 \pm 0.35$ & $0.99 \pm 0.15$ & $0.193 \pm 0.038$ & $4.5 \pm 4.0$ \\
    A2255 & $73.5 \pm 8.8$ & $26.30 \pm 0.23$ & $6.61 \pm 0.35$ & $0.707 \pm 0.073$ & $0.30 \pm 0.11$ & $12 \pm 12$ \\
    A2256 & $119 \pm 30$ & $26.99 \pm 0.51$ & $10.2 \pm 1.7$ & $1.99 \pm 0.11$ & $0.058\ \pm 0.029$ & $1.93 \pm 0.21$ \\
    A2319 & $101.5 \pm 5.9$ & $26.07 \pm 0.12$ & $4.27 \pm 0.11$ & $0.337 \pm 0.054$ & $0.081 \pm 0.067$ & $12 \pm 12$ \\
    A2388 & $9451 \pm 8438$ & $35.6 \pm 1.6$ & $19.8 \pm 2.1$ & $0.841 \pm 0.073$ & $0.362 \pm 0.088$ & $3.14 \pm 0.62$ \\
    A2506 & $53.3 \pm 1.8$ & $24.54 \pm 0.073$ & $5.40 \pm 0.13$ & $0.696 \pm 0.099$ & $0.40 \pm 0.12$ & $22 \pm 17$ \\
    A2665 & $192 \pm 28$ & $27.57 \pm 0.30$ & $7.26 \pm 0.64$ & $0.675 \pm 0.095$ & $0.000^{+0.088}_{-0.000}$ & $1.66\pm 0.44$ \\

    \hline
  \end{tabular}
  \caption{Best-fit Core-S{\'e}rsic parameters. \textit{Col. 1}: BCG name; \textit{Col. 2}: Effective radius; \textit{Col. 3}: Effective surface brightness; \textit{Col. 4}: S{\'e}rsic index; \textit{Col. 5}: break radius, which we use as indicator of the core size; \textit{Col. 6}: central slope $\gamma$; \textit{Col. 7}: parameter $\alpha$ specifying the sharpness of the cutoff from the central region to the outer S{\'e}rsic profile (the higher $\alpha$, the sharper the cutoff).} %\R{Comparison with Matthias: re way too low + some absurd mue and n values. re issue worries me a little, but the fits are good it seems}
  \label{Tab.core_fit}
\end{table*}

\begin{figure*}

\subfloat{\includegraphics[scale=.19]{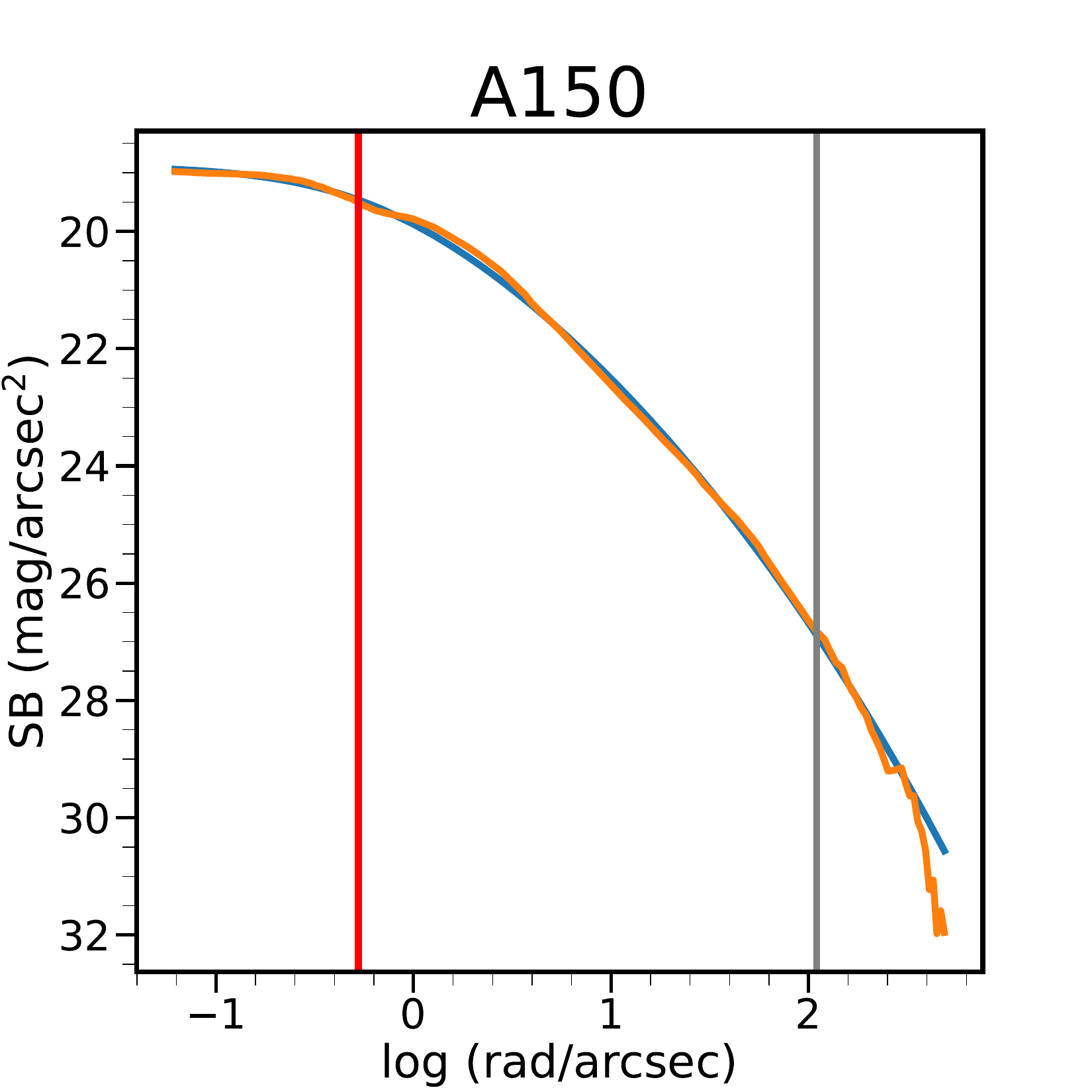}}
\subfloat{\includegraphics[scale=.19]{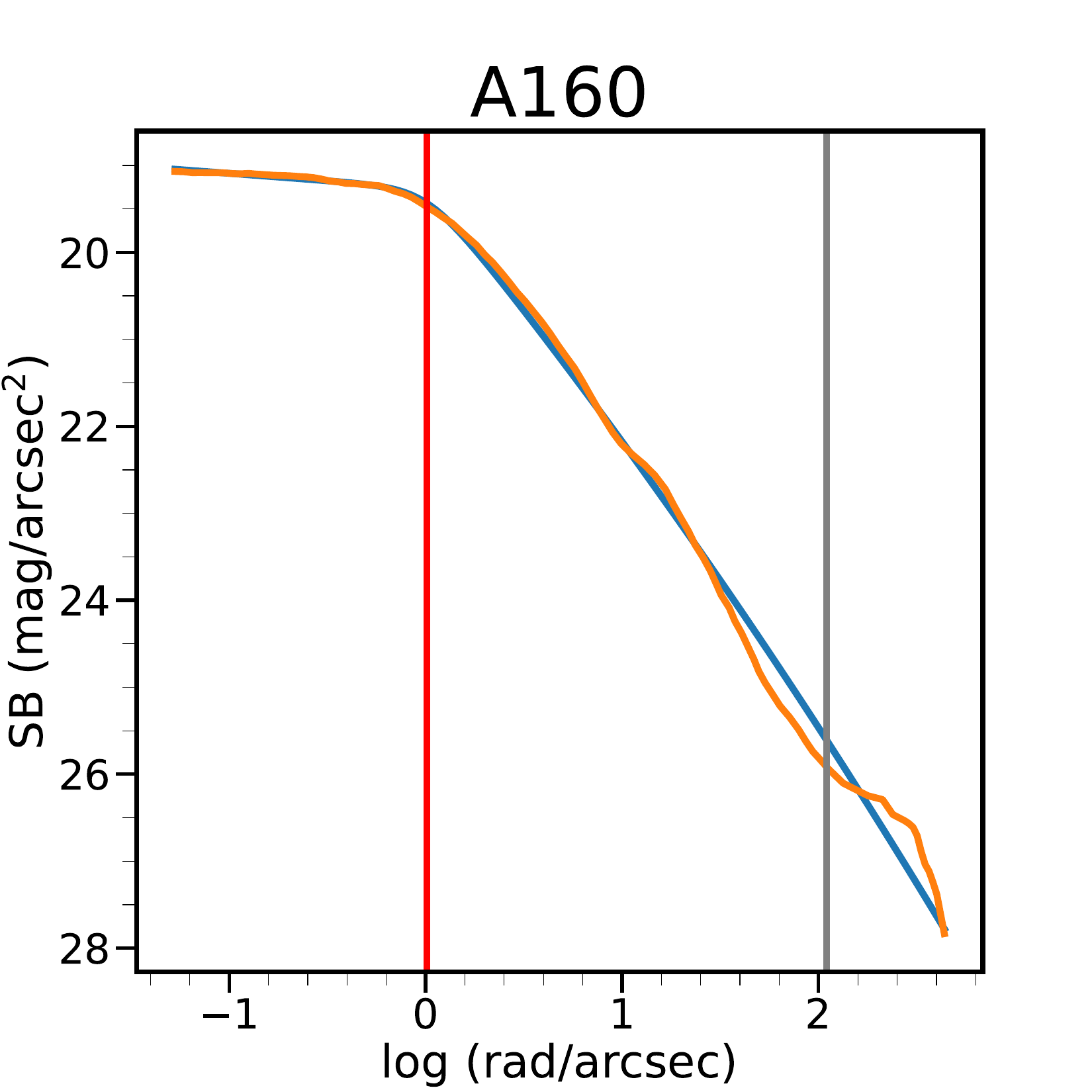}}
\subfloat{\includegraphics[scale=.19]{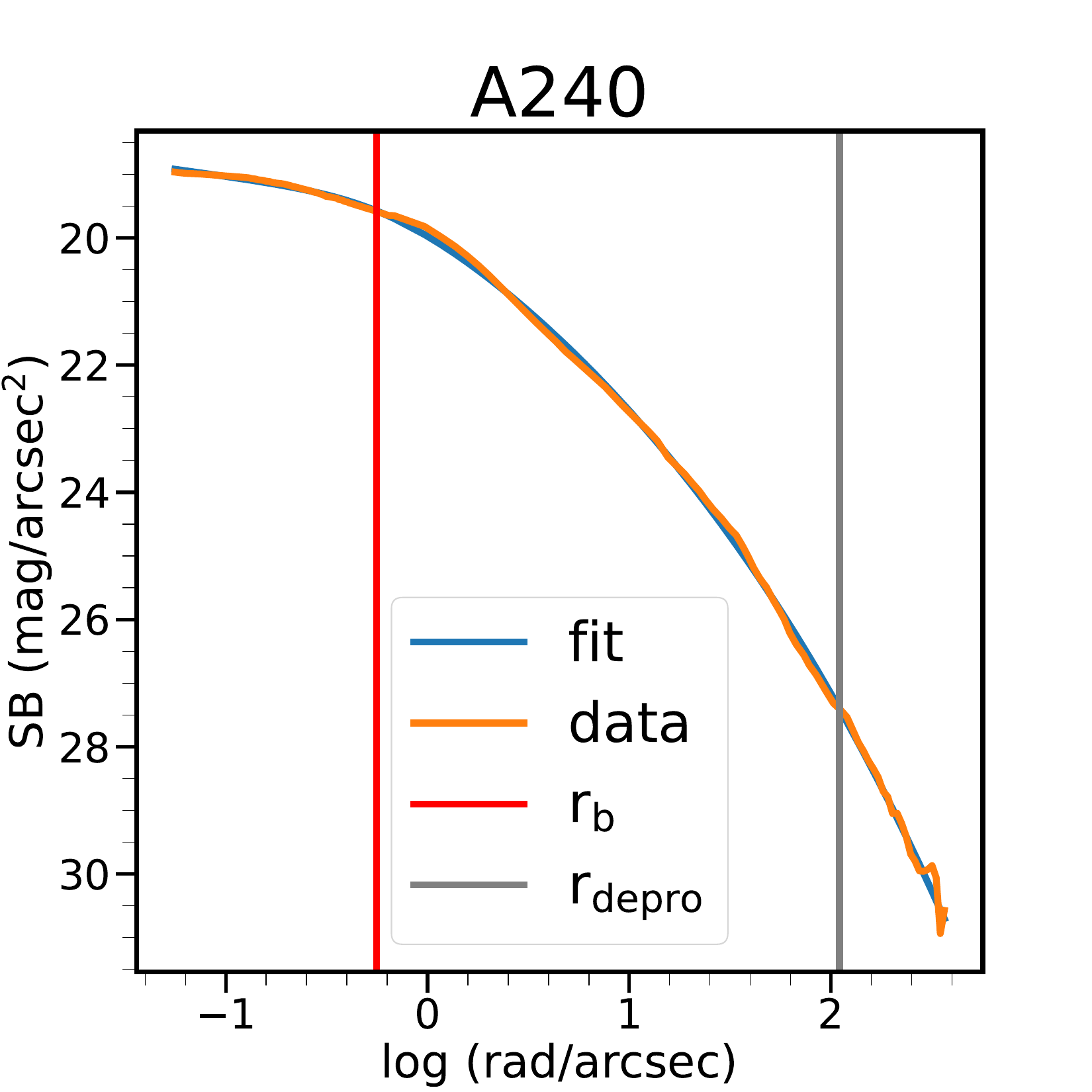}}

\subfloat{\includegraphics[scale=.19]{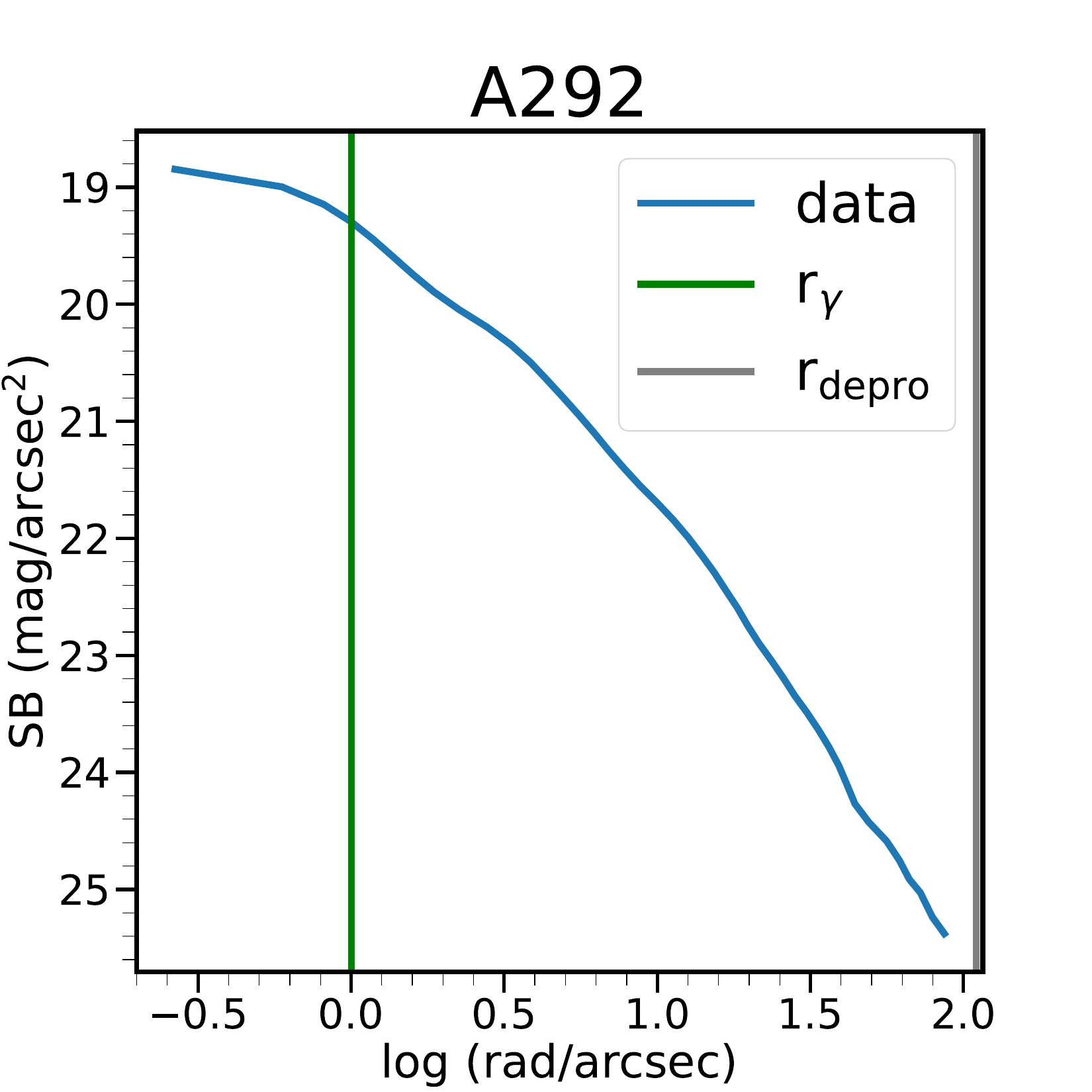}}
\subfloat{\includegraphics[scale=.19]{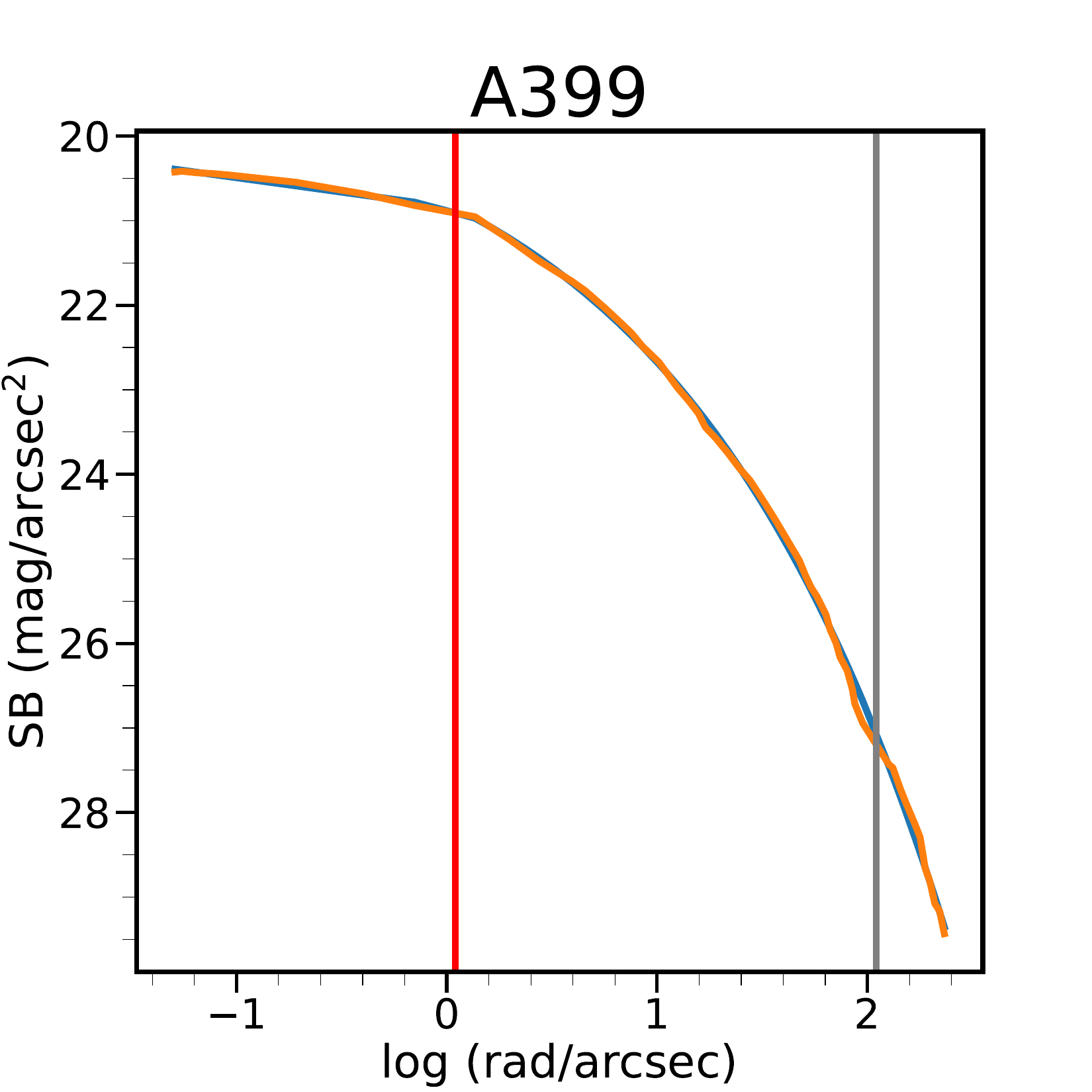}}
\subfloat{\includegraphics[scale=.19]{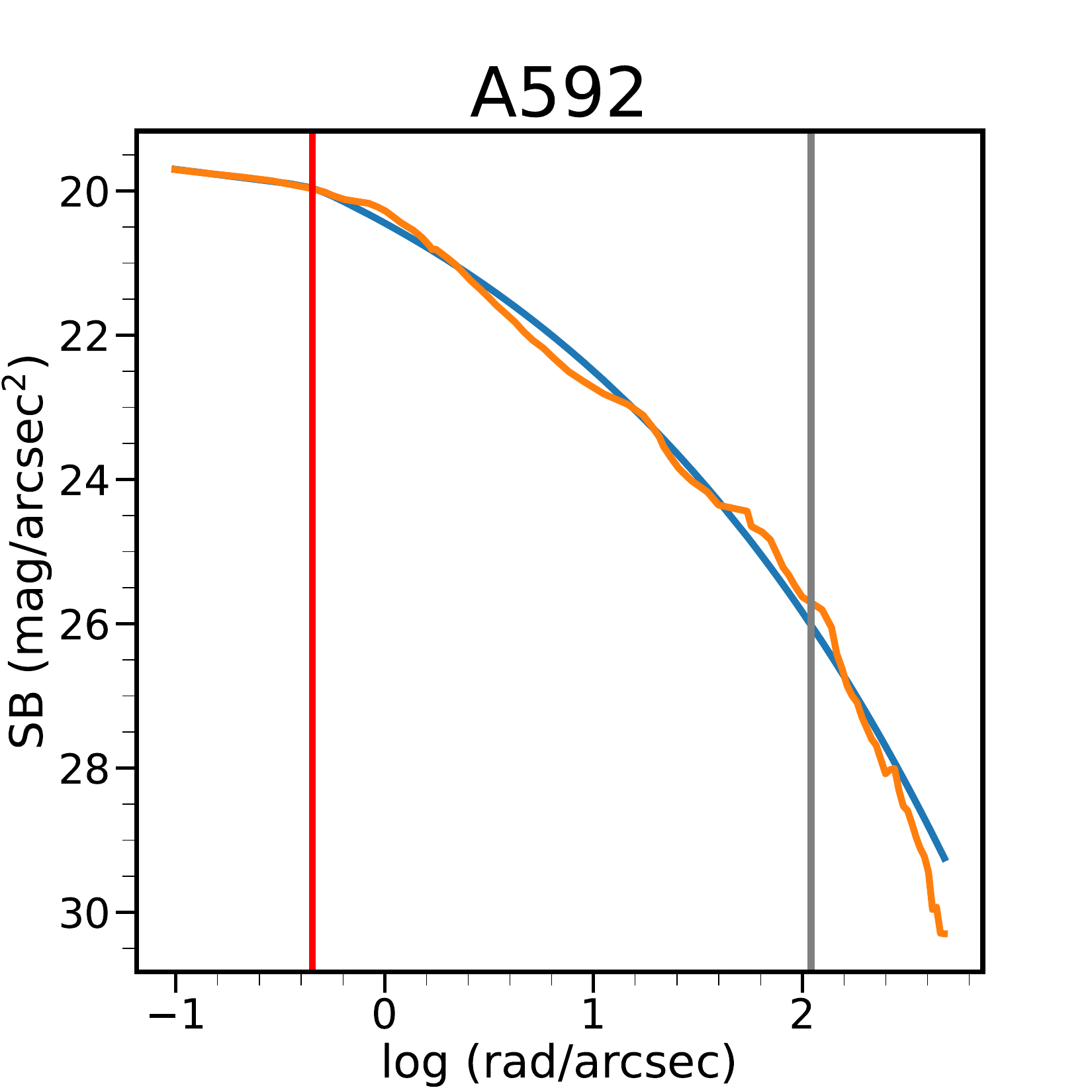}}

\subfloat{\includegraphics[scale=.19]{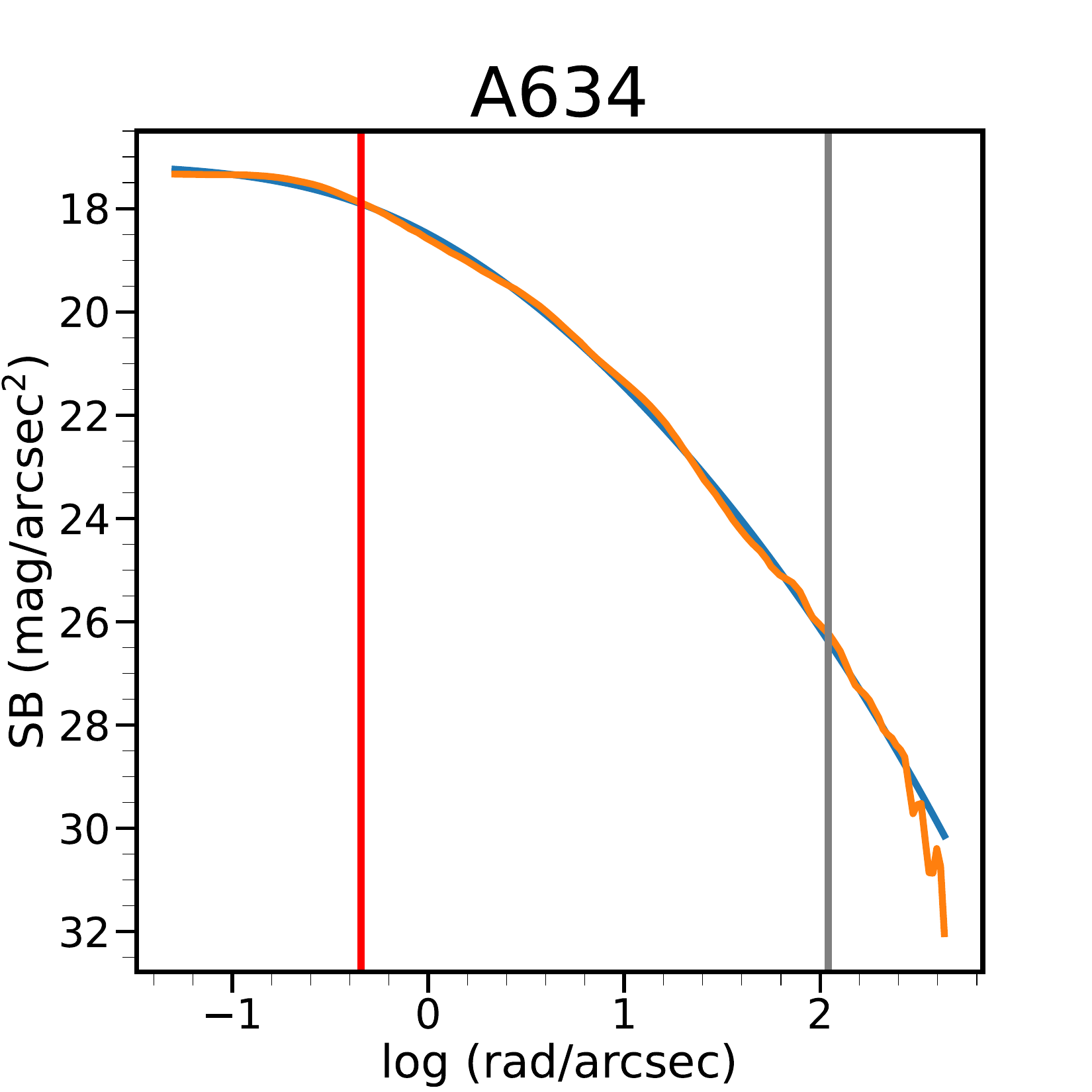}}
\subfloat{\includegraphics[scale=.19]{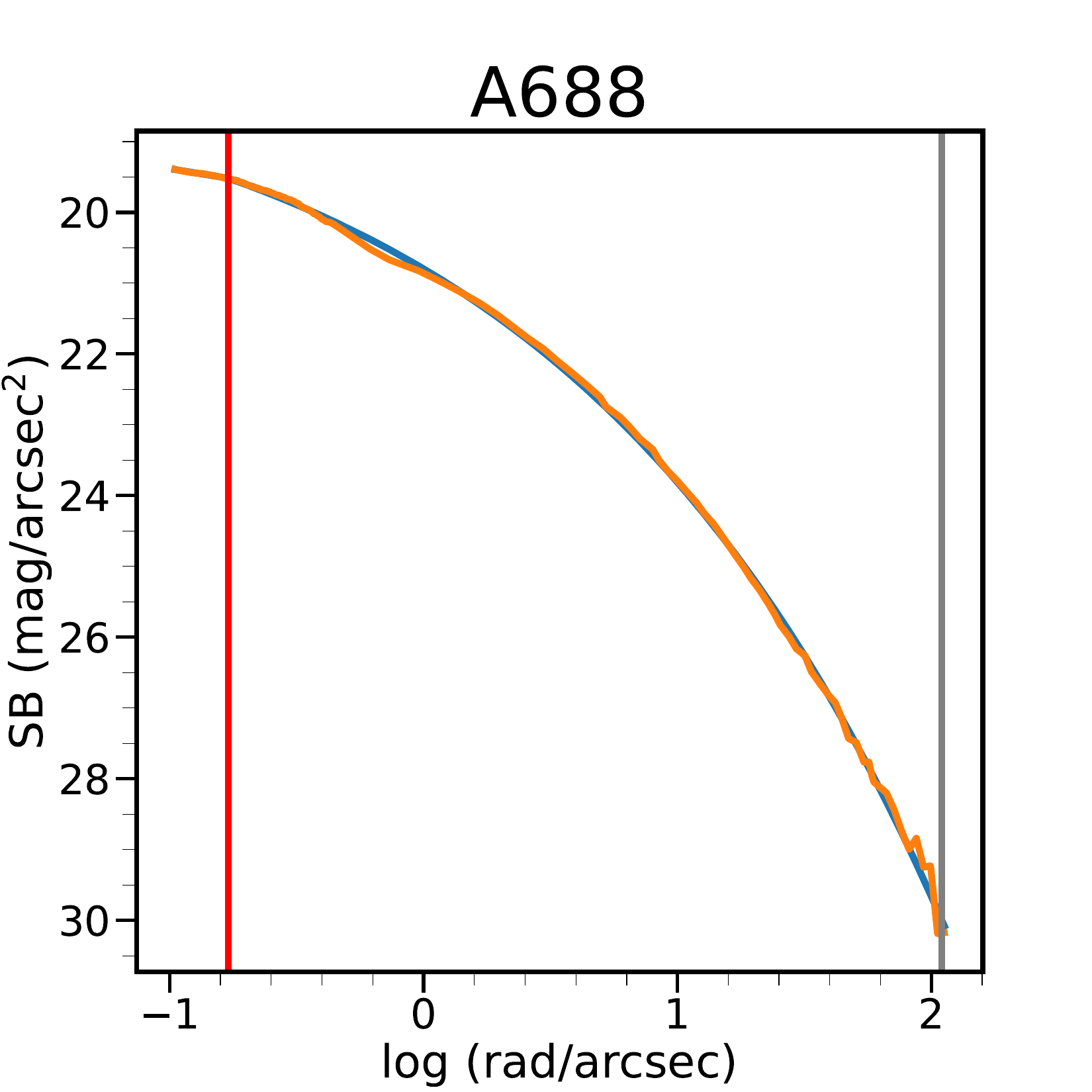}}
\subfloat{\includegraphics[scale=.19]{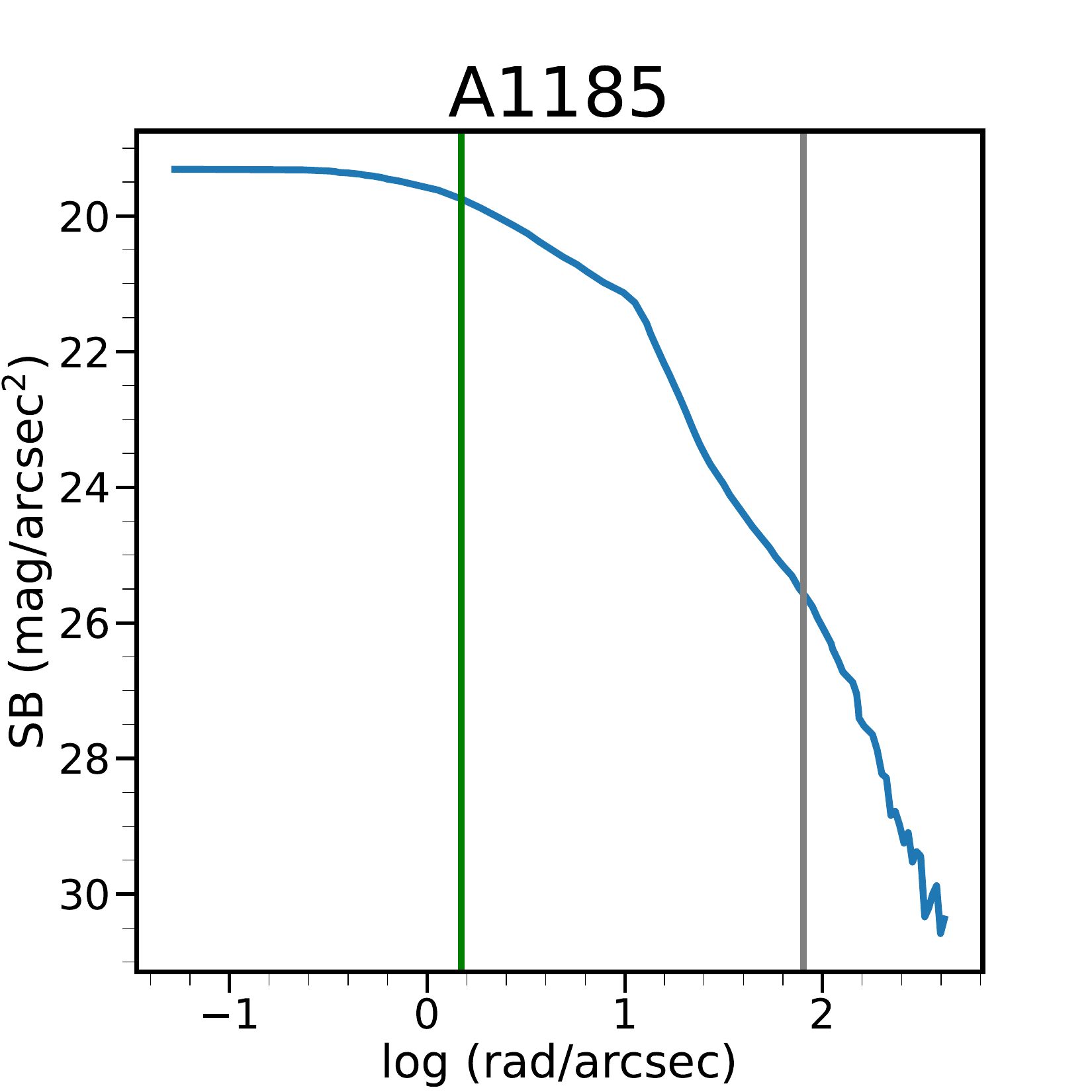}}

\subfloat{\includegraphics[scale=.19]{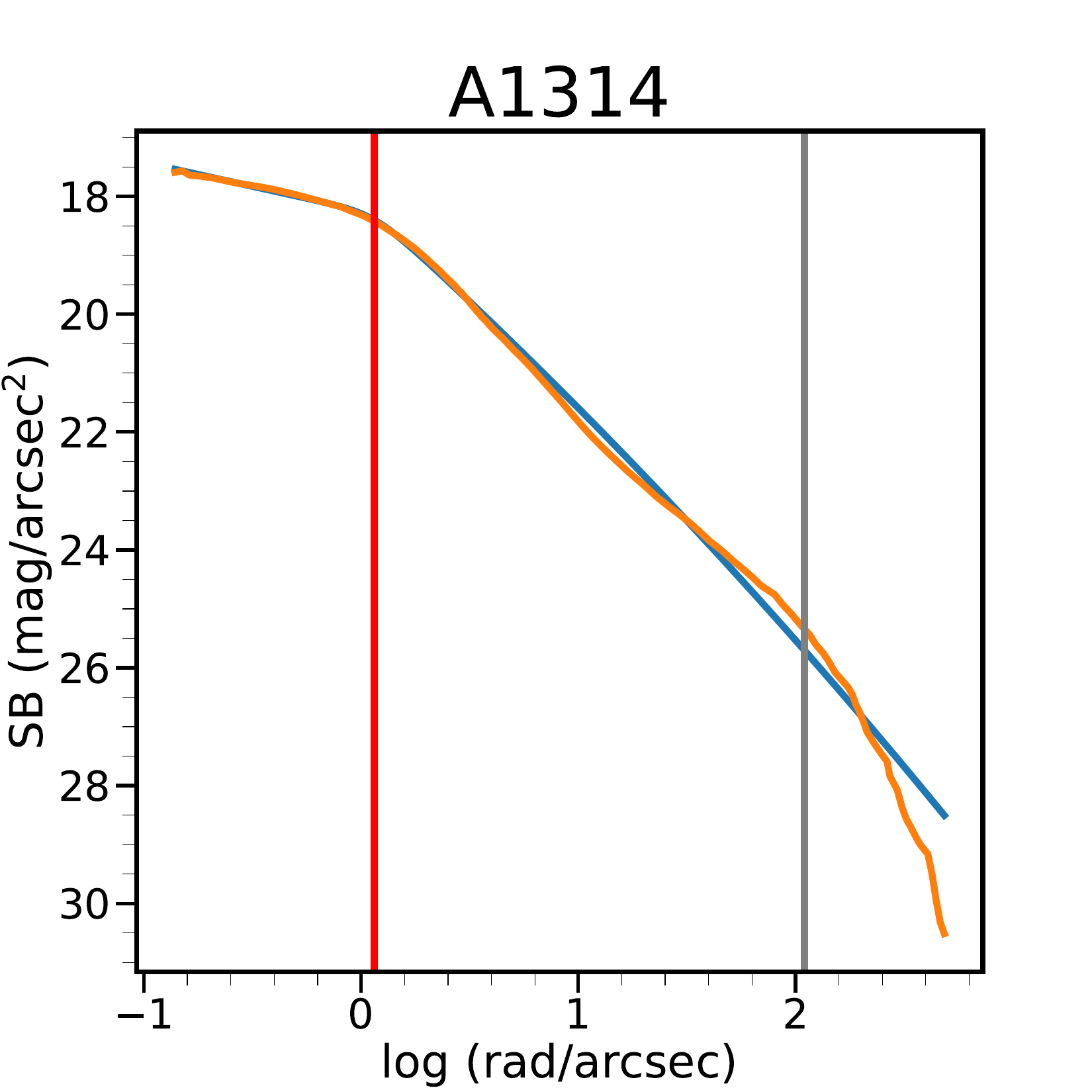}}
\subfloat{\includegraphics[scale=.19]{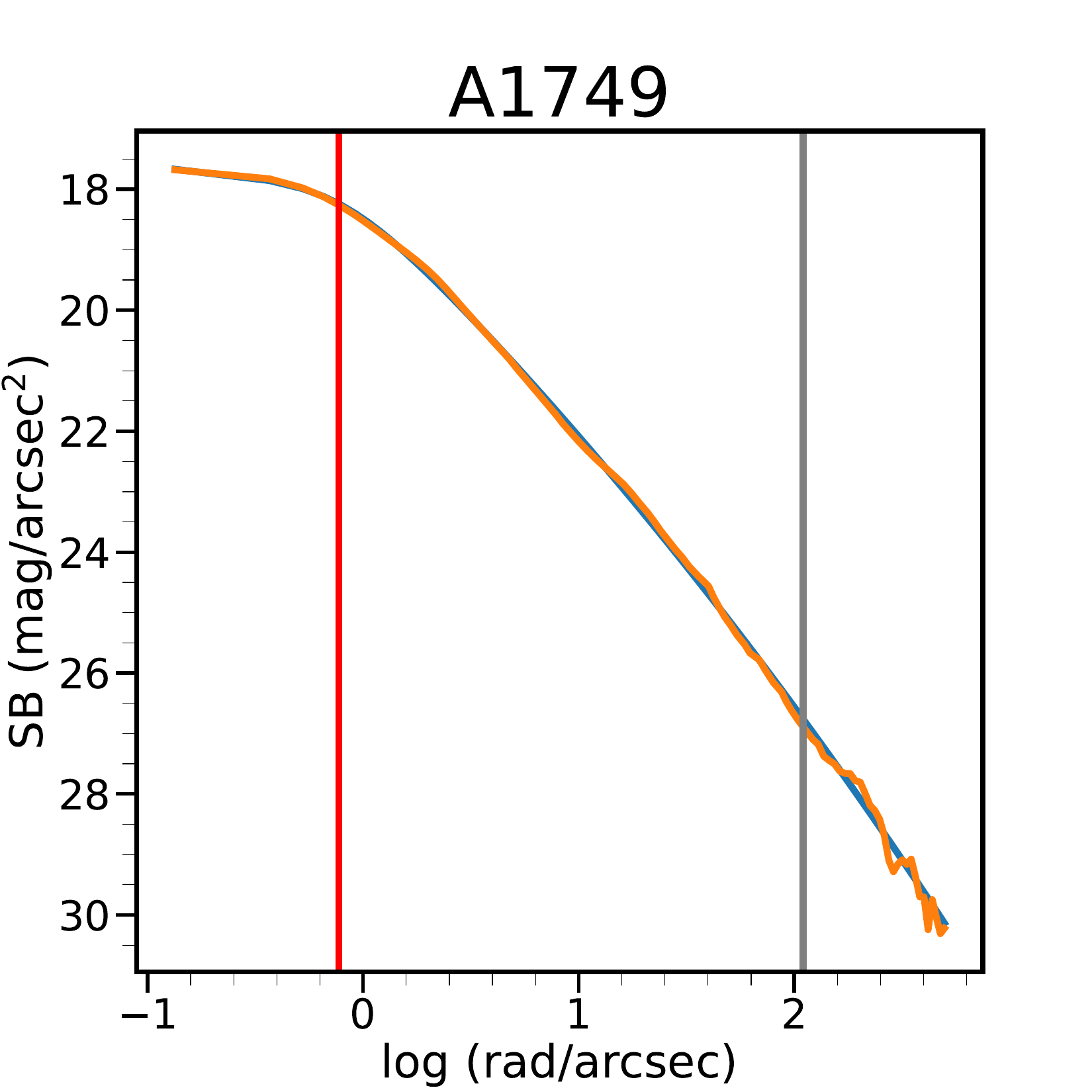}}

    \caption{Surface Brightness profiles for the 21 BCGs of the sample (orange lines) with superimposed Core-S{\'e}rsic best-fit profiles (blue lines). The red lines mark the postion of the break radius r$_\mathrm{b}$, which we use to estimate the core size. For A292 and A1185 we show the cusp radius r$_\gamma$ as vertical green line, while A2107 (see next page) can only be modeled using a single S{\'e}rsic profile. Finally, the gray line denotes the radius where we stop the deprojection.}
    \label{Fig.cores}
\end{figure*}

\begin{figure*}

\subfloat{\includegraphics[scale=.19]{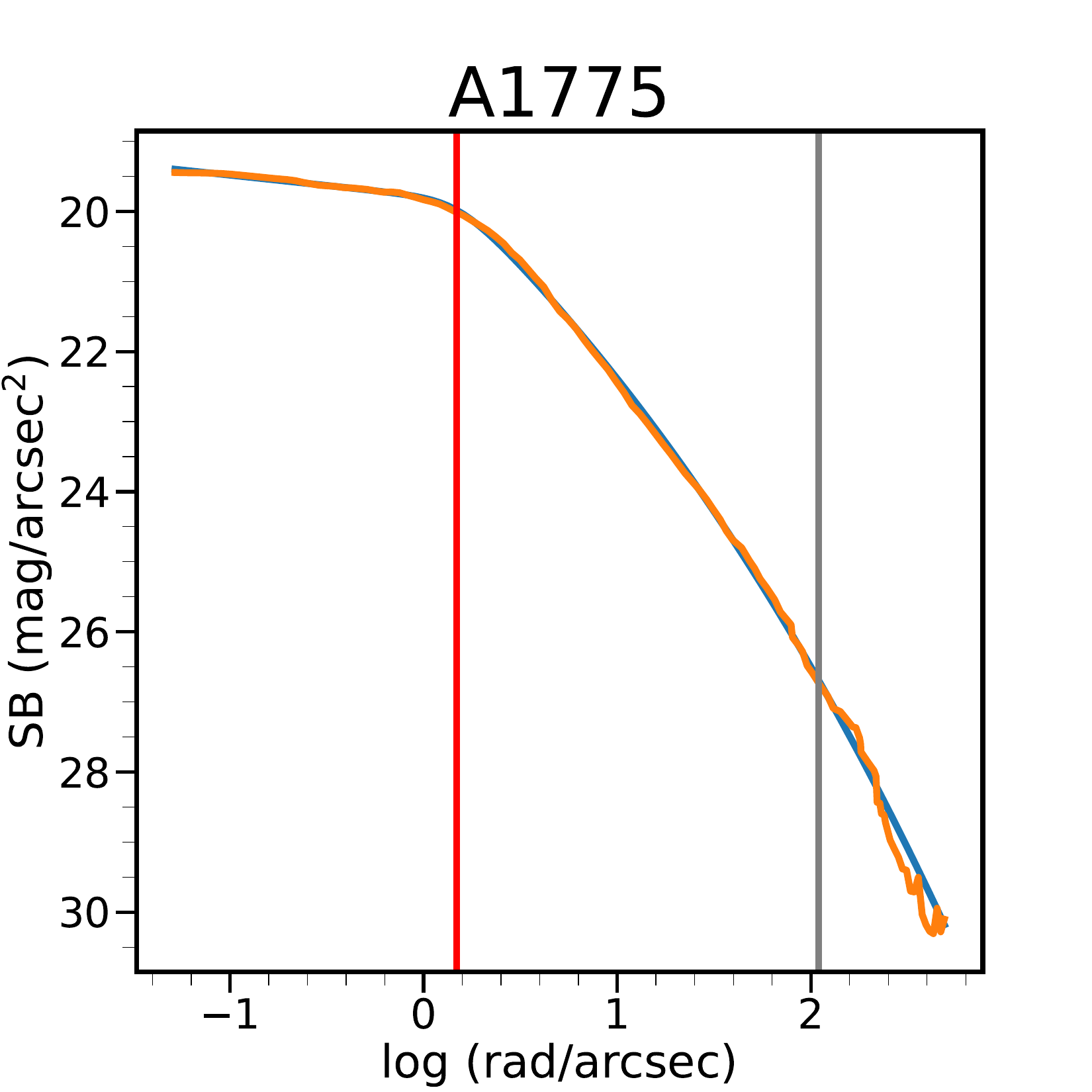}}
\subfloat{\includegraphics[scale=.19]{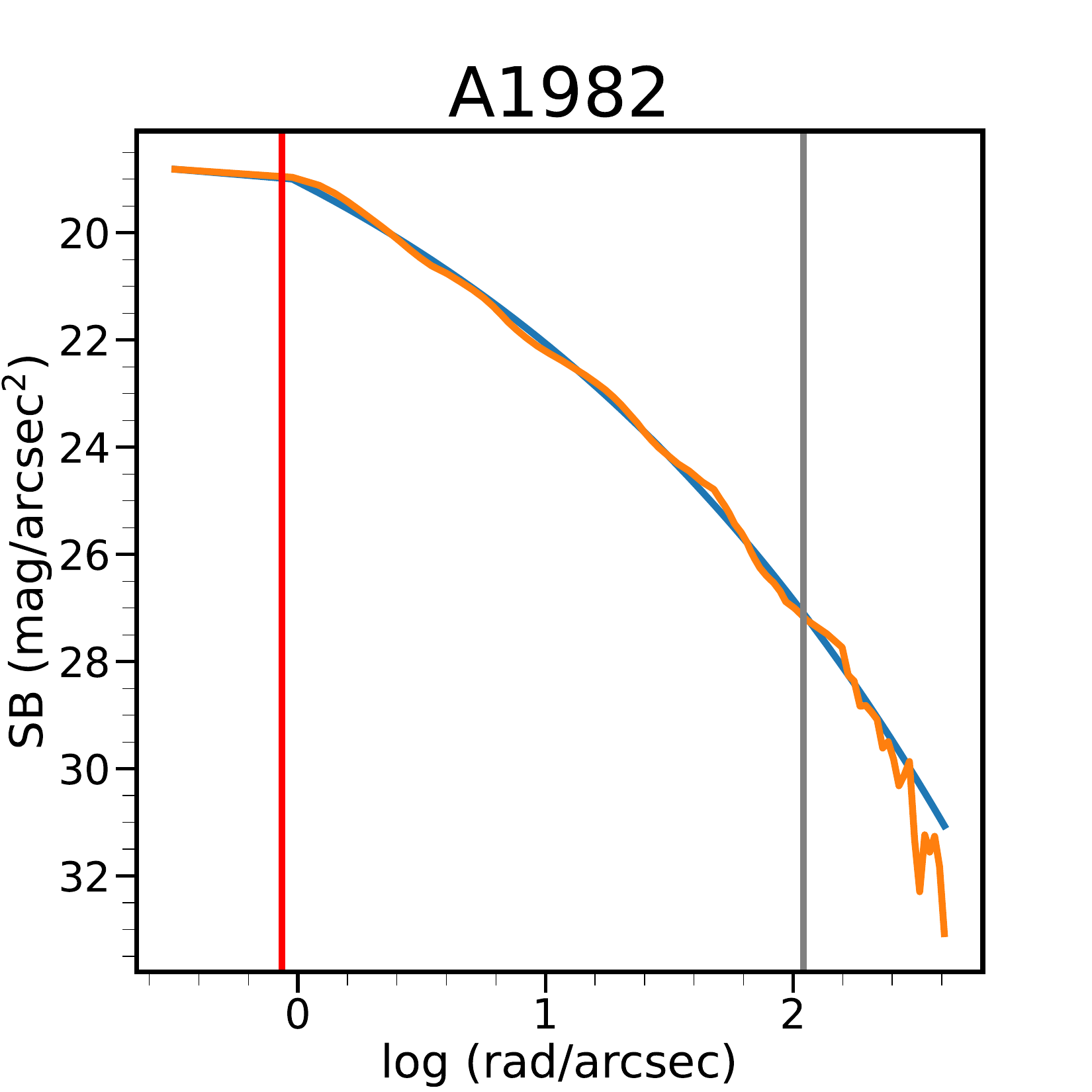}}
\subfloat{\includegraphics[scale=.19]{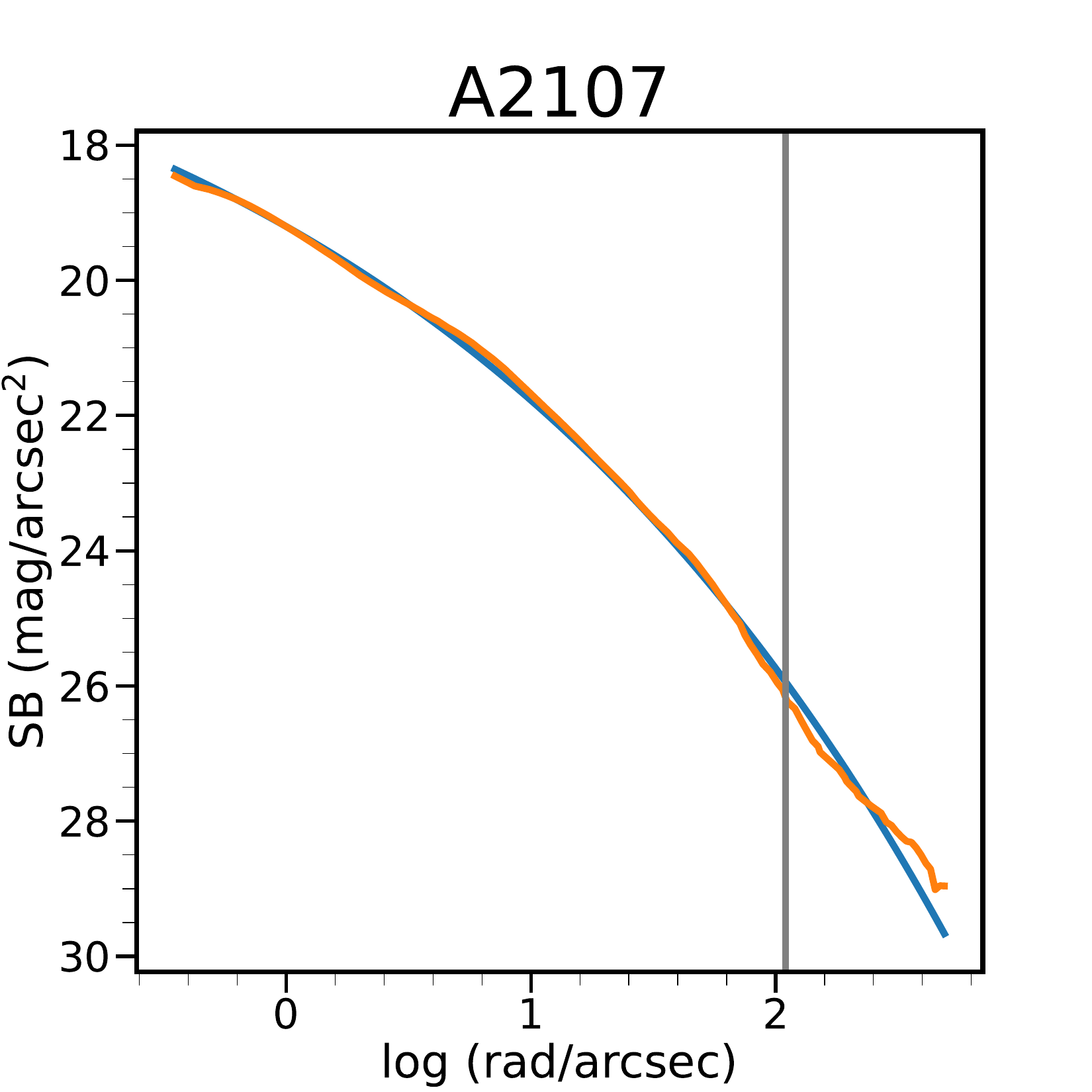}}

\subfloat{\includegraphics[scale=.19]{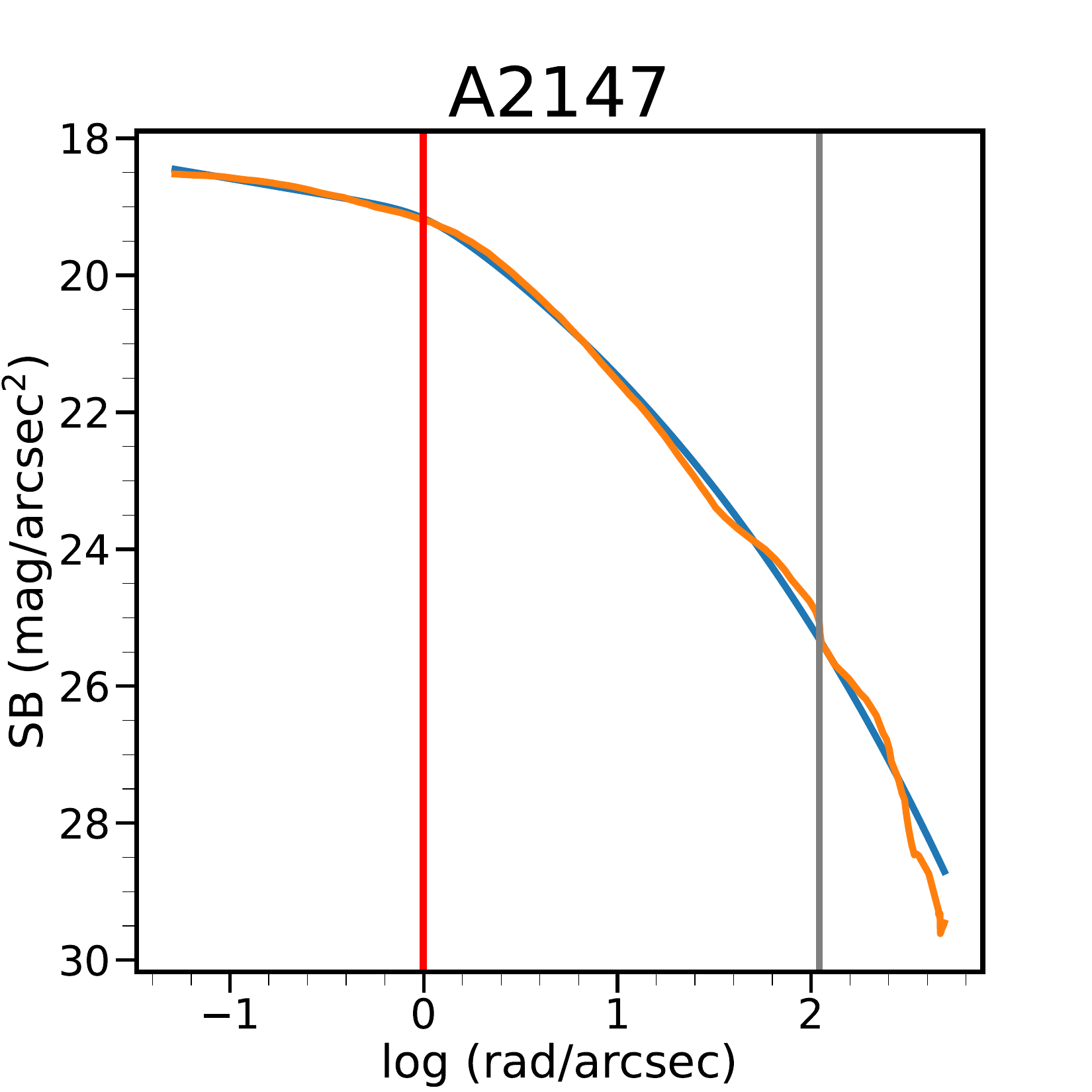}}
\subfloat{\includegraphics[scale=.19]{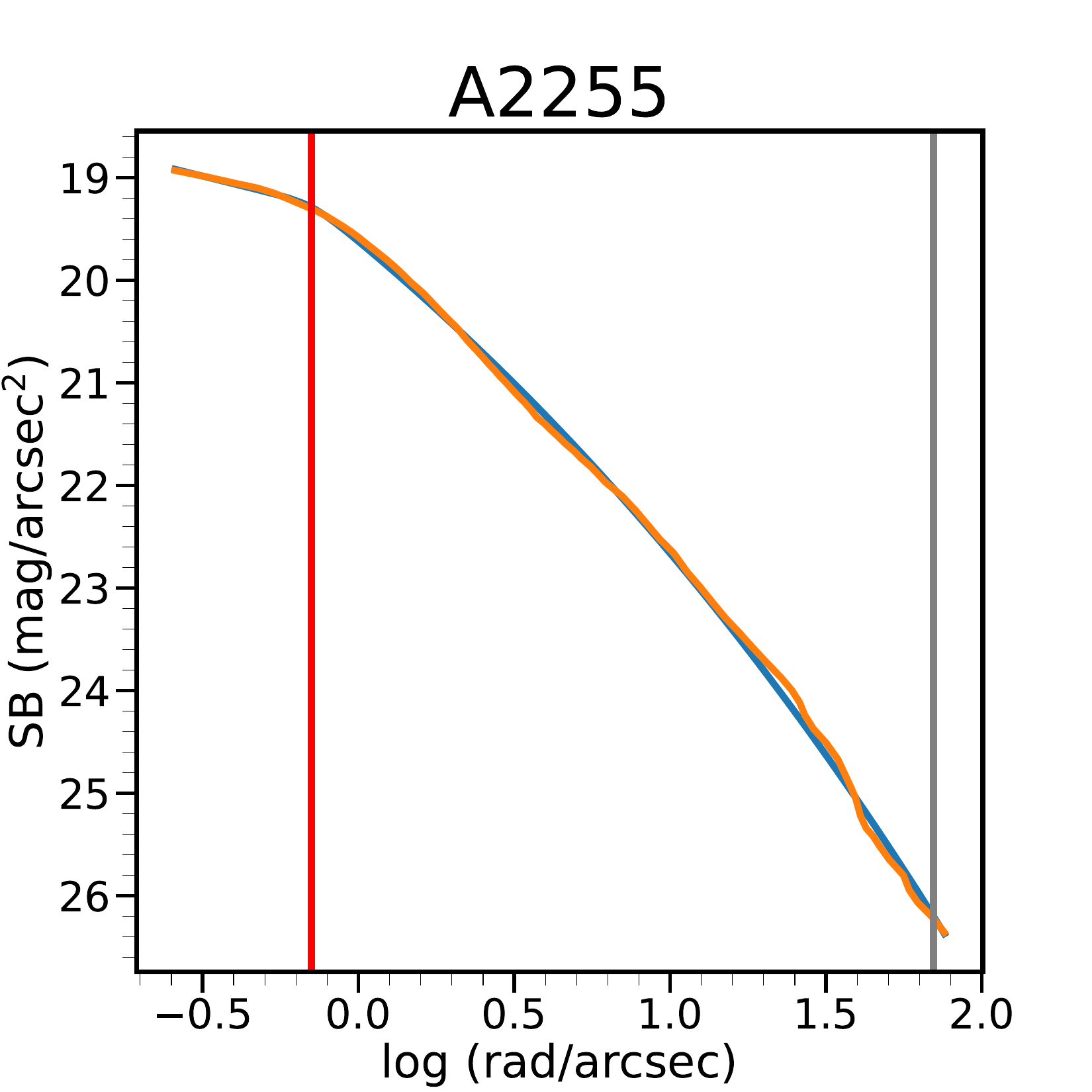}}
\subfloat{\includegraphics[scale=.19]{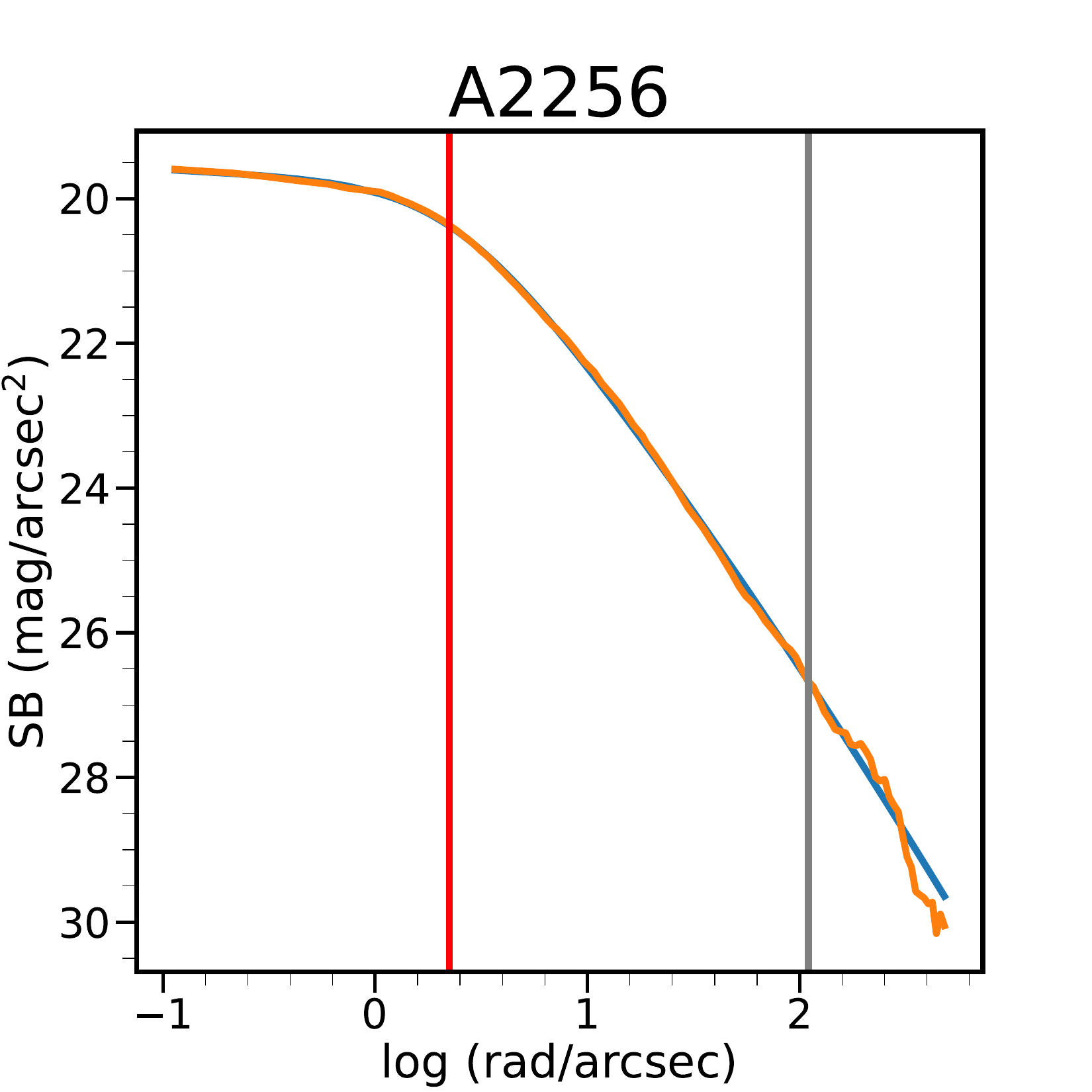}}

\subfloat{\includegraphics[scale=.19]{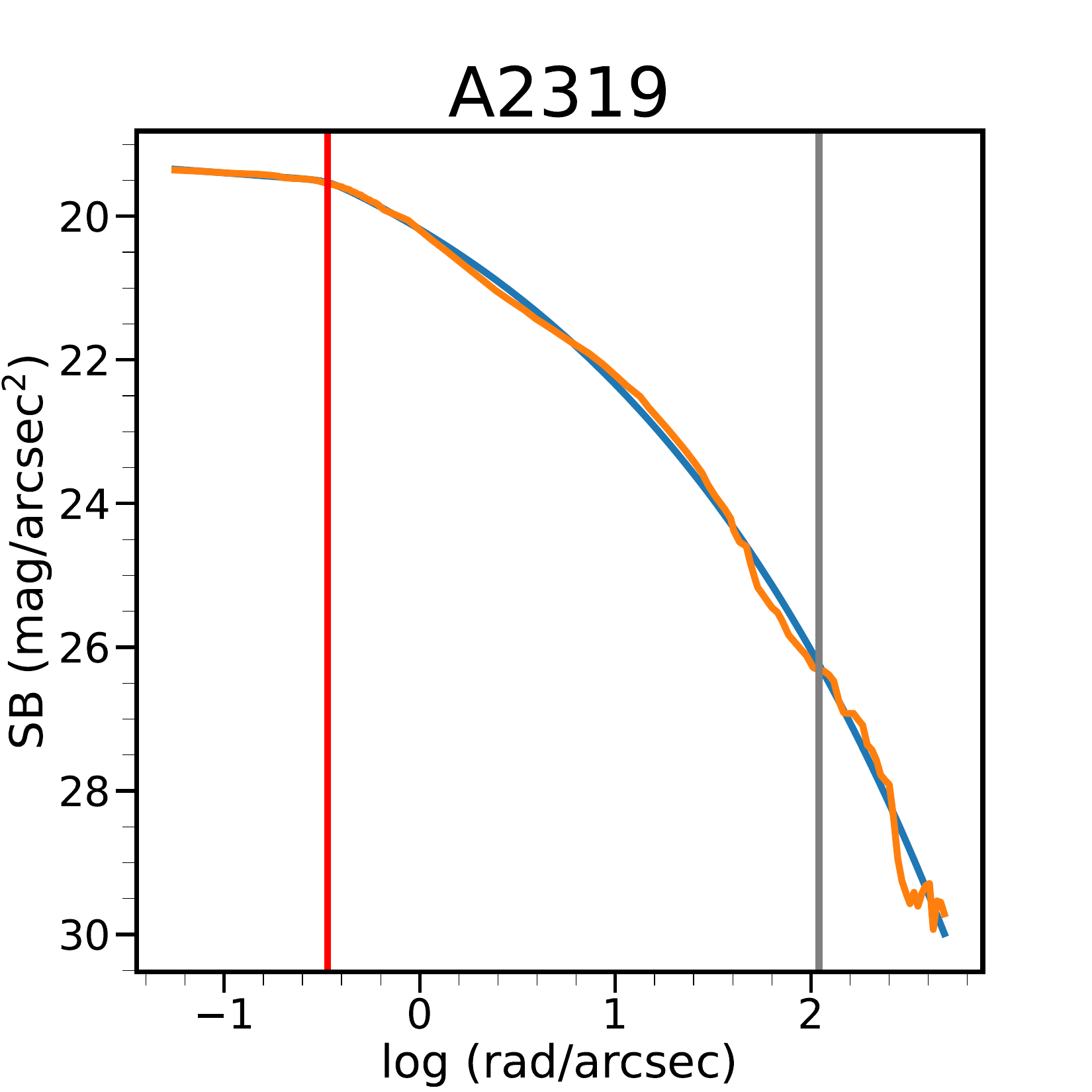}}
\subfloat{\includegraphics[scale=.19]{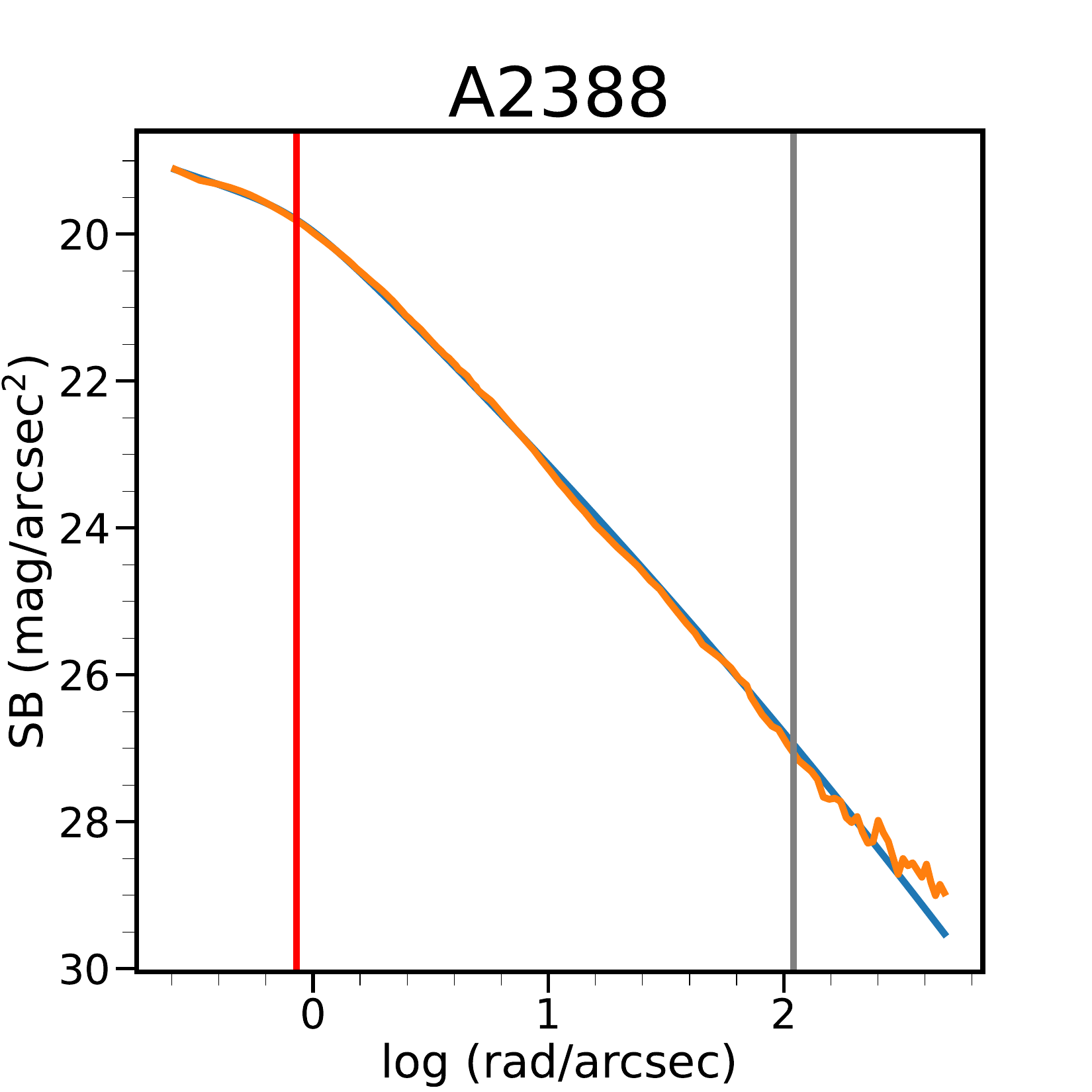}}
\subfloat{\includegraphics[scale=.19]{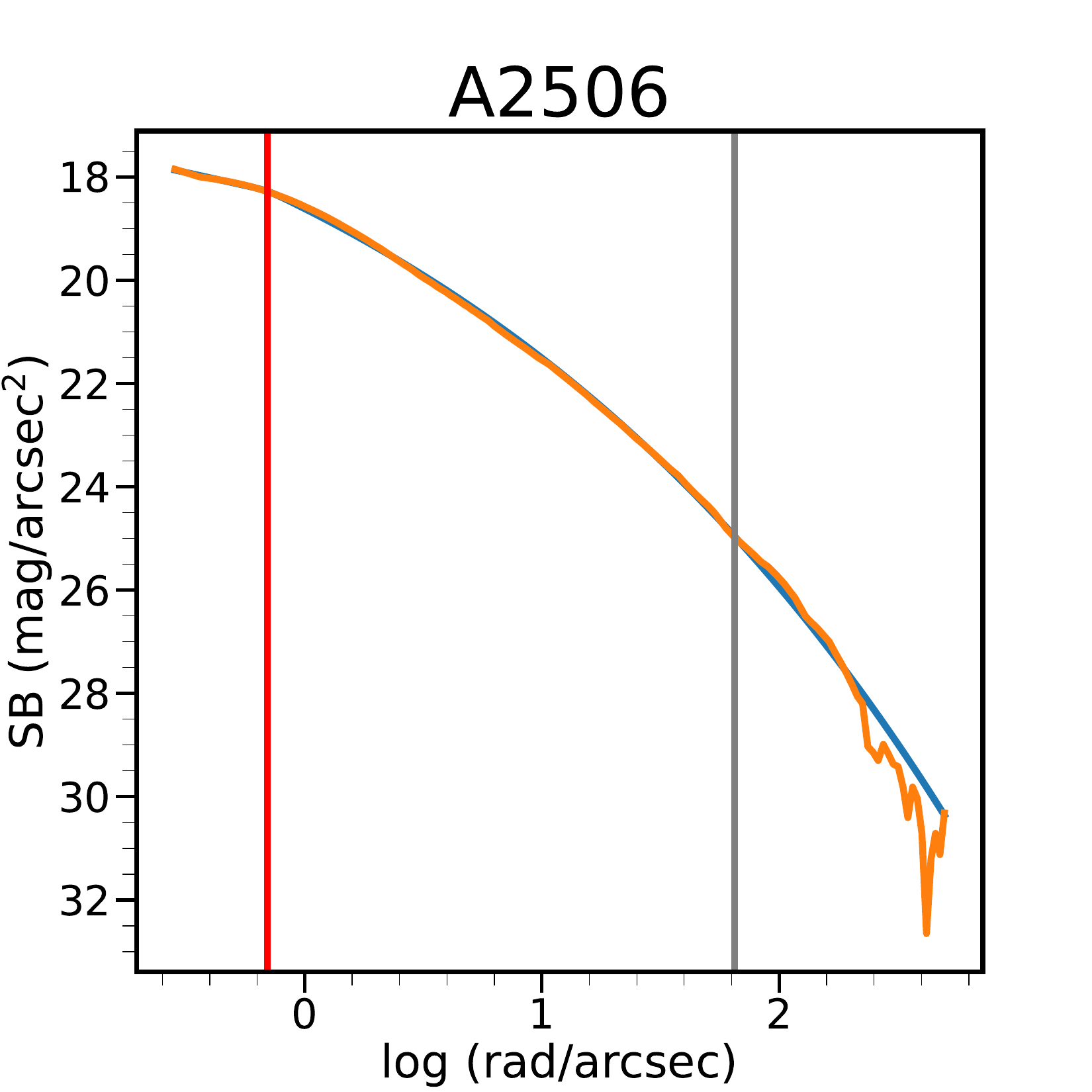}}

\subfloat{\includegraphics[scale=.19]{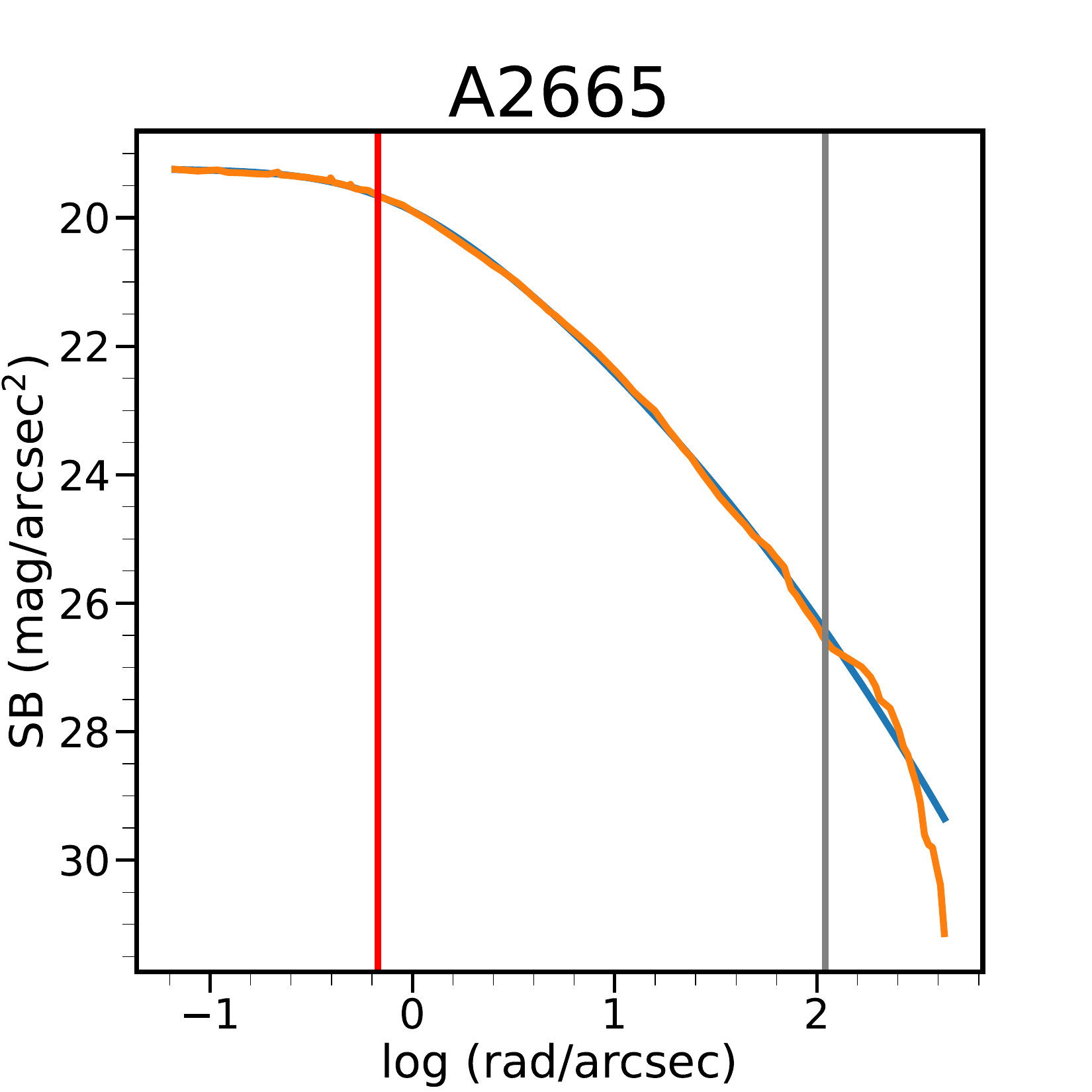}}

    \caption*{continued.}
    \label{Fig.cores_p2}
\end{figure*}

\subsection{Deprojection} \label{Ssec.depro}
We use our semi-parametric triaxial deprojection code SHAPE3D \citep{dN20} to recover the three-dimensional galaxy light density $\rho_*$ starting from the observed photometry, assuming the galaxy to be stratified on concentric ellipsoids:

   \begin{equation}
        m^{2-\xi(x)} = x^{2-\xi(x)} + \left[\frac{y}{p(x)}\right]^{2-\xi(x)} + \left[\frac{z}{q(x)}\right]^{2-\xi(x)}.
        \label{eq.def_ellips}
    \end{equation}

\begin{figure}
    \centering
    \includegraphics[width=1.\linewidth]{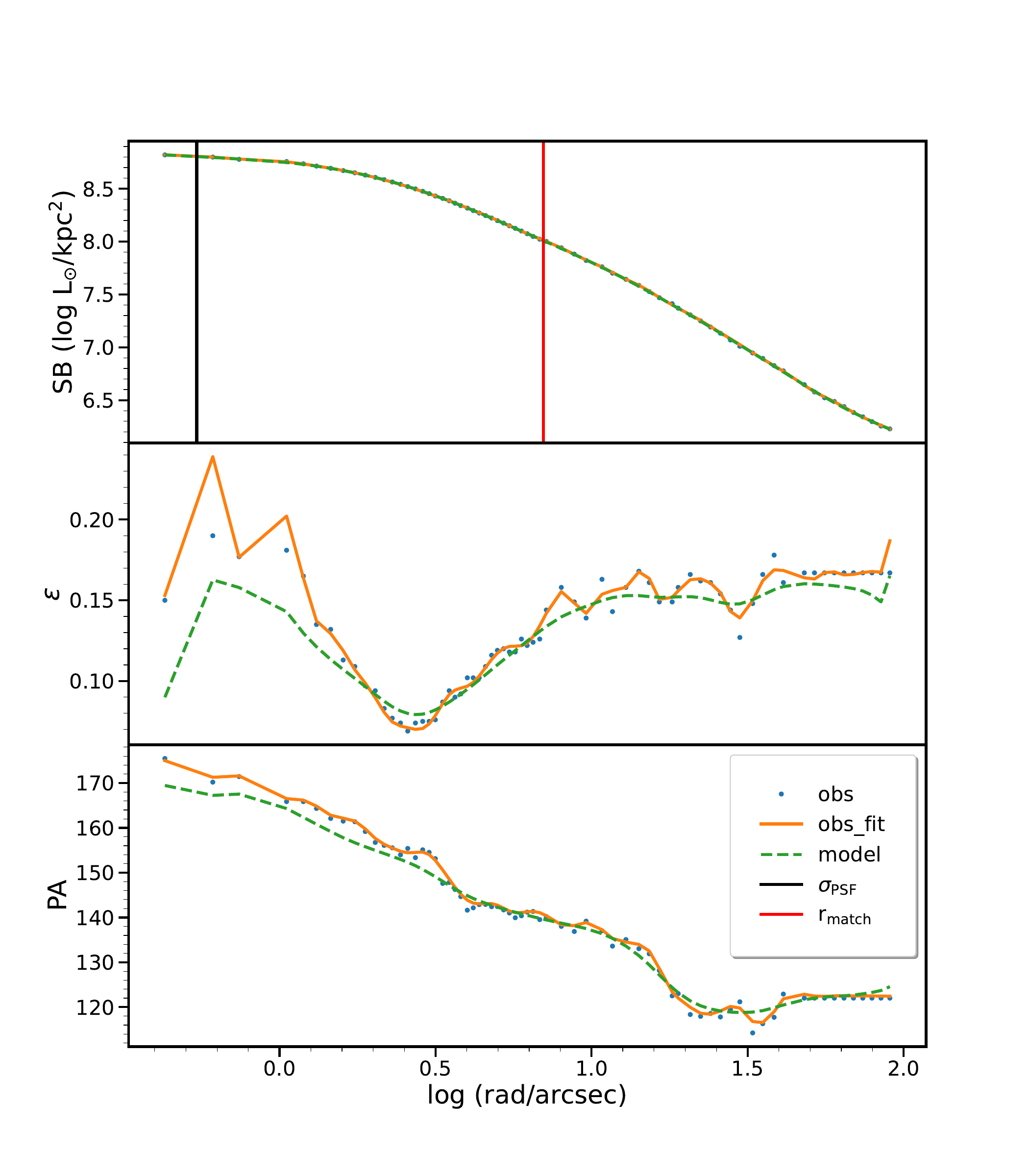}
    \caption{Isophotes (from top to bottom: Surface Brightness, ellipticity, PA) of the BCG of A2256. The blue points represent the observed photometry, while the orange lines are computed by placing the SB on the grid and performing isophotal fits. The green lines are obtained from the projection of the deprojected Surface Brightness; the orientation is (56, 127, 37)$^\circ$. The black vertical line is placed at $\sigma_\mathrm{PSF} \equiv \mathrm{FWHM}_\mathrm{PSF} / 2.35$, whereas the red line is at r$_\mathrm{match}$: from this radius outwards we take WWFI photometry.}
    \label{Fig.photometry}
\end{figure}

The strategy is the same as that used in \citet{dN22BCGs, dN24}. In brief, we sample the viewing angles \va, in 10$^\circ$ steps, with $\theta, \phi \in \left[0, 90\right]^\circ$ and $\psi \in \left[0, 180\right]^\circ$. The first two angles specify the position of the line of sight (LOS), while $\psi$ is a rotation about the LOS itself (see Fig. 2 of \citealt{dN20}). \\
Details about the deprojections are reported in Table 2 of \citet{dN22BCGs} and in Tab.~\ref{Tab.depro_parameters}. We chose the grid radii to ensure proper sampling of the central core as well as to extend far beyond the outermost point with kinematical coverage, verifying that the surface brightness placed on the grid reproduces the observed photometry well. We estimate the flattening $\eta$ of the SB grid (eq. 10 of \citealt{dN20}) from the isophotal shape, and compute the two flattenings $P, Q$ of the $\rho_*$ grid (eq. 11 of \citealt{dN20}) by calculating the expected $p(r), q(r)$ using eqs. A8 of \citet{deZeeuwFranx1989}. Finally, we ensure that the inequality $1 \geq \langle p(r) \rangle \geq \langle q(r) \rangle$ holds for each deprojection, as this is not necessarily the case if only one octant is sampled.  \\
In addition to selecting all deprojections with RMS $\leq 1.2\,\text{RMS}_\mathrm{MIN}$, where $\text{RMS}_\mathrm{MIN}$ is the smallest RMS between the observed and modeled surface brightness (eqs. 8–9 of \citealt{dN20}), as done in \citet{dN22BCGs}, we also require that $\text{RMS}_{\text{SB}} < 2$\%, $\text{RMS}_\varepsilon < 0.05$, and $\text{RMS}_\text{PA} < 5^\circ$ (see Fig.~\ref{Fig.photometry} for an example). App.~\ref{App.depro} reports details on the deprojection for those galaxies not present in \citet{dN22BCGs}. \\
Finally, we inspect the resulting intrinsic shapes to ensure that all possible geometries (e.g., triaxial, oblate, prolate) yielding a good fit to the observed photometry (see above) are represented when performing the dynamical modeling. \\
In cases where orientations lie along or near the principal axes, we sample additional deprojections at the same orientation to explore degeneracies. If a galaxy is particularly round - i.e., with $\varepsilon < 0.3$ at all radii (as in A240, A634, A2255, A2256, A2319) - the number of deprojections yielding a good fit can become very large ($> 100$). In this case, following \citet{Rob24}, we select one deprojection whose $p(r), q(r)$ profiles are closest to the average shape among all acceptable deprojections \citep{dN22Nbody}, as well as one close to the major axis. \\
In Fig.~\ref{Fig.photometry} we show the isophotes - whose parameters are derived using the software of \citet{Bender87} - of A2256 with superimposed results when projecting the recovered intrinsic density yielded by the reprojection code at the best-fit angles, showing that we recover the observed photometry well. Here, we also show r$_\mathrm{match}$ (see Sec.~\ref{Ssec.phot_data} above) and the PSF, which shows that the central core is well resolved by our high-resolution photometry.

%https://scienceops.lbto.org/mods/instrument-characteristics/grating-and-prisms/
\section{Spectroscopy} \label{Sec.spectroscopy}

This section presents the kinematical data and the observations performed to acquire them, along with a description of notable results.

%Other observers: C. Saulder, J. Thomas, M. Fabricius, M. Lipka, D. Delley, M. Anetjärvi
\subsection{Data}   \label{Ssec.spec_data}
\begin{figure}
    \centering
    \includegraphics[scale=.2]{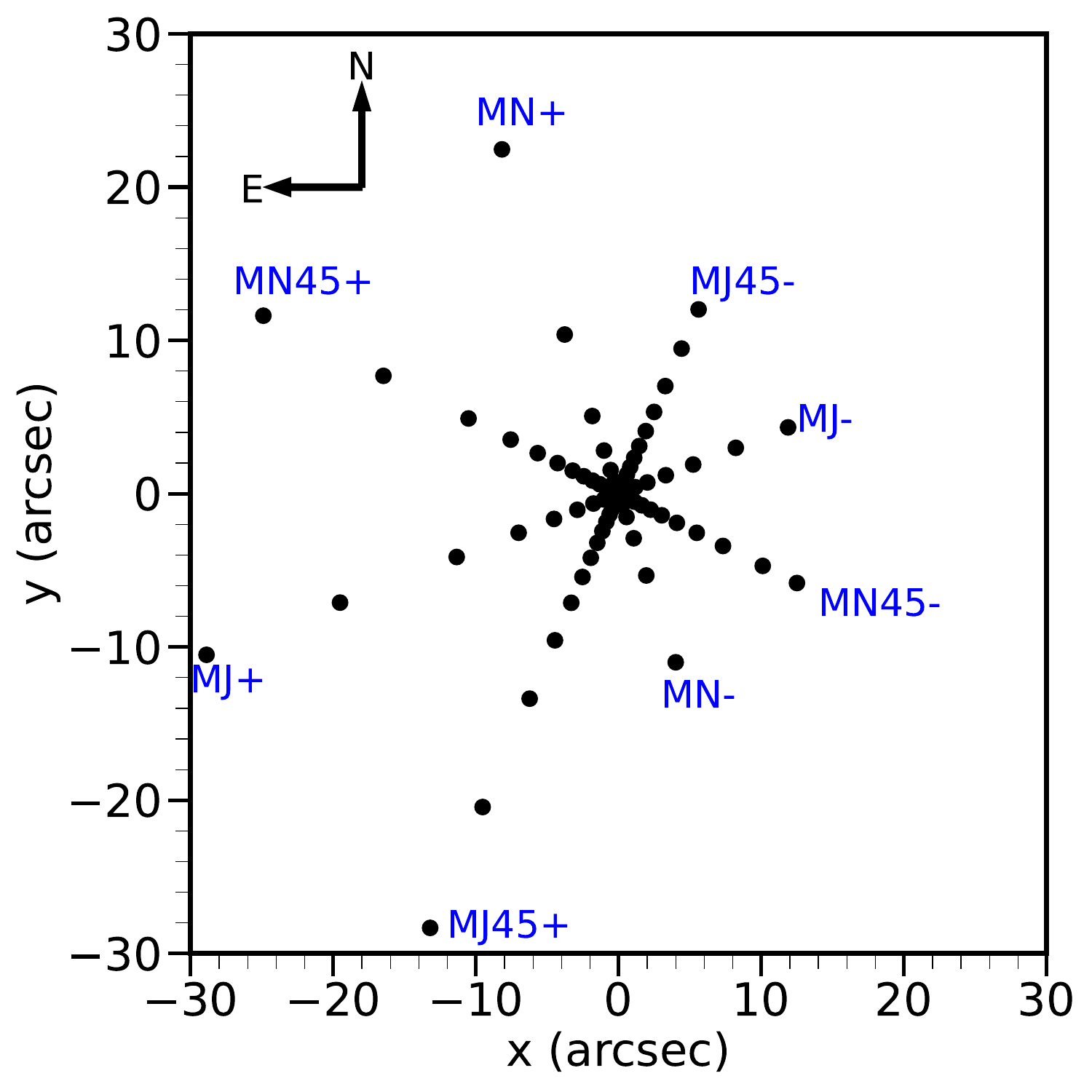}
    \caption{Example of slit placement on the plane of the sky for A2107. The negative radii are always North for PA = 0$^\circ$. The PA of the individual slits correspond to [MJ, MN, MJ45, MN45] = [290,200,335,245]$^\circ$ (see also Tab.~\ref{Tab.kinematics}).}
    \label{Fig.slits}
\end{figure}

Spectroscopy is the second key ingredient to perform dynamical modeling. In our case, we do not simply fit moments but exploit the full kinematical information by fitting non-parametric LOSVDs (like in our previous studies, e.g. \citealt{Rusli13DM, dN22Nbody, Bianca23Nbody, Kianusch24,Mathias24data}).\\ %using our code WINGFIT (\citealt{Kianusch23}, Thomas in prep.). \\
The data for all BCGs in our sample have been obtained using the Multi-Object Double Spectrograph (MODS, \citealt{Pogge10}) at the LBT Observatory. The observations span a 6-year period and have been carried out both in situ and remotely (P.I. R. Saglia, observers J. Snigula, S. de Nicola, R. Saglia).  \\
The observations are performed either in binocular or monocular mode (see Tabs.~\ref{Tab.kinematics} and App.~\ref{App.kin} for details). In the first case, following the same strategy as that presented in \citet{dN24}, we choose 0.8" as slit width and first execute a run placing one slit on the galaxy major axis and the other slit on the minor axis (MJ and MN configurations), then rotate both slits by 45$^\circ$ (MJ+45 and MN+45 configurations), to allow for fair coverage of the galaxy. An example is shown in Fig.~\ref{Fig.slits}. Instead, for the monocular case we only have one slit observed at the time, with the same slit width.\\
We use single exposures of 30' for each configuration, repeated a certain number of times (see Tab.~\ref{Tab.kinematics} for the total exposure times for each object), with pixel scale of 0.12 arcsec/pixel for the left MODS instrument and 0.123 arcsec/pixel for the right one. We employ the G400L (400 line/mm) reflection grating, with linear dispersion 0.515 \AA/pix and 5200 spectral pixels. \\
A sky image needs to be acquired afterwards to subtract the background. In case of unfavorable observing conditions (e.g., poor weather, bad seeing), we attempt to always complete the MJ+MN setup before starting the new one. Therefore, not all galaxies have all four slits at our disposal (see Tab.~\ref{Tab.kinematics} and Sec.~\ref{Sec.comments}). Finally, we image a spectrophotometric star needed for flux calibration. The images are corrected for bias, dark, flat field and wavelength calibrated using the Python and IDL pipelines\footnote{The Python pipeline is available \href{https://cgi.astronomy.osu.edu/MODS/Software/modsCCDRed/}{here}, whereas the IDL scripts can be found \href{https://cgi.astronomy.osu.edu/MODS/Software/modsIDL/}{here}.} provided by LBTO. To perform this last step, we use MJ as reference and match the wavelength of prominent sky lines of the other three slits to those observed along MJ. Given that for the dynamical modeling it is crucial to have spectra with reasonable S/N ($\geq 40 $/\AA), we bin our spectra to reach such S/N in all bins that we model. \\
To determine the LOSVDs we use the blue part of the spectrum ($\lambda \in [3200-5700]$ \AA). However, only in the range [4000–5400] \AA\ is the S/N good enough for an accurate determination of the LOSVDs. The exact intervals are reported in Tab.~\ref{Tab.kinematics}; these depend on factors such as contaminated absorption lines and/or the continuum (usually due to a second nearby galaxy) and the presence of strong emission lines. This can vary strongly from one bin to another, requiring individual masking for each bin. In some cases, we find external objects on one (or more) slit(s), due to the fact that BCGs often interact with other galaxies in the cluster. In such cases, we try to disentangle the spectra by masking the neighboring object and/or contaminated spectral regions, performing a kinematical fit with two components, or, if this is not possible, we simply discard the affected bin(s).

%Kinematics
\subsection{Methodology} \label{Ssec.kinematics}

\begin{figure}
    \centering
    \includegraphics[width=.7\linewidth]{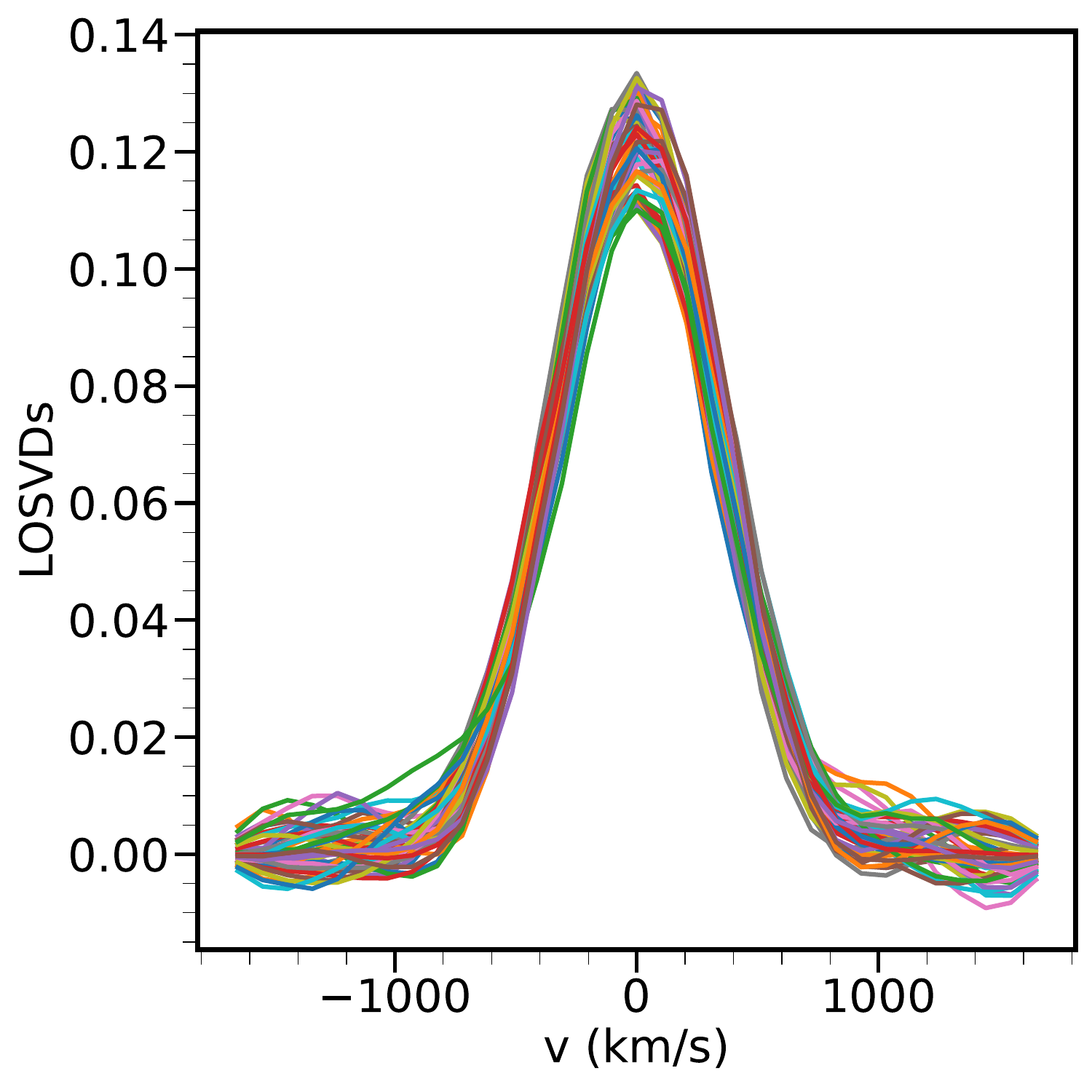}
    \caption{Example of non-parametric LOSVDs of A2107 recovered by WINGFIT, which we use as input for the dynamical model. The LOSVD is sampled up to five times the central velocity dispersion of the galaxy.}
    \label{Fig.all_LOSVDs_A2107}
\end{figure}

The non-parametric reconstruction of the LOSVDs is done using our non-parametric code WINGFIT (Thomas in prep.; see \citealt{Kianusch23} and \citealt{Mathias24data} for extensive applications). \\
The fits are performed using stellar templates from the MILES \citep{SB06, FB11} library. We perform a preliminary fit assuming a parametric LOSVD using PPXF \citep{Cappellari04, Cappellari17} to determine the optimal template set for each galaxy. Typically, fewer than 20 template spectra are needed. \\
A non-parametric LOSVD fitting algorithm can yield unsmooth LOSVDs. To prevent this, the code uses the model selection technique based on a generalization of the classical Akaike Information Criterion (AIC, \citealt{Akaike74}) described in \citet{Mathias21, Jens22} to optimize the smoothing of the LOSVDs and, at the same time, find a solution that fits the data well. This consists in minimizing the generalized Akaike Information Criterion AIC$\mathrm{p} = \chi^2 + 2 \mathrm{m}_\mathrm{eff}$ instead of $\chi^2$, where m$_\mathrm{eff}$ is the effective number of free parameters each model has.\footnote{The subscript \textit{p} denotes a penalized variant, as AIC$\mathrm{p}$ extends the classical AIC to penalized models.} \\
Despite the smoothing, some LOSVD values in the bins found in the wings can exhibit negative values, which are nonetheless always consistent with 0 within the uncertainties (see, e.g., Tab.~\ref{Tab.LOSVD}). We tested this by comparing the results obtained for A2107 with a run in which we enforced positivity in the LOSVDs, finding no significant differences. \\
We limit ourselves to third-order multiplicative polynomials and do not use additive polynomials to avoid generating artificially enhanced wings \citep{Kianusch23}. Finally, we recover the LOSVD moments v, $\sigma$, h$_3$ and h$_4$ by fitting Gauss-Hermite polynomials to the recovered LOSVD. To subtract the velocity of the galaxy barycentre $\mathrm{v}_\mathrm{bary}$, we take the redshift values published in \citet{Matthias20} and use v = c $\times \ln$(1+z) to derive a first-order estimate of $\mathrm{v}_\mathrm{bary}$. We then average the resulting velocities of the three innermost bins, repeat the fits and subtract this value so that the galaxy has v $\sim$ 0 at the centre. As an example, we show in Tab.~\ref{Tab.kin_2D} the moments of A2107 along MAJOR\footnote{All other slits/galaxies are available electronically.}. \\
The resulting non-parametric LOSVDs are used for the dynamical modeling. Examples are shown in Tab.~\ref{Tab.LOSVD} for Bin 9 of A2107 along MAJOR and in Fig.~\ref{Fig.all_LOSVDs_A2107}, where we plot all the modeled LOSVDs in a single figure. These are sampled up to velocity values of $\sim 5\sigma_0$, where $\sigma_0$ is the central velocity dispersion of the galaxy\footnote{The dispersion in the outskirts can be higher by a factor of 2, thus implying that these LOSVDs here are only sampled up to $\sim 2.5\sigma_0$. However, sampling up to larger velocity values would lead to include, for the central bins, LOSVD tails where the noise dominates over the signal from the LOSVDs themselves.}. We select the number of velocity bins for the dynamical models so as to have the velocity resolution as close as possible to that of MODS. While having more bins increases the computational time, too coarse a sampling likely results in very wide AIC$_\text{p}$ minima (see Sec.~\ref{Ssec.dynamics_methodology} below) when estimating the mass distribution. Details on individual objects are reported in Sec.~\ref{Sec.comments}. \\
Given that our galaxies are often located at distances $>$ 200 Mpc, we typically have good enough spectra only within 10", sometimes even 5". However, for all galaxies we have a fair number of bins inside the BH SOI, allowing us to resolve the region where the potential is dominated by the BH itself. Instead, the coverage out to 10" allows for robust sampling of the region where stars contribute the most to the total mass and, for most galaxies, for a robust determination of the DM density at 10 kpc (see eq.~\ref{eq.Zhao_models}). \\

\begin{table*}
\centering
\begin{tabular}{c c c c c c c}
\hline
Bin & x (arcsec) & y (arcsec) & v $\pm$ $\Delta$v (km/s) & $\sigma$ $\pm$ $\Delta\sigma$ (km/s) & h$_3$ $\pm$ $\Delta$h$_3$ & h$_4$ $\pm$ $\Delta$h$_4$ \\
\hline
1  & 11.6 & 4.22 & $-33.7 \pm 73.4$ & $456 \pm 72$ & $-0.054 \pm 0.069$ & $0.013 \pm 0.068$ \\
2  & 8.1  & 2.94 & $-5.6 \pm 43.4$  & $393 \pm 36$ & $-0.026 \pm 0.046$ & $0.067 \pm 0.055$ \\
3  & 5.2  & 1.89 & $-0.4 \pm 32.2$  & $356 \pm 27$ & $0.028 \pm 0.049$  & $0.067 \pm 0.046$ \\
4  & 3.3  & 1.21 & $-7.3 \pm 13.4$  & $365 \pm 25$ & $-0.028 \pm 0.037$ & $0.0080 \pm 0.031$ \\
5  & 2.0  & 0.74 & $-27.1 \pm 16.3$ & $315 \pm 15$ & $0.030 \pm 0.038$  & $0.019 \pm 0.044$ \\
6  & 1.2  & 0.44 & $-45.8 \pm 6.2$  & $319 \pm 11$ & $0.064 \pm 0.026$  & $0.055 \pm 0.031$ \\
7  & 0.65 & 0.24 & $-67.8 \pm 9.0$  & $298 \pm 13$ & $0.102 \pm 0.043$ & $-0.010 \pm 0.032$ \\
8  & 0.26 & 0.095& $-41.0 \pm 6.3$  & $306.2 \pm 8.4$ & $0.057 \pm 0.025$  & $0.064 \pm 0.024$ \\
9  & -0.073& -0.027& $-4.5 \pm 7.0$ & $317.6 \pm 8.0$ & $0.040 \pm 0.023$  & $0.024 \pm 0.018$ \\
10 & -0.46 & -0.17 & $40.8 \pm 7.7$  & $328 \pm 11$ & $-0.043 \pm 0.021$ & $0.033 \pm 0.018$ \\
11 & -0.96 & -0.35 & $19.4 \pm 9.8$  & $325 \pm 13$ & $-0.054 \pm 0.018$ & $0.026 \pm 0.029$ \\
12 & -1.67 & -0.61 & $2.7 \pm 10.7$  & $362 \pm 14$ & $-0.018 \pm 0.024$ & $0.027 \pm 0.024$ \\
13 & -2.79 & -1.02 & $-11.8 \pm 18.0$ & $390 \pm 22$ & $-0.046 \pm 0.037$ & $-0.027 \pm 0.030$ \\
14 & -4.40 & -1.60 & $-23.3 \pm 24.4$ & $314 \pm 33$ & $-0.049 \pm 0.047$ & $0.020 \pm 0.053$ \\
15 & -6.82 & -2.48 & $-25.5 \pm 36.2$ & $332 \pm 28$ & $-0.081 \pm 0.048$ & $-0.025 \pm 0.059$ \\
16 & -11.0 & -3.99 & $-36.1 \pm 55.9$ & $490 \pm 42$ & $-0.104 \pm 0.069$ & $-0.062 \pm 0.050$ \\
17 & -19.1 & -6.95 & $-44.3 \pm 71.9$ & $506 \pm 54$ & $-0.164 \pm 0.098$ & $-0.042 \pm 0.056$ \\
18 & -28.3 & -10.3 & $17.1 \pm 54.6$  & $851 \pm 74$ & $-0.0098 \pm 0.030$ & $-0.131 \pm 0.0099$ \\
\hline
\end{tabular}

\caption{2D kinematics along MAJOR for A2107. \textit{Col. 1}: bin number. \textit{Cols. 2-3}: coordinates on the plane of the sky in a NW frame of reference (see Fig.~\ref{Fig.slits}). \textit{Cols. 4-7}: moments v, $\sigma$, h$_3$, h$_4$ and their uncertainties, obtained by fitting 4th-order GH polynomials to the non-parametric LOSVDs reconstructed by WINGFIT.}
\label{Tab.kin_2D}
\end{table*}

\begin{table*}
\centering
\begin{tabular}{r r r r}
\hline
v (km/s) & LOSVD & $\sigma_\mathrm{LOSVD,low}$ & $\sigma_\mathrm{LOSVD,up}$ \\
\hline
-1650 & -0.00071 & 0.00194 & 0.00236 \\
-1547 & -0.00017 & 0.00380 & 0.00268 \\
-1444 & 0.00178 & 0.00412 & 0.00185 \\
-1341 & 0.00312  & 0.00309 & 0.00239 \\
-1238 & 0.00456  & 0.00376 & 0.00252 \\
-1134 & 0.00498 & 0.00366 & 0.00277 \\
-1031 & 0.00415  & 0.00385 & 0.00187 \\
-928.1 & 0.00235 & 0.00400 & 0.00184 \\
-825.0 & 0.00205 & 0.00358 & 0.00229 \\
-721.9 & 0.00625 & 0.00336 & 0.00266  \\
-618.8 & 0.01705 & 0.00330 & 0.00269 \\
-515.6 & 0.03372 & 0.00426 & 0.00230 \\
-412.5 & 0.05862 & 0.00323 & 0.00191 \\
-309.4 & 0.08160 & 0.00394 & 0.00209 \\
-206.3 & 0.10560  & 0.00291 & 0.00225\\
-103.1 & 0.11980  & 0.00294 & 0.00207\\
0.0    & 0.12410  & 0.00355 & 0.00199\\
103.1  & 0.11690  & 0.00424 & 0.00260 \\
206.3  & 0.10040  & 0.00391 & 0.00265 \\
309.4  & 0.07628  & 0.00256 & 0.00242 \\
412.5  & 0.05636  & 0.00245 & 0.00265 \\
515.6  & 0.03623  & 0.00338 & 0.00221\\
618.8  & 0.02353  & 0.00405 & 0.00262 \\
721.9  & 0.01453  & 0.00442 & 0.00248 \\
825.0  & 0.00894  & 0.00428 & 0.00251\\
928.1  & 0.00573 & 0.00474 & 0.00277 \\
1031   & 0.00373  & 0.00452 & 0.00294 \\
1134   & 0.00253  & 0.00321 & 0.00272 \\
1238   & 0.00105  & 0.00263 & 0.00240 \\
1341   & -0.00079 & 0.00205 & 0.00203\\
1444   & -0.00235 & 0.00201 & 0.00268 \\
1547   & -0.00234 & 0.00292 & 0.00245 \\
1650   & -0.00130 & 0.00194 & 0.00132 \\
\hline
\end{tabular}
\caption{Example of non-parametric LOSVD for Bin 9 of MAJOR for A2107 (see Tab.~\ref{Tab.kin_2D}. This what we use as input for the dynamical modeling. \textit{Col. 1}: Velocity. \textit{Col. 2}: LOSVD value. \textit{Col. 3-4}: Lower and upper LOSVD errors.}
\label{Tab.LOSVD}
\end{table*}

\subsection{Kinematics results} \label{Ssec.kin_res}

\begin{figure*}

\subfloat[A2506, major-axis rotation. \label{Fig.kin_A2506}]{\includegraphics[width=.3\linewidth]{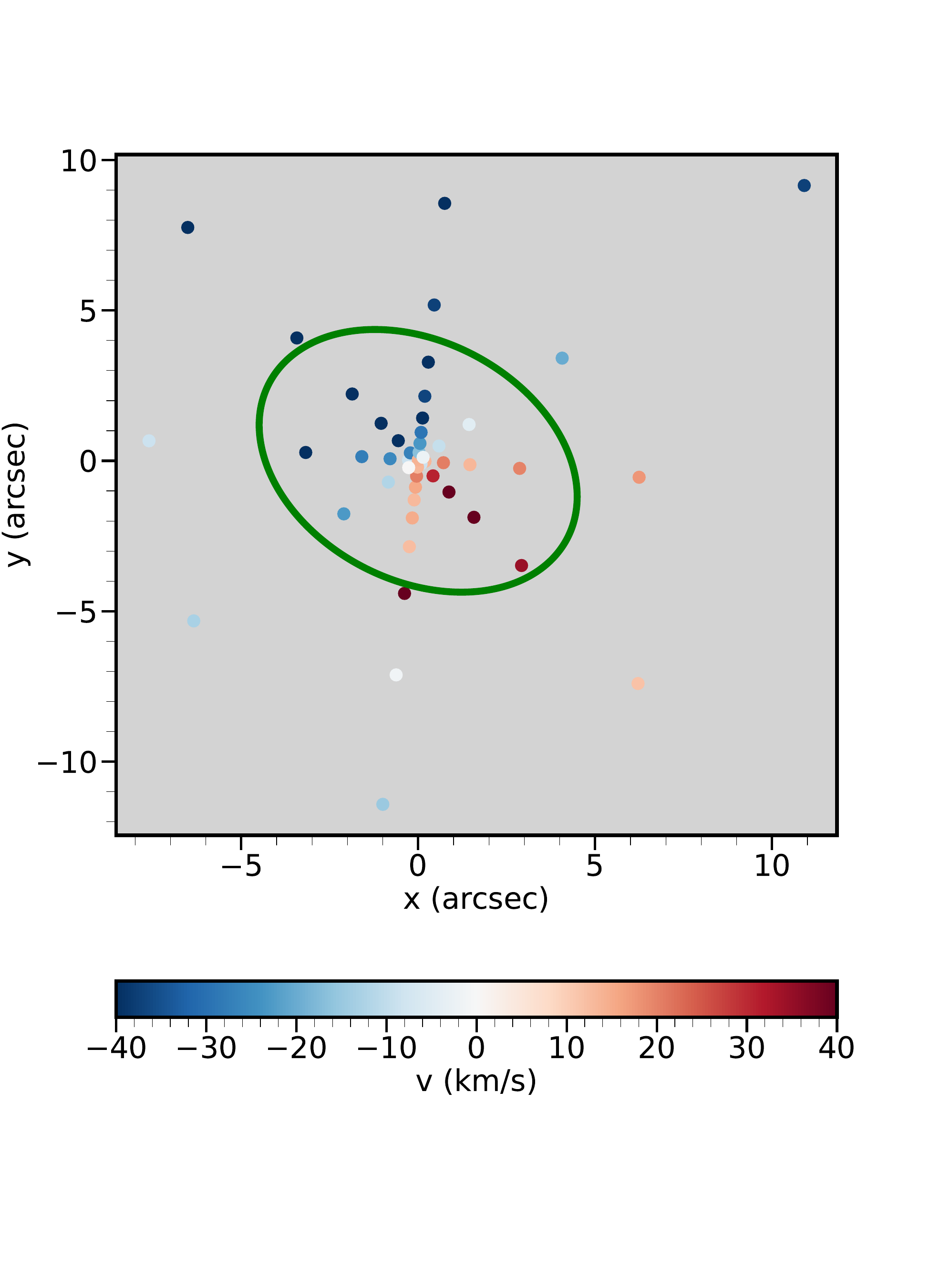}}
\subfloat[A1749, major-axis rotation.  \label{Fig.kin_A1749}]{\includegraphics[width=.3\linewidth]{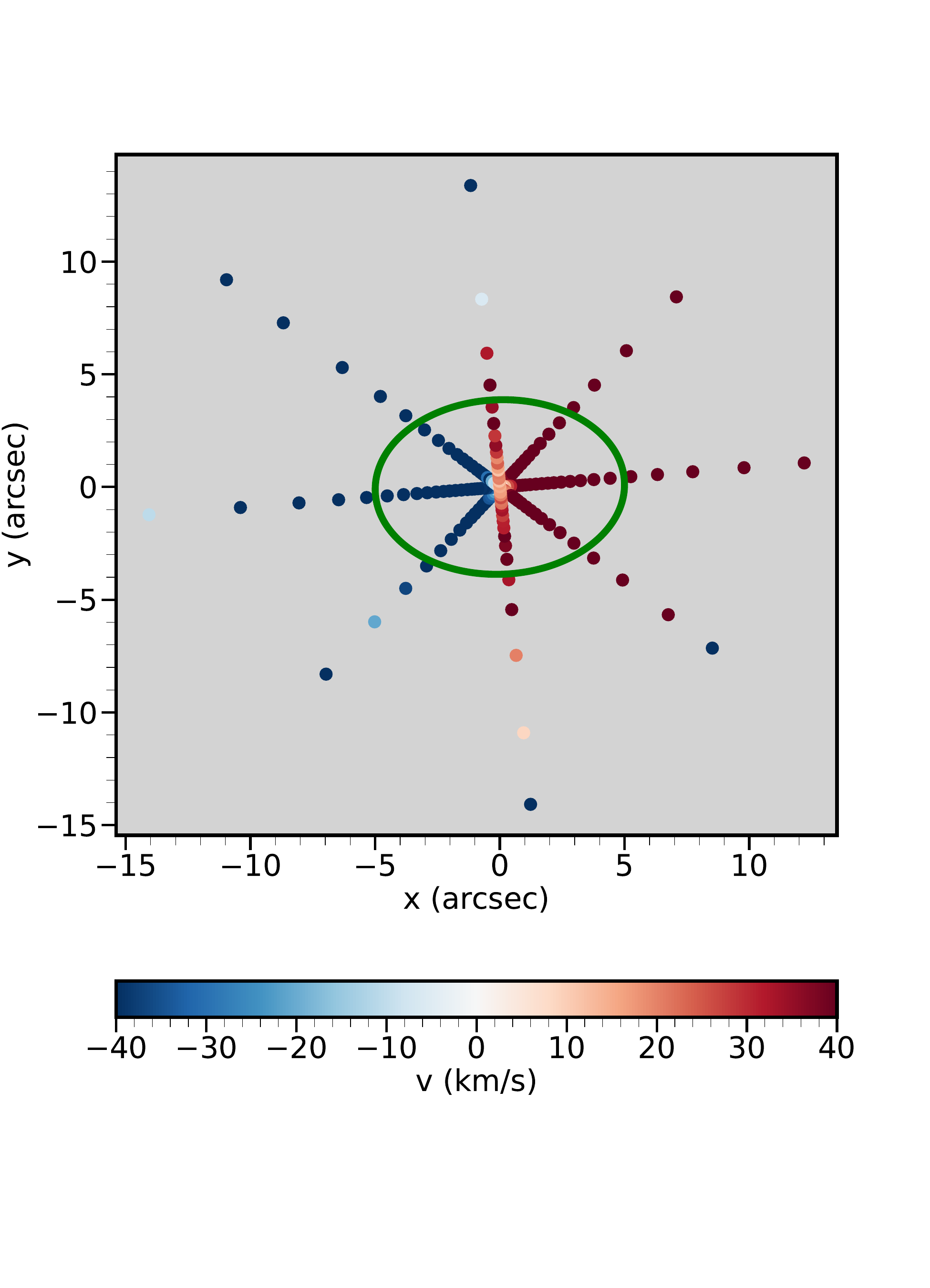}}

\subfloat[A1314, minor-axis rotation.  \label{Fig.kin_A1314}]{\includegraphics[width=.3\linewidth]{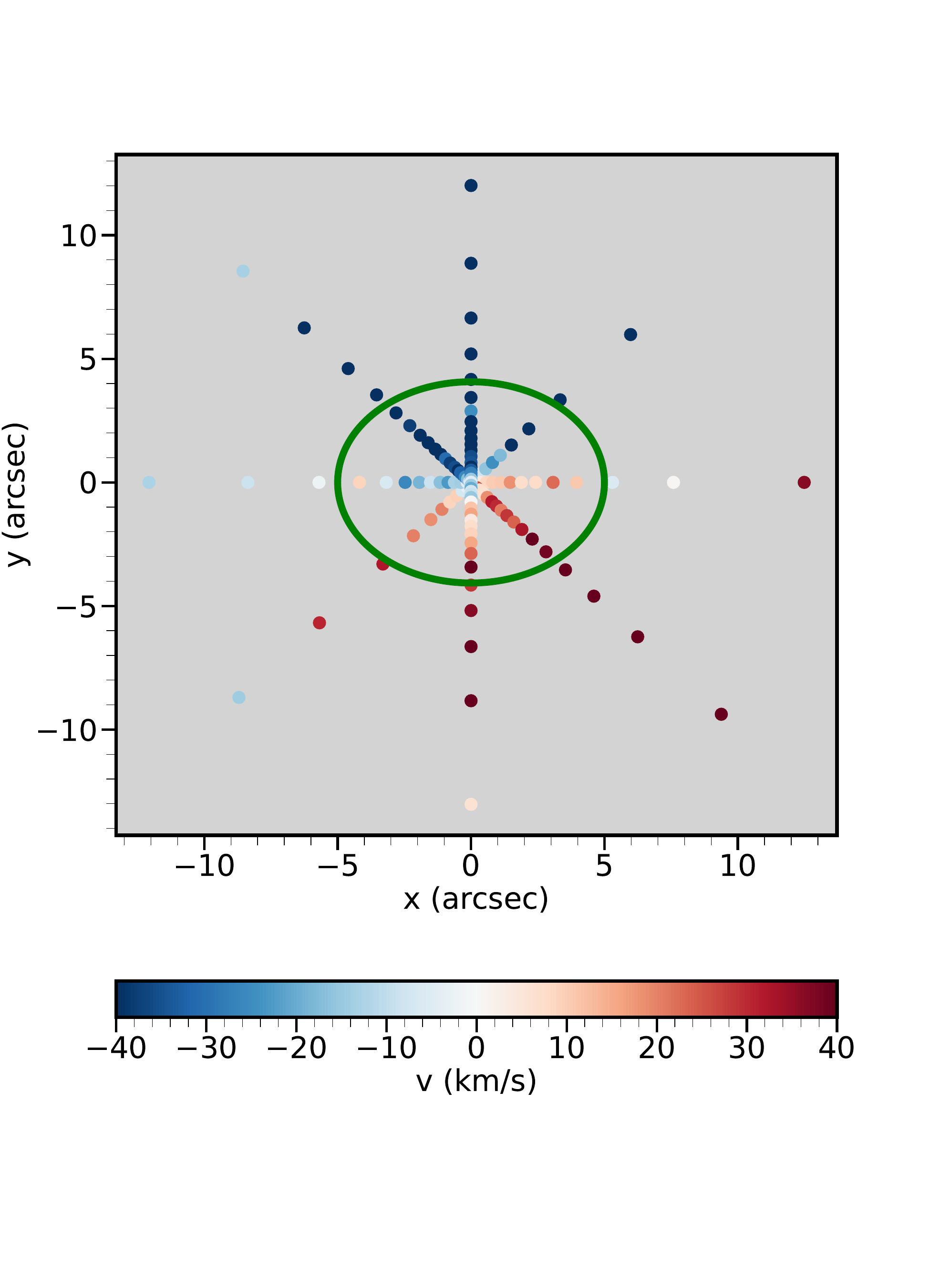}}
\subfloat[A2107, possible KDC.  \label{Fig.kin_A2107}]{\includegraphics[width=.3\linewidth]{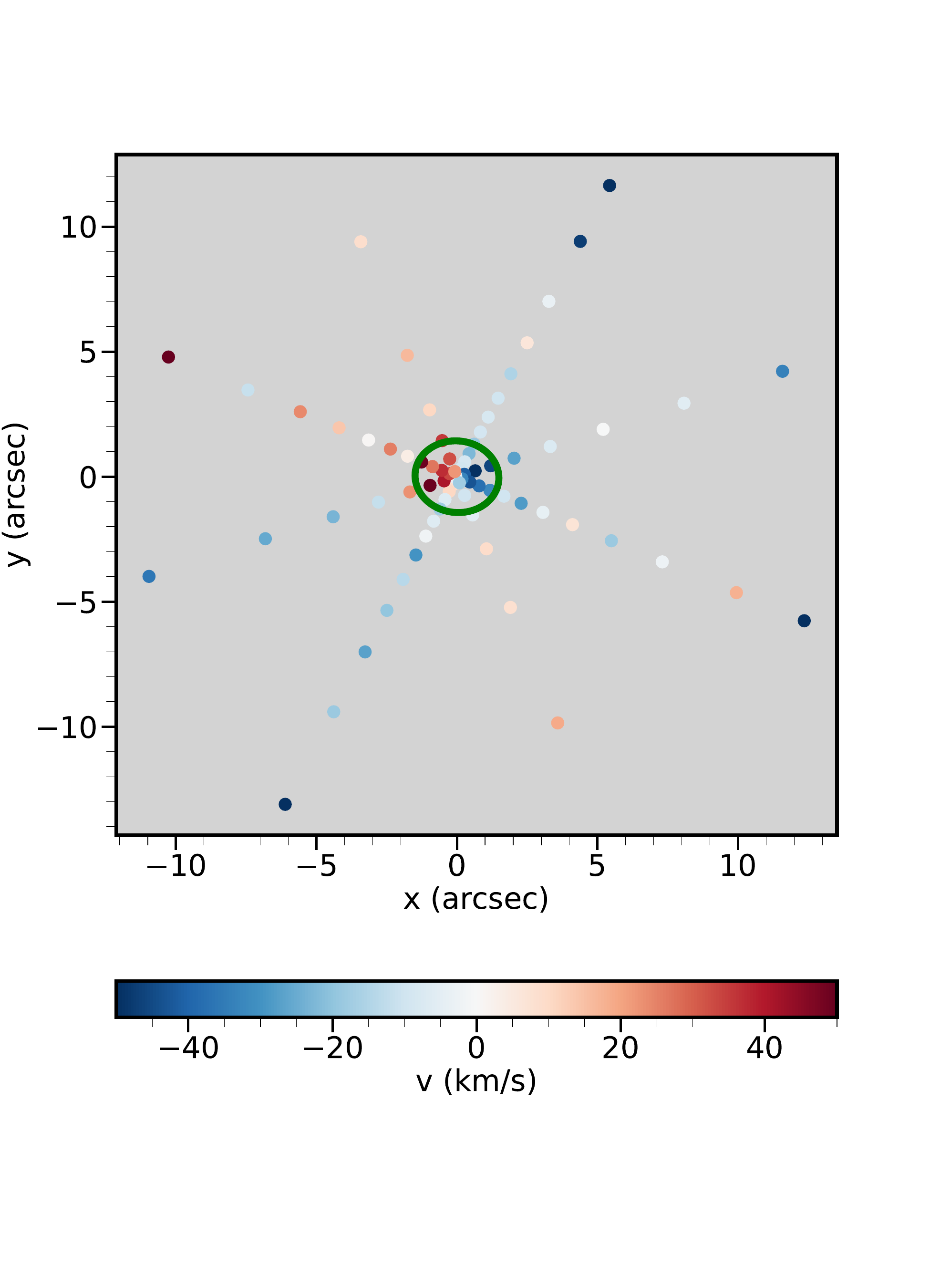}}

    \caption{Velocity maps for notable cases in our sample. For A2506 and A1749 (top panels) we observe major-axis rotation. For A1314 (bottom left) we observe minor-axis rotation, which implies triaxiality or at most a prolate geometry. For A2107 (bottom right) we observe a possible KDC with significant rotation inside 2", while the rotation signal is much weaker at larger radii. In all plots, the green ellipse shows the typical orientation of the isophotes — all four galaxies have only weak twists.}
    \label{Fig.kinematics}
\end{figure*}

\begin{figure*}

\subfloat[\label{Fig.sigma0}]{\includegraphics[height=.2\paperheight]{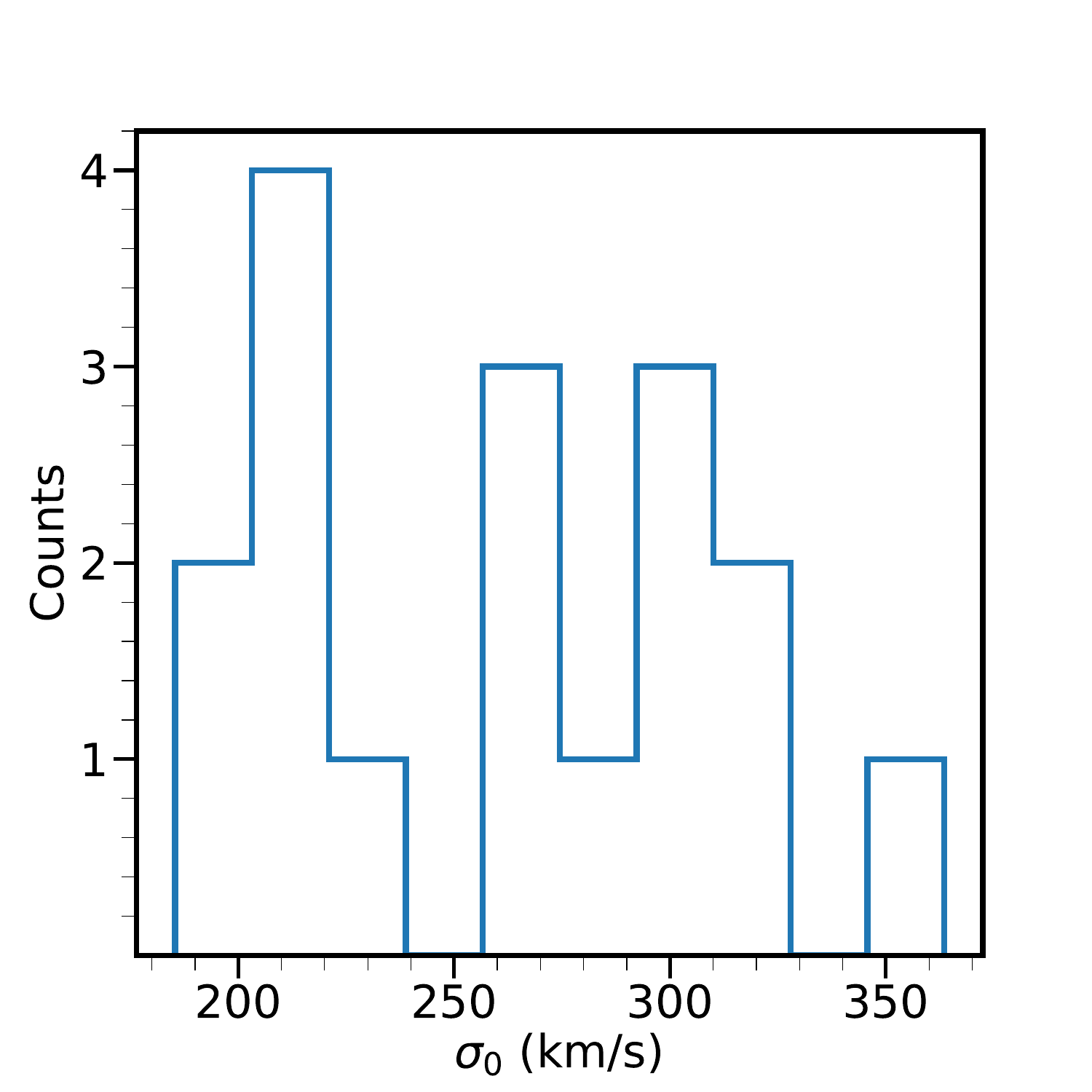}}
\subfloat[\label{Fig.sigmaout}]{\includegraphics[height=.2\paperheight]{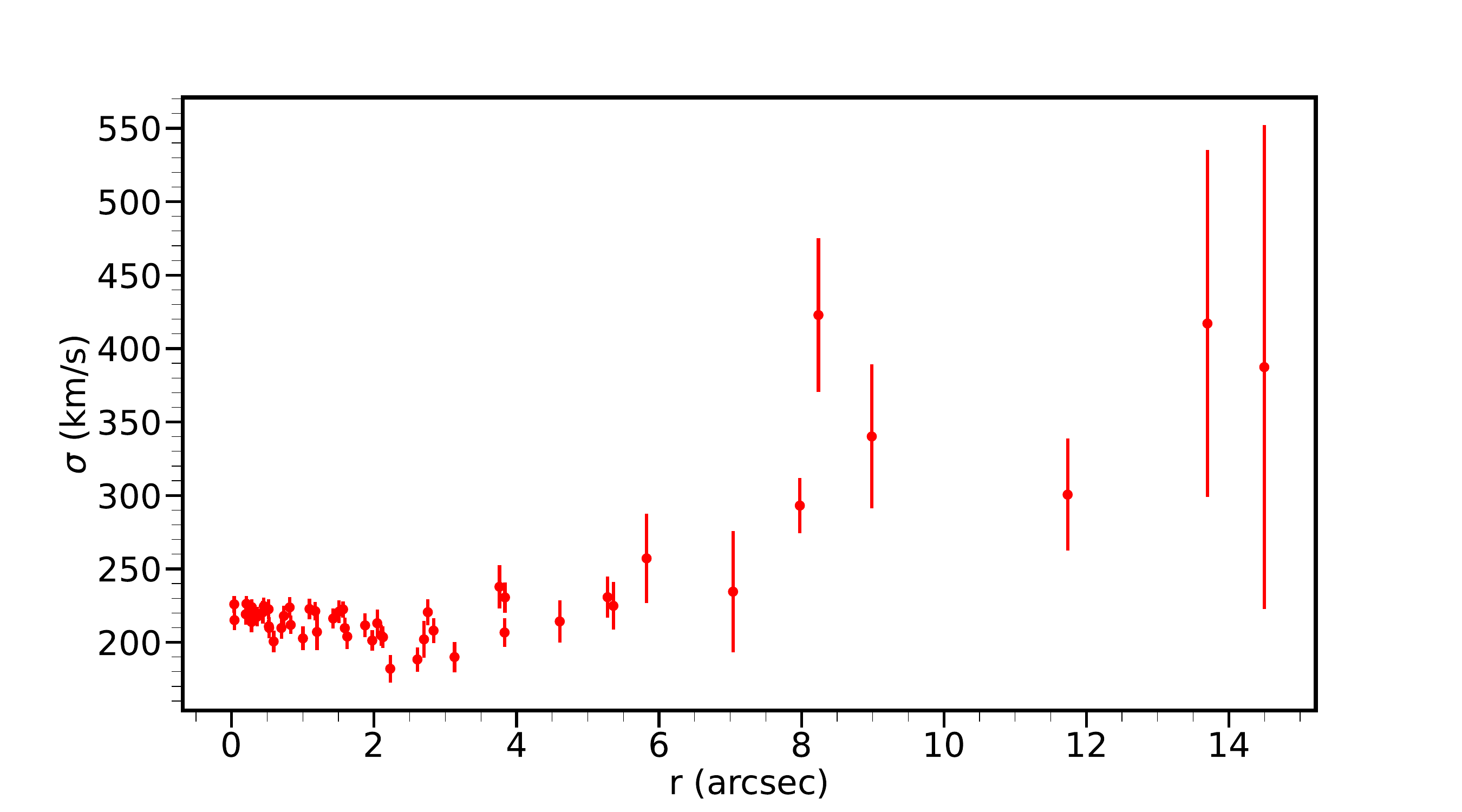}}

    \caption{Left: histogram of the central velocity dispersion $\sigma_0$ for our BCGs. A value of 300 km/s would predict $\mbhm = 10^9$ M$_\odot$ according to the \mbh-$\sigma$ relation of \citet{Rob16}, implying that we should not find any UMBH in our sample. Right: dispersion profile for A2388 (all four slits), showing that $\sigma$ starts increasing from 2" already.}
    \label{Fig.sigma}
\end{figure*}

The most important points can be summarized as follows:
\begin{itemize}
    \item We find one galaxy (A1314, Fig.~\ref{Fig.kin_A1314}) with minor-axis rotation, two galaxies (A1749, A2506, Figs.~\ref{Fig.kin_A1749} and \ref{Fig.kin_A2506}) with major-axis rotation (A1749 also shows a $v$–$h_3$ anti-correlation), and a possible kinematically decoupled core (A2107, Fig.~\ref{Fig.kin_A2107}), with a $\sim$50 km/s rotation in the first 2" and no rotation at larger radii. All other galaxies do not show significant rotation (v/$\sigma < $ 0.1).
    \item As already noted by \citet{Matthias23BCGs}, many BCGs have low central velocity dispersion, sometimes even below 200 km/s (e.g., A1185 and A1982). As shown in Fig.~\ref{Fig.sigma0}, this is indeed the case for our sample as well. Such low values are incompatible with the measured black hole masses (a value of $\sigma = 300$ km/s is needed for $\mbhm = 10^9$ M$_\odot$ according to the \mbh–$\sigma$ relation), stressing the need to use scaling relations involving core properties (core size, central surface brightness, amount of missing light) to estimate \mbh.
    \item An accurate non-parametric reconstruction of the LOSVD wings is crucial to break the mass–anisotropy degeneracy (see e.g., Sec. 4.9 of \citealt{BinneyTremaine2008}). 
    %Quantifying this parametrically via e.g. the h$_4$ coefficient, which measures the kurtosis of the LOSVD itself, does not work in the central regions where one can observe different h$_4$ profiles (see e.g. Fig. 3 of \citealt{Baes05}). 
    Interestingly, in this work we find that this is the case also at larger radii: while the majority of our BCG have positive h$_4$ values, some of them (A240, A399, A634, A2255, A2256, A2388, A2506) show negative h$_4$ values at all radii.
    \item Interactions and/or superpositions along the line of sight of BCGs with other cluster members manifest as LOSVDs with multiple peaks at different velocity ranges. In these cases, after identifying those peaks at $|\mathrm{v}| \gg 0$, we do not take them into account when deriving the moments and discard these bins for the dynamical modeling.
    \item In Fig.~\ref{Fig.sigmaout} we show an example of a velocity dispersion profile that starts increasing outward from 2" already. This feature is observed, although not always so extremely, in most of our BCGs: for A160, A240, A592, A1185, A2255, A2256, A2388, and A2506 the dispersion starts increasing at distances larger than 5", while for A688, A1775, A1982, and A2319 this occurs from 10" on. This could be explained by the fact that BCGs tend to accrete most of their mass in their outer regions, leaving the central regions - and thus the velocity dispersion - largely unaffected, while increasing it in the outskirts. Thus, in practice this increasing profile may reflect the effect of the ICL on the kinematics, as suggested for NGC 6166 by \citet{Bender15}. Another possibility would be the presence of massive DM halos tracing the cluster potential, which is indeed what we find for most BCGs (see Sec.~\ref{Ssec.dyn_res} below). 
    
    %\R{Several galaxies have (asymmetric) rotation signals at large radii but velocity dispersions as high as 800 km/s are almost never reached.}

    %Other options are ruled out: the typical dispersion of a cluster is $\sim$800-1000 km/s \R{Bender?}, and such high values are not reached (see also \R{Matthias?}). Companion galaxies cannot play a role either, since they get discarded when deriving GH moments as described above.

\end{itemize}

%Maybe histogram of sigma_out too (like ell)

\begin{table}
    \centering

    \begin{tabular}{c c c}
    \hline
    Variable & Interval & Number of values \\
    \hline
    \mbh~($\times 10^9$ M$_\odot$) & [1-50] & 150 \\
    $\Gamma$ & [2-20] & 200 \\
    log ($\rho_0$ / M$_\odot$ kpc$^{-3}$)  & [6.5,8.3] & 100 \\
    p$_\mathrm{DM}$ & [0.4-1.0] & 7 \\
    q$_\mathrm{DM}$ & [0.3-1.0] & 8 \\
    $\gamma$ & [0.0-1.4] & 8 \\
    \hline
    \end{tabular}

    \caption{The different parameters (black hole mass, mass-to-light ratio, halo normalization at 10 kpc, halo flattenings and inner logarithmic slope) used to build the potential (eq.~\ref{eq.dens_SMART}), along with the probed range and the number of sampled values. These are generated using linear spacing.}
    \label{Tab.NOMAD_sampling}
\end{table}

\section{Dynamics}

Having described the photometry and the spectroscopy, we can now turn to the dynamical models of the BCGs, presenting general results in Sec.~\ref{Ssec.dyn_res} (comments on individual objects can be found in Sec.~\ref{Sec.comments}).

\subsection{Methodology}  \label{Ssec.dynamics_methodology}
The dynamical models are performed using our triaxial Schwarzschild code SMART \citep{Bianca21}. SMART can handle both parametric and non-parametric stellar and dark matter (DM) density profiles. The best-fitting model is identified by minimizing AIC$\mathrm{p} = \chi^2 + 2 m_\text{eff}$ \citep{Mathias21, Jens22}. The gravitational potential itself is constructed as

\begin{equation}
\rho_\text{TOT} = \mbhm \times \delta (r) + \Gamma \times \rho_* + \rho_\text{DM}
    \label{eq.dens_SMART}
\end{equation}

with \mbh\,the central point-like contribution from the SMBH, $\rho_*$ is the deprojected light density and $\rho_\text{DM}$ is the triaxial DM density, parametrized using a \citet{Zhao1996} model

\begin{equation}
  \rho_\text{DM} (m) = \frac{\rho_0}{\text{p}_\text{DM} \cdot \text{q}_\text{DM} \cdot \left(\frac{m}{m_0}\right)^{\gamma} \left[1 + \left(\frac{m}{m_0}\right)^{1/\alpha} \right]^{(\beta - \gamma)/\alpha}}
    \label{eq.Zhao_models}
\end{equation}

\noindent with flattenings p$_\text{DM}$ and q$_\text{DM}$ and ellipsoidal radius 
\begin{equation}
m=\sqrt{x^2+\frac{y^2}{p_{\mathrm{DM}}^2}+\frac{z^2}{q_\mathrm{DM}^2}}.
\end{equation}
While we fix $\alpha = 1$, $\beta = 3$ and m$_0$ = 150 kpc we leave the inner logarithmic slope $\gamma$ as free parameter, along with the normalization $\rho_0$ specifying the DM density at 10 kpc. In eq.~\ref{eq.dens_SMART}, the stellar mass is obtained by multiplying $\rho_*$ by the parameter $\Gamma$, implying that the light density is not fitted but treated as a constraint by the code. Therefore, to fully specify the gravitational potential, we need to assume six free parameters (\mbh, $\Gamma$, p$_\text{DM}$, q$_\text{DM}$, $\gamma$, $\rho_0$). Together with the three viewing angles this leads to a total of nine unknown parameters. \\
Once the potential is constructed, a time-averaged orbit library is generated within this potential. 
%The shape of the potential depends of the radius: while in the central region the potential is Keplerian due to the BH, in the region where the stars dominate the phase space can be 5-dimensional. 
The orbits are then projected onto the plane of the sky, the predicted kinematics are computed, and the stellar orbital weights are iteratively adjusted to maximize an entropy-like function, ensuring a smooth and physically plausible orbital distribution and at the same time minimizing kinematic differences:

\begin{equation}
\hat{S} = S - \alpha\chi^2
    \label{eq.Shannon}
\end{equation}

\noindent where $S$ is related to the Shannon entropy, $\alpha$ is a smoothing term and $\chi^2$ compares the differences between fitted and observed LOSVDs:

\begin{equation}
       \chi^2 = \sum_{i=1}^{N_{\text{losvd}}} \sum_{j=1}^{N_{\text{vel}}} \left(     \frac{\text{LOSVD}_{\text{model}}^{i,j} - \text{LOSVD}_{\text{data}}^{i,j} }{\Delta\text{LOSVD}_{\text{data}}^{i,j}} \right)^2.
        \label{eq.chi2_Schw}
    \end{equation}

\noindent and it is then used to compute AIC$_\mathrm{p}$ and find the best-fit model.  \\
The ranges of the modeled parameters along with the number of sampled values are reported in Tab.~\ref{Tab.NOMAD_sampling}. The search for the minimum is performed using the optimizer NOMAD \citep{Audet06, LeDigabel11}, with each model being run in parallel on 12 cores on our MPCDF machines using 23000 stellar orbits. For \mbh, $\Gamma$, $\rho_{\mathrm{DM},0}$ and $\gamma$ we sample a broad range to account for a large number of potentials. To select the lower limit for the DM halo flattenings, we evaluate the smallest $p$ and $q$ from the deprojection starting from $\sim$50 kpc and use these values since the deprojection shape is expected to trace the underlying DM distribution \citep{dN22BCGs}, whereas the upper limits are set to 1.0 to allow for spherical halos. The orientations are determined as in \citet{dN24}, thus first launching a NOMAD run using the deprojection whose shape profiles are closest to the average ones, finding the best-fit mass parameters and subsequently launching a second NOMAD run, this time with \textit{all} mass parameters fixed and only fitting for the orientation. A third, final NOMAD run is then launched to find the best-fit mass distribution using the viewing angles recovered in the second run. \\
To determine the uncertainties on the mass parameters, we follow \citet{Bianca21} by splitting the kinematic bins into two halves and modeling them independently. The best-fit result is then taken as the average of the two results, while the uncertainty is half the width of the interval. If this is not possible due to a low number of kinematic bins, we model all bins together and adopt the relative errors from \citet{dN24}. In that paper, we used mock kinematics derived from the N-body simulations of \citet{Rantala18, Rantala19} to reproduce the long-slit MODS data used to model NGC 708 and estimate uncertainties on the mass parameters using nine different mock runs.\\
For each galaxy, the orbital initial conditions are sampled from a radius r$_\mathrm{min}$ = 0.1" up to a variable radius r$_\mathrm{max}$. This is chosen to be a factor 1.1 larger than the outermost radius\footnote{It is important to have photometric coverage up at least 5 times the largest kinematics radius to allow for the integration of the most eccentric stellar orbits of the library.} at which the deprojection is sampled. We report the exact values in Tab.~\ref{Tab.dyn_details}, along with the number of bins each galaxy has.

\begin{table*}
    \centering
\begin{tabular}{c c c c c}
\hline
Cluster & N$_\mathrm{LOSVD}$ & N$_\mathrm{vel}$ & r$_\mathrm{depro}$ (arcsec) & r$_\mathrm{kin}$ (arcsec) \\
\hline
%A150 &  &  &  &   \\
A160 & 32 & 29 & [0.563 - 103] & [0.11 - 7.3]   \\
A240 & 37 & 25 & [0.675 - 81.8] & [0.059 - 6.4]   \\
A292 & 32 & 29 & [0.512 - 80.4] &  [0.079 - 7.6] \\
A399 & 17 & 31 & [0.632 - 58.5] & [0.41 - 7.9]    \\
A592 & 31 & 27 & [0.572 - 43.5] & [0.063 - 4.9]   \\
A634 & 36 & 27 & [0.483 - 101] & [0.20 - 10]   \\
%A688 &  &  &  &    \\
A1185 & 20 & 27 & [0.600 - 104] & [0.13 - 12]   \\
A1314 & 92 & 35 & [0.504 - 103] & [0.061 - 6.7]   \\
A1749 & 55 & 33 & [0.384 - 55.7] & [0.17 - 5.5]   \\
A1775 & 67 & 29 & [0.563 - 63.8] &  [0.085 - 6.2]   \\
%A1982 &  &  &  &    \\
A2107 & 49 & 33 & [0.379 - 124] & [0.046 - 11]    \\
A2147 & 54 & 27 & [0.374 - 91.0] & [0.085 - 9.6]   \\
%A2255 &  &  &  &    \\
A2256 & 19 & 29 & [0.635 - 70.3] & [0.37 - 8.2]   \\
A2319 & 28 & 31 & [0.642 - 76.5] & [0.15 - 6.5]   \\
A2388 & 36 & 23 & [0.368 - 84.0] & [0.21 - 5.8]  \\
A2506 & 38 & 25 & [0.318 - 106] & [0.036 - 18]   \\
%A2665 &  &  &  &    \\
\hline
\end{tabular}

    \caption{Details about the dynamical models. \textit{Col. 1}: Cluster; \textit{Col. 2}: Number of modeled LOSVD for each galaxy; \textit{Col. 3}: Number of velocity bins used to sample the LOSVD; \textit{Cols. 4-5}: Radial extensions of the deprojection and the kinematics.}
    \label{Tab.dyn_details}
\end{table*}

%For 9 galaxies, we have derived M/L profiles from SSP fitting \citep{Thomas03, Maraston05} assuming a Kroupa IMF, using Lick indices also obtained with MODS observations at LBT. By comparing these SSP-based values with our dynamical estimates—specifically in the region where stellar mass dominates the potential (e.g., see Fig.~\ref{Fig.mass_profile}) we infer the IMF shape. We find a non-universal IMF: most galaxies exhibit a Kroupa IMF (A160, A292, A1185, A1314, A2147), while A592, A634, and A2506 show a Salpeter-like IMF.\\
\subsection{General results}  \label{Ssec.dyn_res}
\begin{figure*}

\subfloat{\includegraphics[width=.3\linewidth]{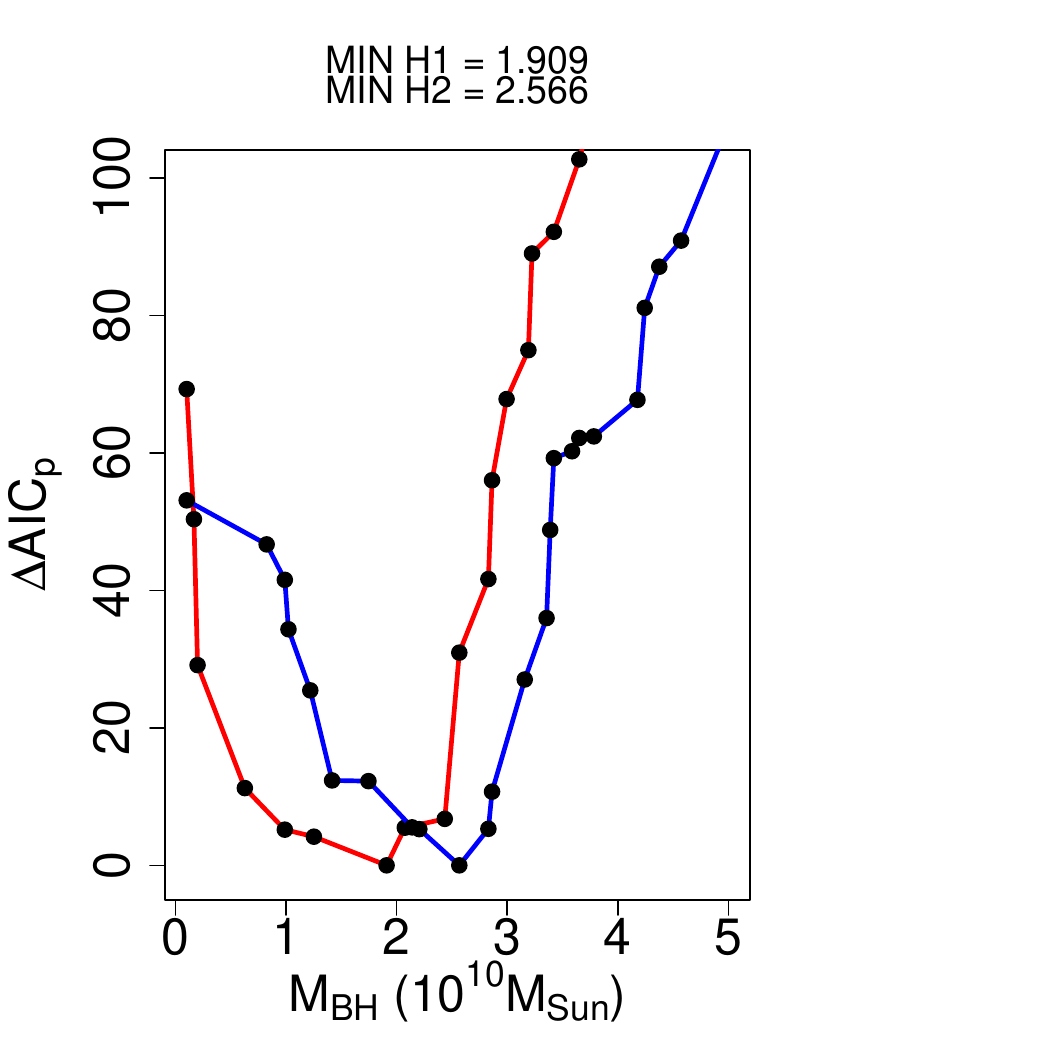}}
\subfloat{\includegraphics[width=.3\linewidth]{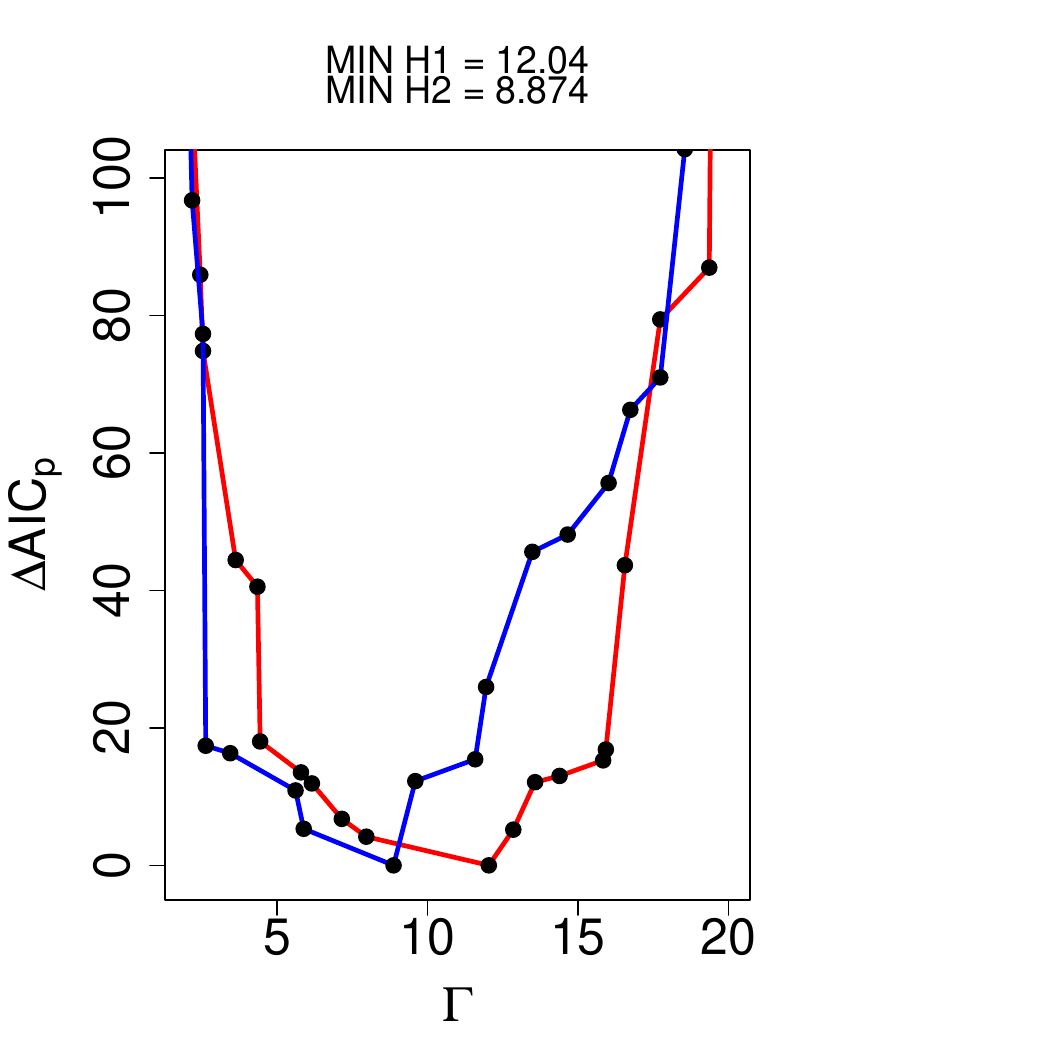}}
\subfloat{\includegraphics[width=.3\linewidth]{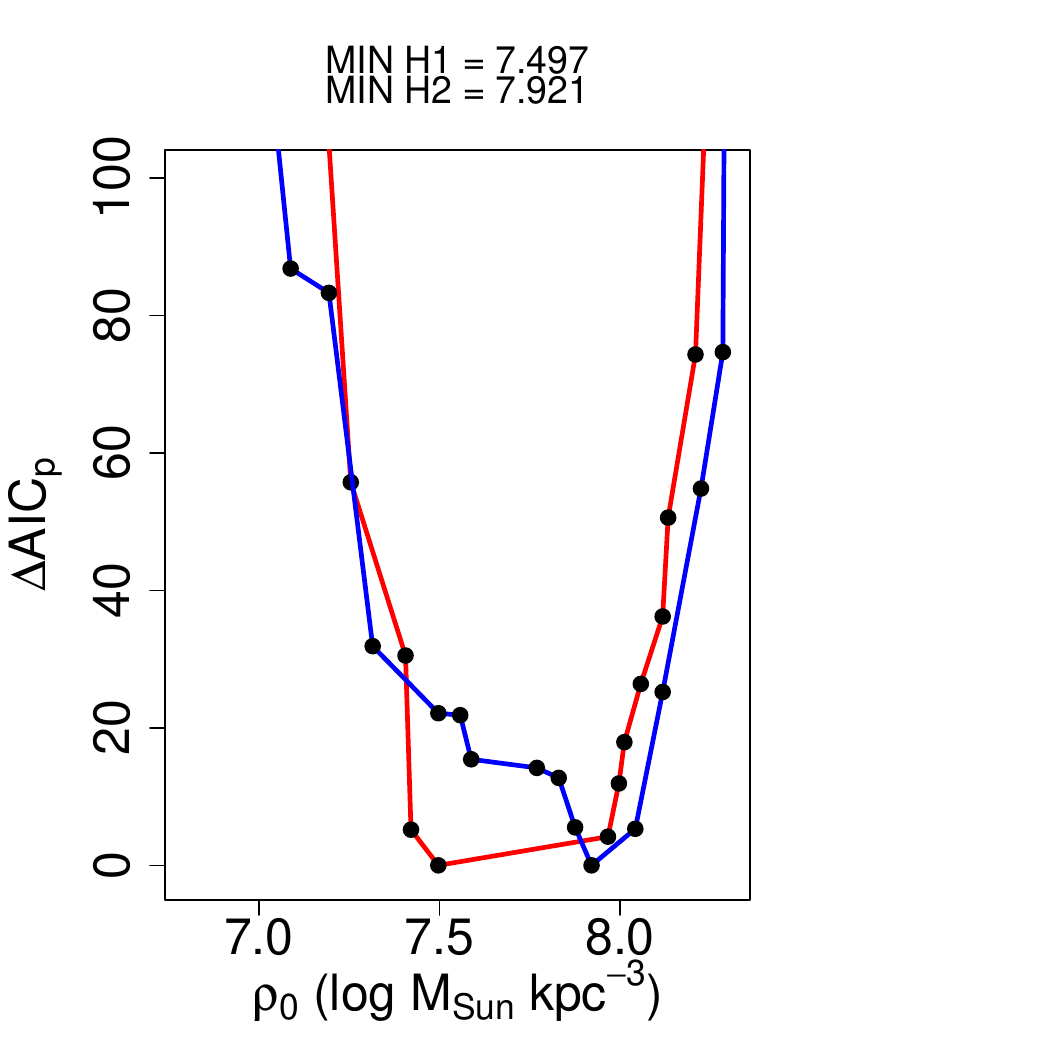}}

\subfloat{\includegraphics[width=.3\linewidth]{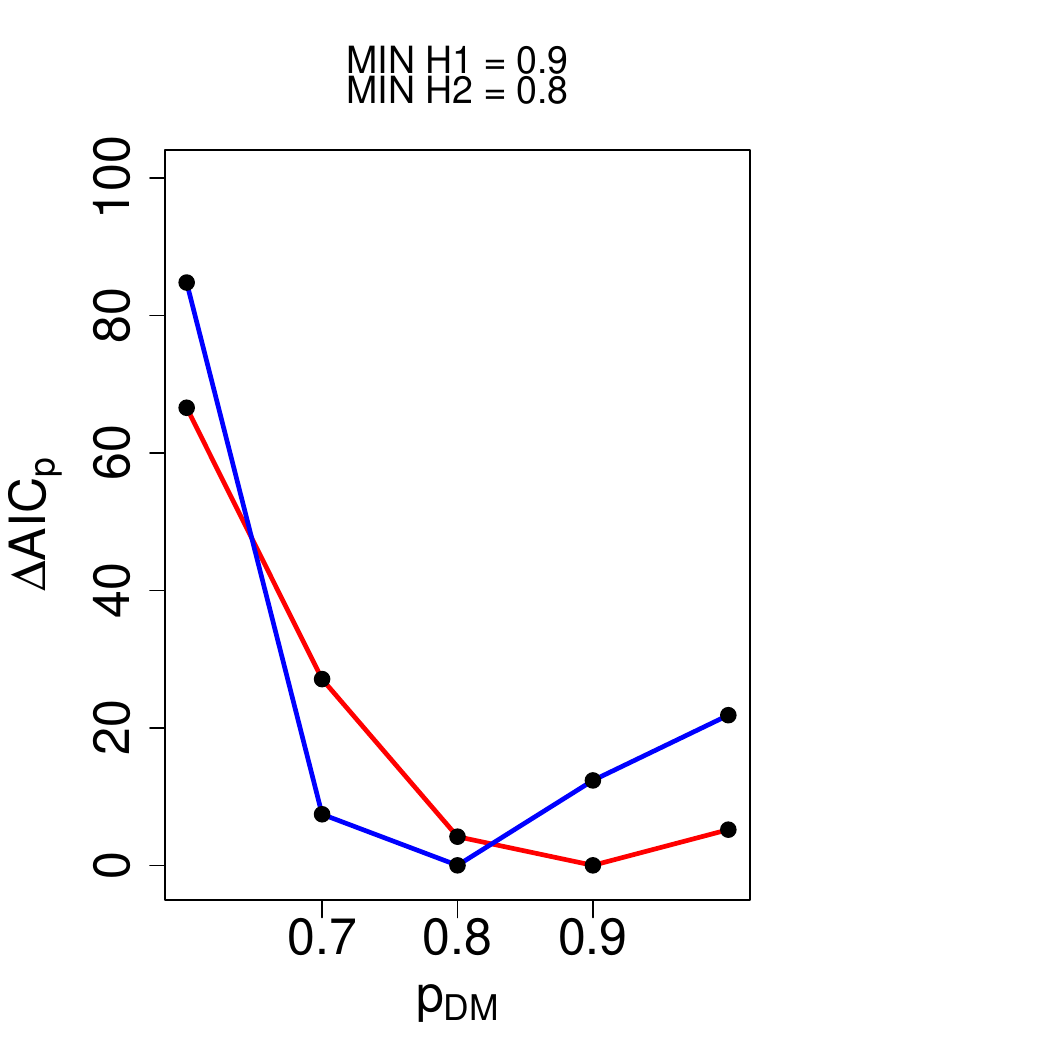}}
\subfloat{\includegraphics[width=.3\linewidth]{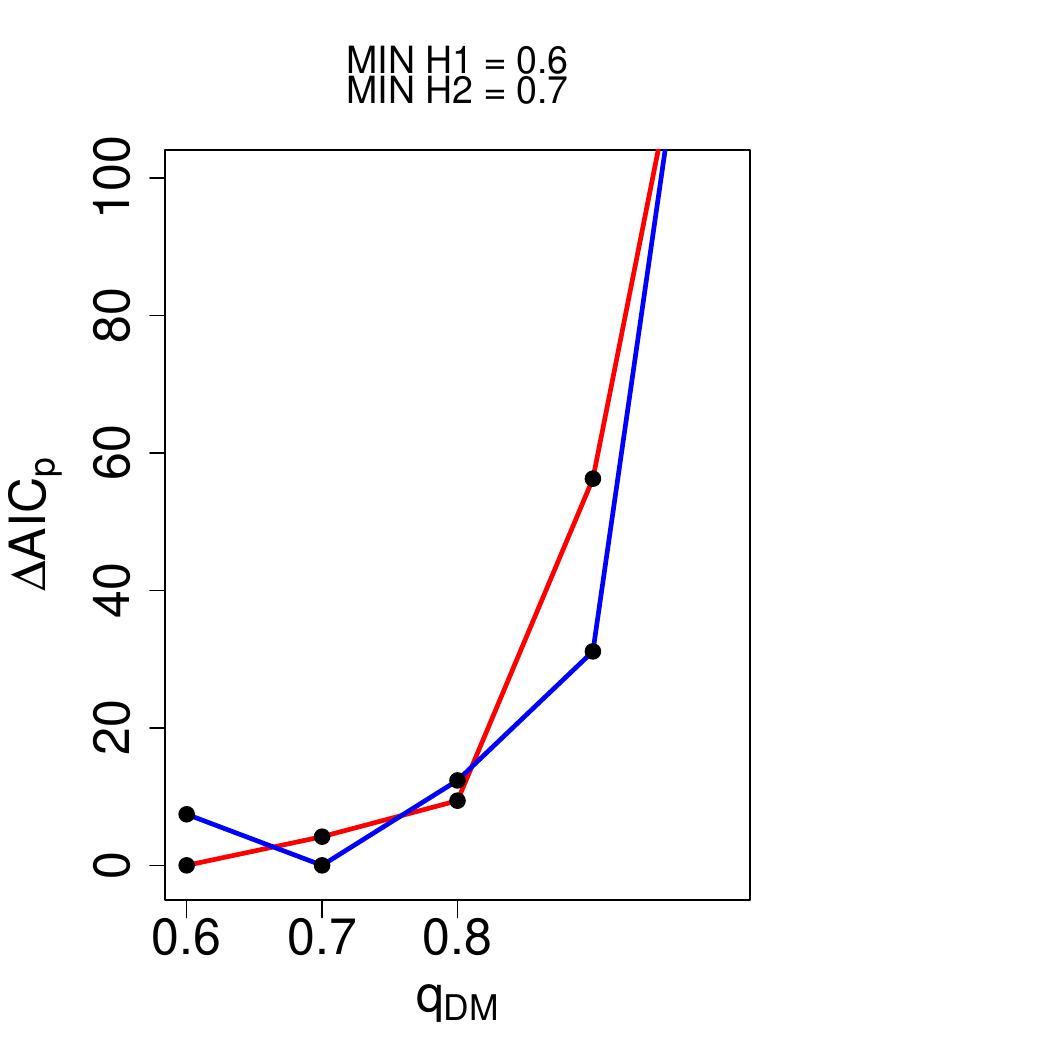}}
\subfloat{\includegraphics[width=.3\linewidth]{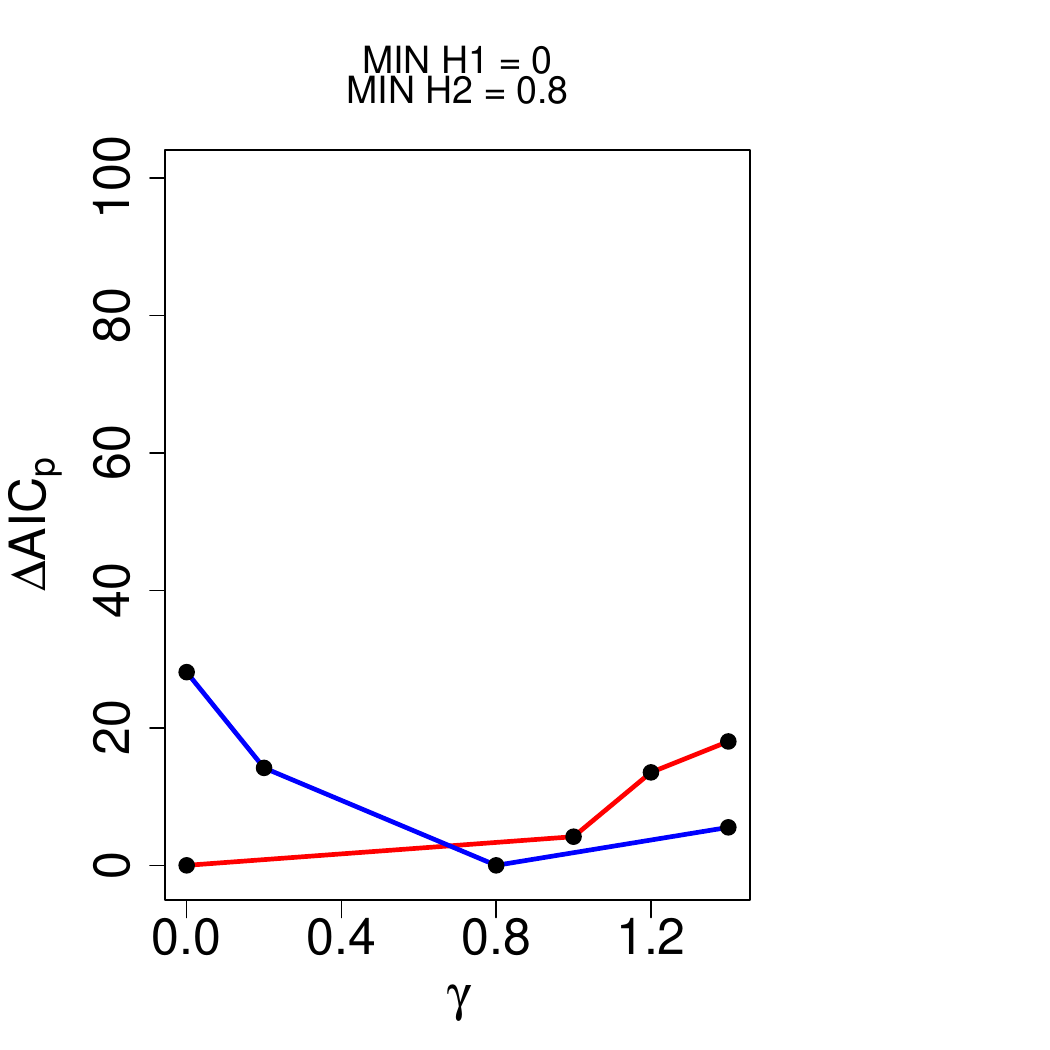}}

\caption{AIC$_\text{p}$ values plotted against the six parameters fitted in our final NOMAD run for A2107. Red and blue points represent the individual models for each half of the galaxy (H1 and H2), while the solid lines trace the best-fit model at each tested parameter value. \textit{Left to right, top to bottom:} \mbh, $\Gamma$, $\rho_0$, p$_\mathrm{DM}$, q$_\mathrm{DM}$, $\gamma$.}
\label{Fig.Schw_results}
\end{figure*}

\begin{figure}
    \centering
    \includegraphics[width=\linewidth]{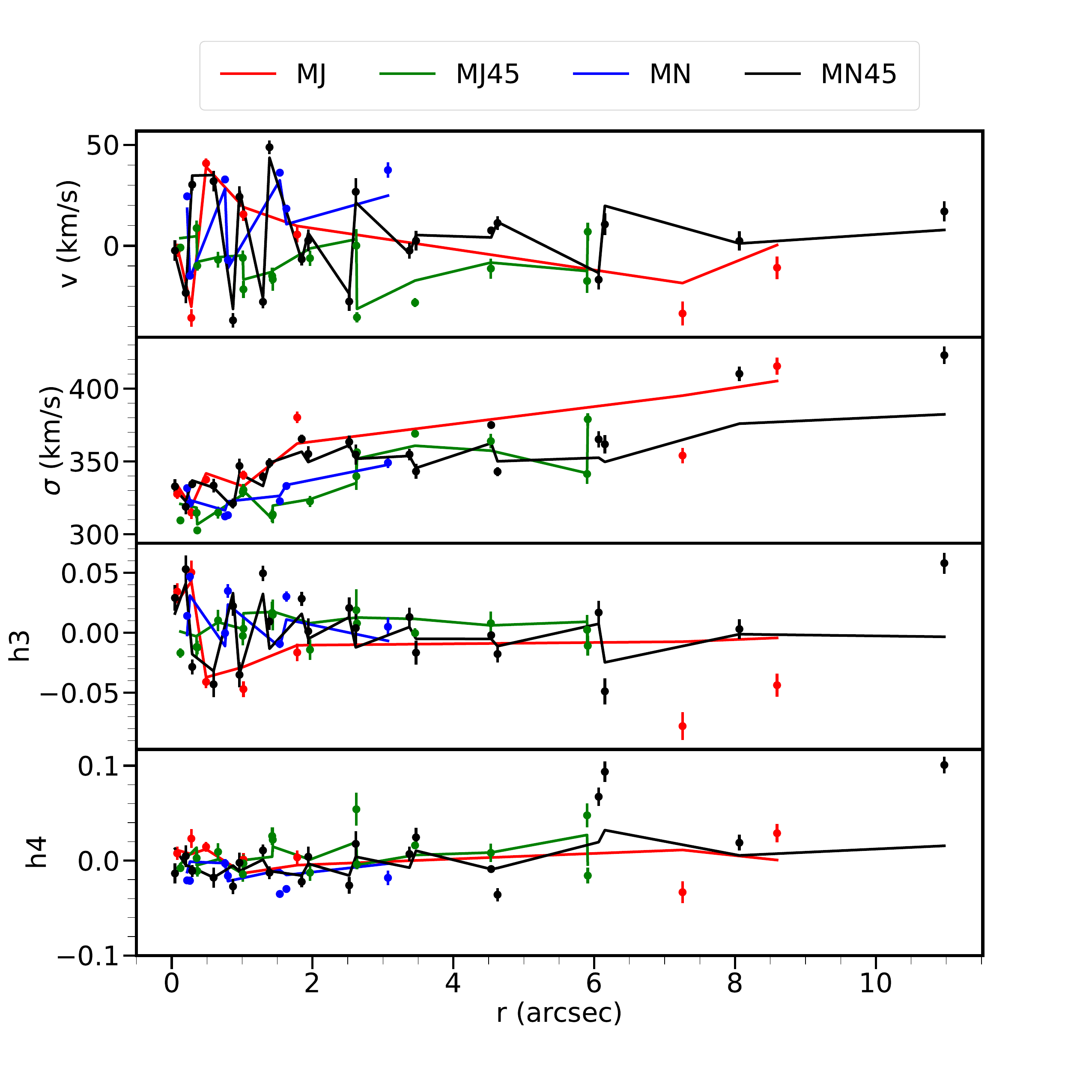}
    \caption{Radial profiles of the Gauss-Hermite moments (v, $\sigma$, h$_3$, h$_4$) for A2107, shown for all four slits (colored points), with the best-fit Schwarzschild model overplotted. Different colors indicate different slit orientations.}
    \label{Fig.kinfit}
\end{figure}

\begin{figure}
    \centering
    \includegraphics[width=0.8\linewidth]{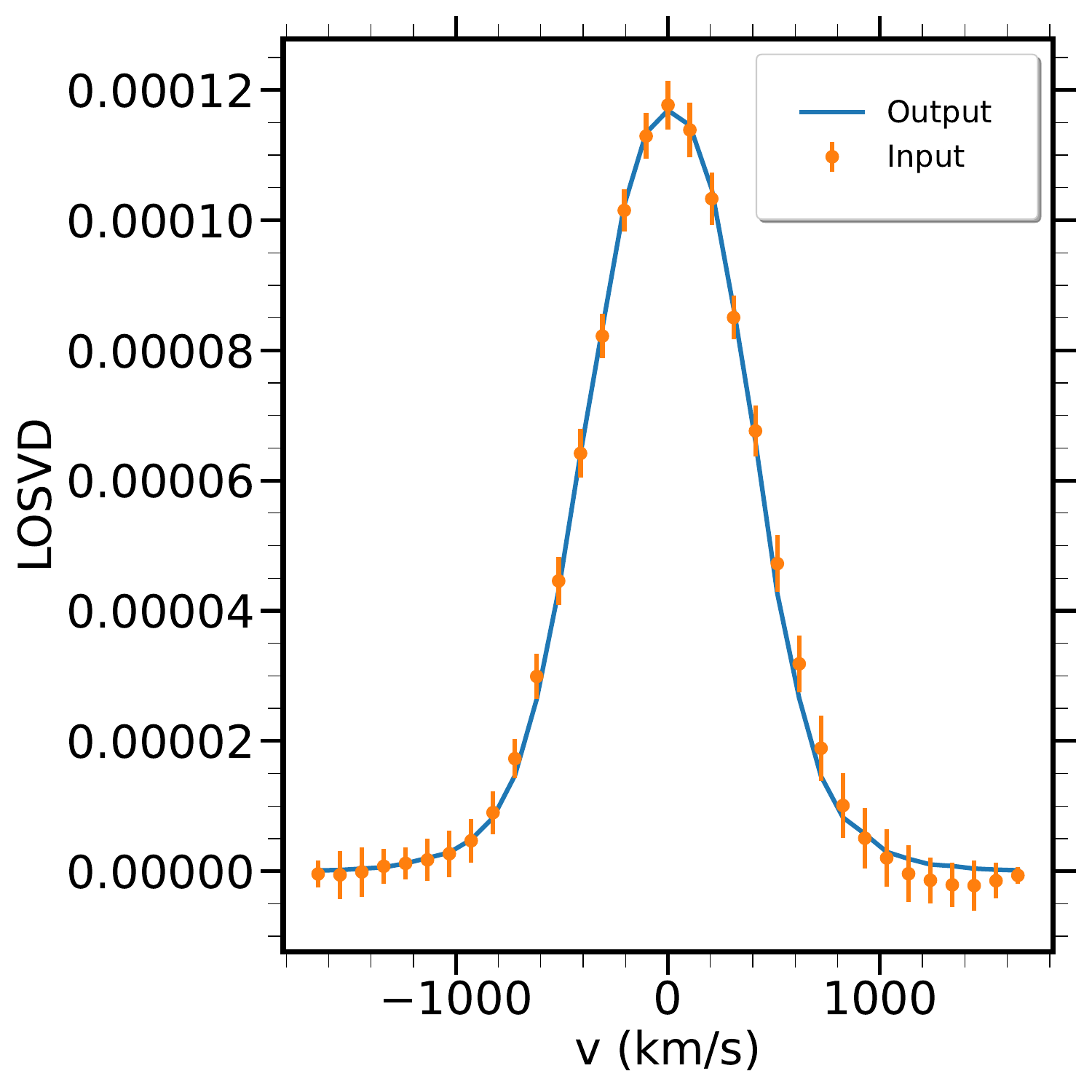}
    \caption{Non-parametric LOSVD determined using our code WINGFIT (orange points) along with the best-fit parametric LOSVD (solid blue line) found by SMART. The bin belongs to A2107 along MN45 (see Sec.~\ref{Ssec.spec_data}) and its coordinates are (4.11139, -1.91717) in a NW frame of reference.}
    \label{Fig.LOSVD_comparison}
\end{figure}

\begin{figure*}
    \centering
    \subfloat[\label{Fig.mass_profile}]{\includegraphics[width=.35\linewidth]{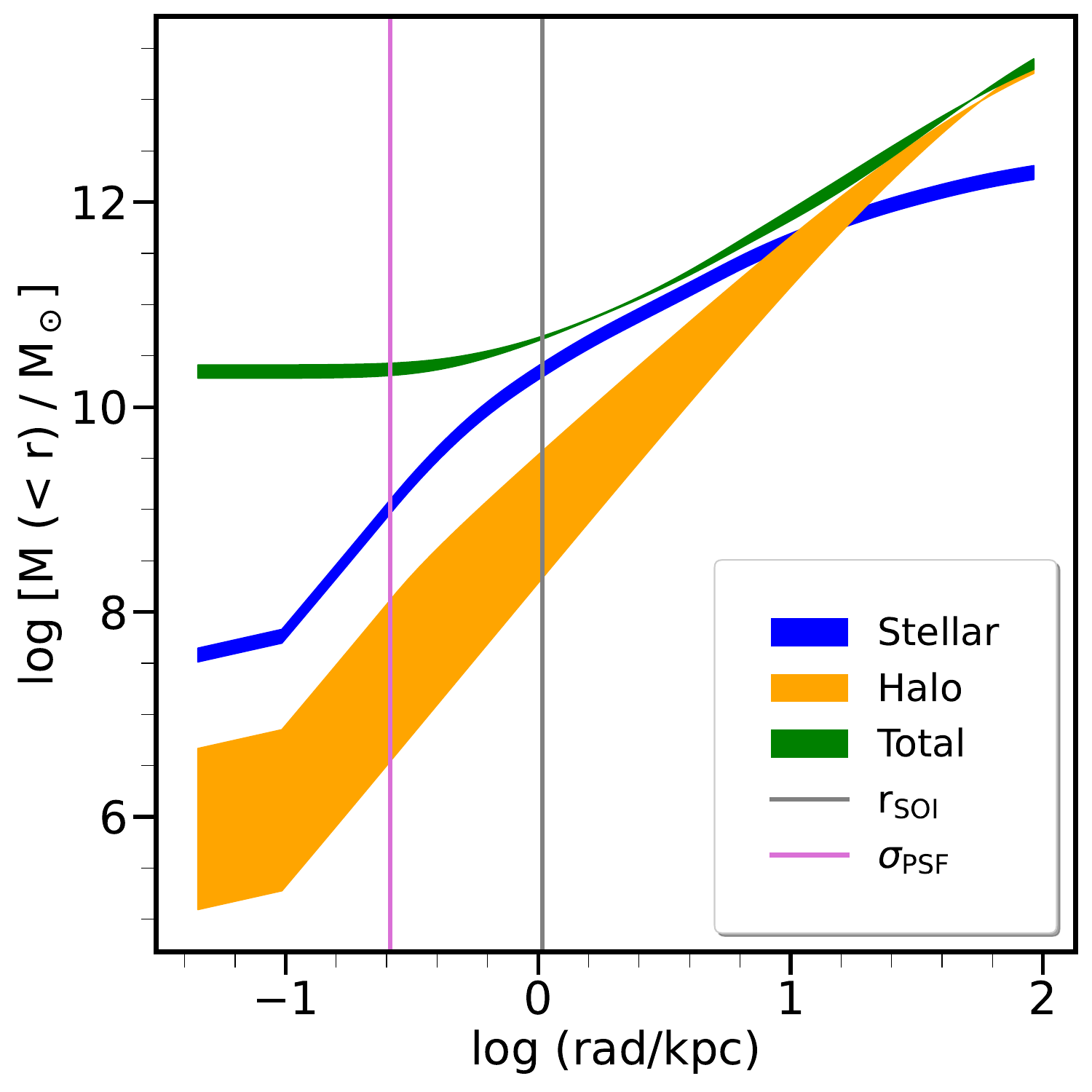}}
    \subfloat[\label{Fig.tot_star}]{\includegraphics[width=.35\linewidth]{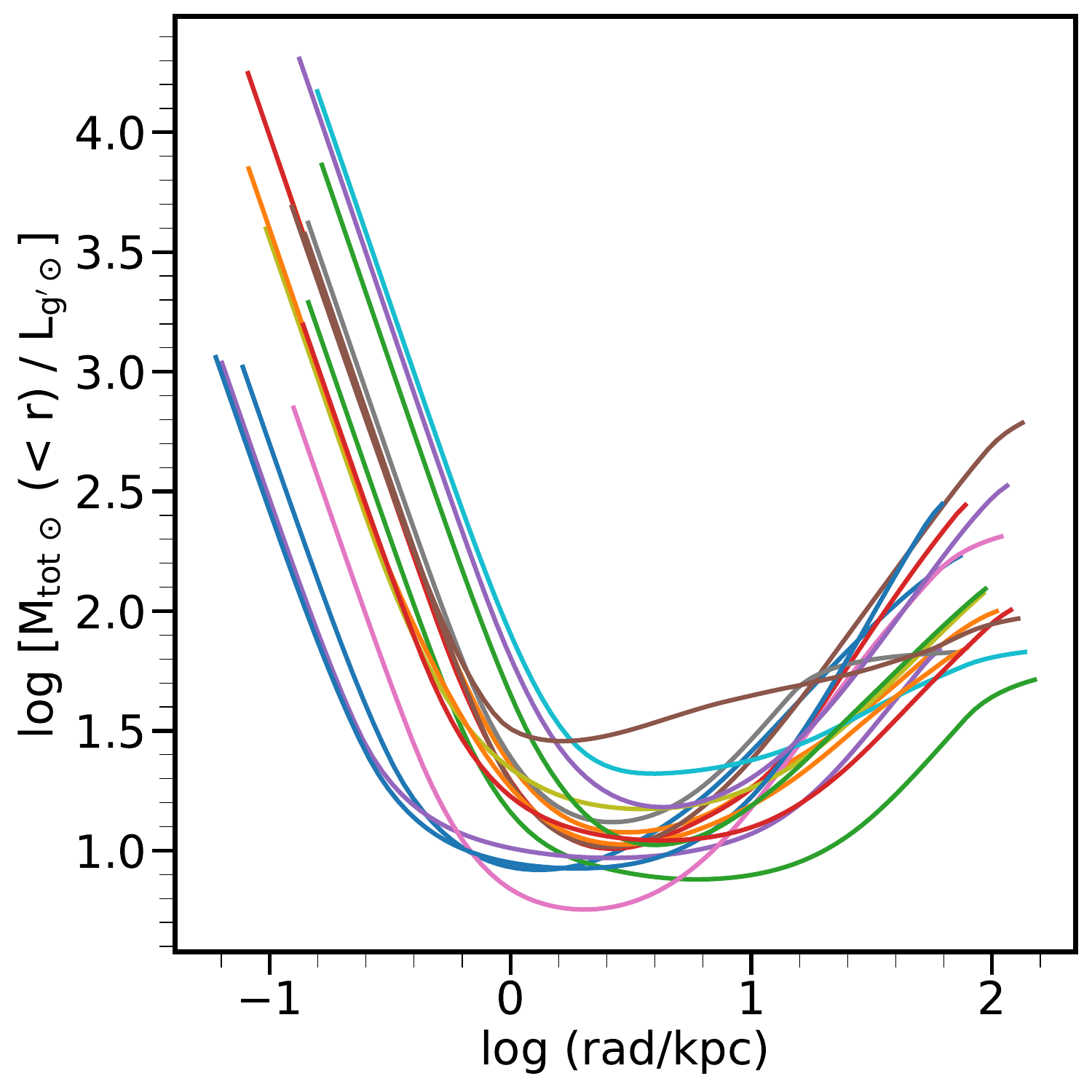}}

    \caption{Left: Integrated dynamical mass profiles of A2107. The shaded region encloses the results from the two galaxy halves. The grey vertical line marks the black hole sphere of influence (SOI), whereas the purple line is the PSF from the kinematics $\sigma_\mathrm{PSF}$ = FWHM/2.35. Right: Total mass-to-light (M$_\text{tot} = \mbhm + \text{M}_* + \text{M}_\text{DM}$) profiles for all BCGs in our sample. This provides an immediate visual insight into which of the three mass components — BH, stars, or DM — dominates the potential.}
    \label{Fig.mass_total_stars}
\end{figure*}

\begin{figure*}

\subfloat[\label{Fig.rSOI_PSF_hist}]{\includegraphics[width=.3\linewidth]{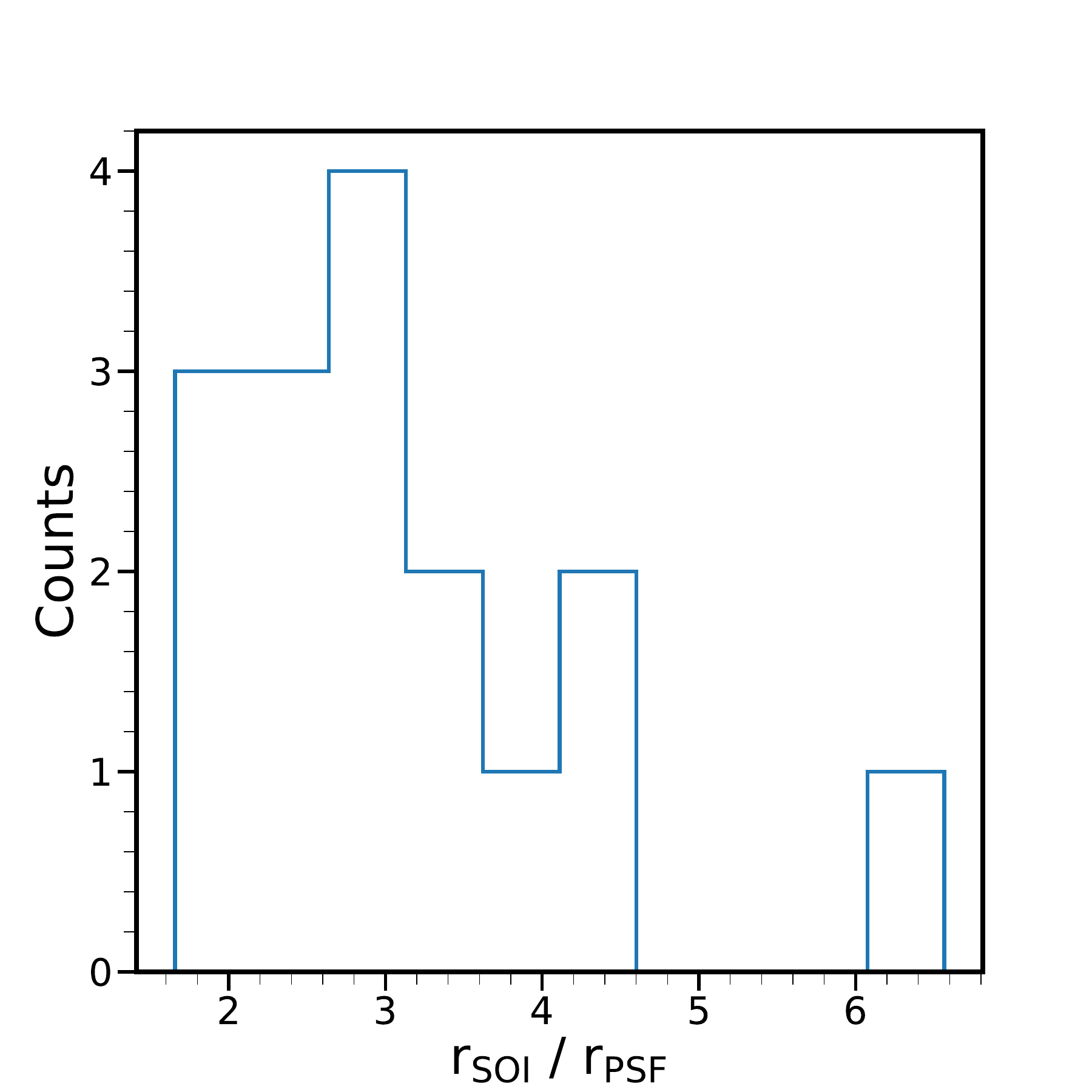}}
\subfloat[\label{Fig.rmax_rDM_hist}]{\includegraphics[width=.3\linewidth]{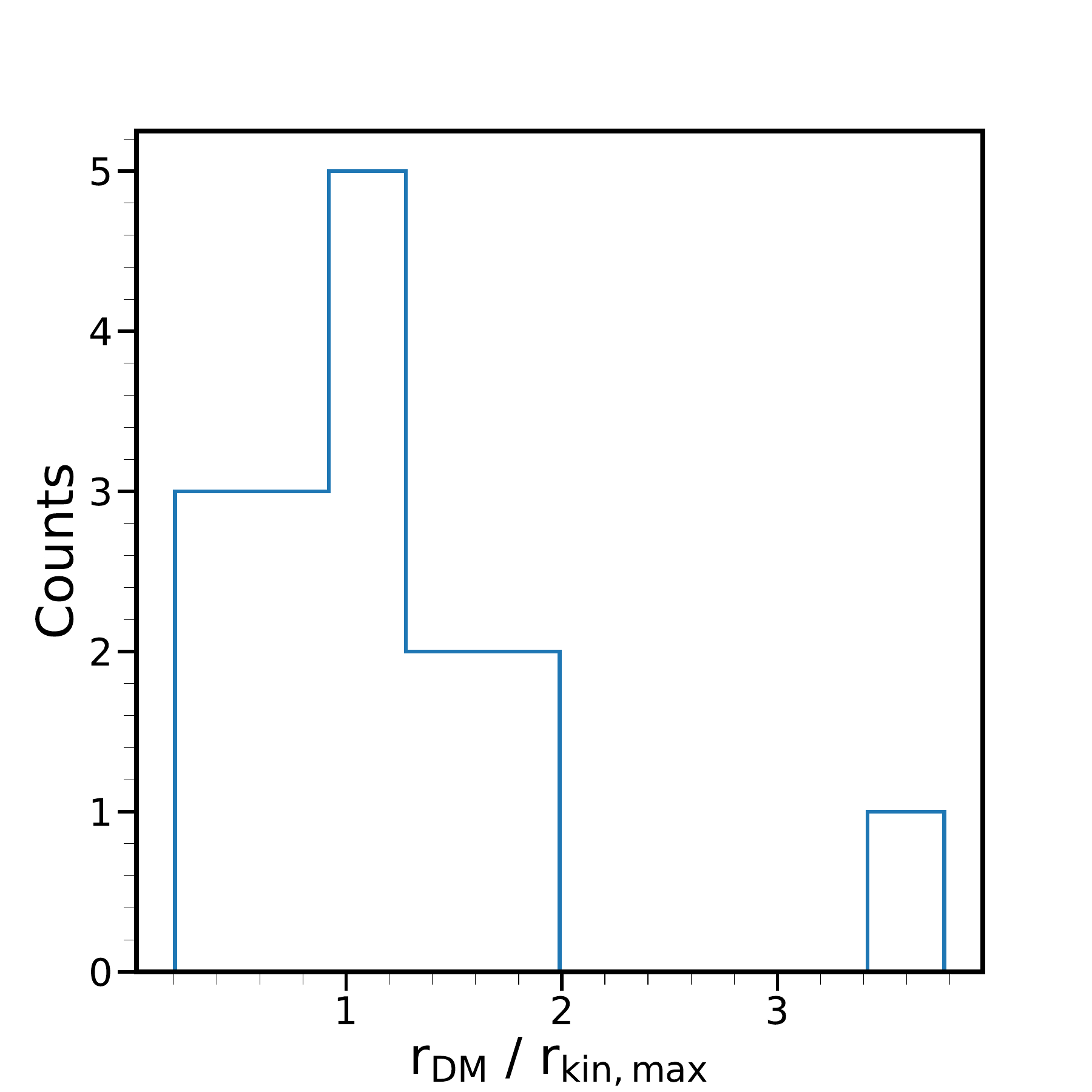}}

    \caption{Left: Histogram of the ratio between the size of the BH sphere of influence and the kinematics PSF. The fact that this ratio is well above unity shows that we robustly resolve the sphere of influence for all galaxies. Right: Histogram of the ratio between the radius of the farthest modeled kinematics bin and the radius from which DM dominates the potential. For five galaxies, this is smaller than unity, implying that the properties of the DM halo might not be well constrained.}
    \label{Fig.dyn_hists}
\end{figure*}

\begin{figure*}

\subfloat[A1749, tangential anisotropy. \label{Fig.beta_tang}]{\includegraphics[width=.3\linewidth]{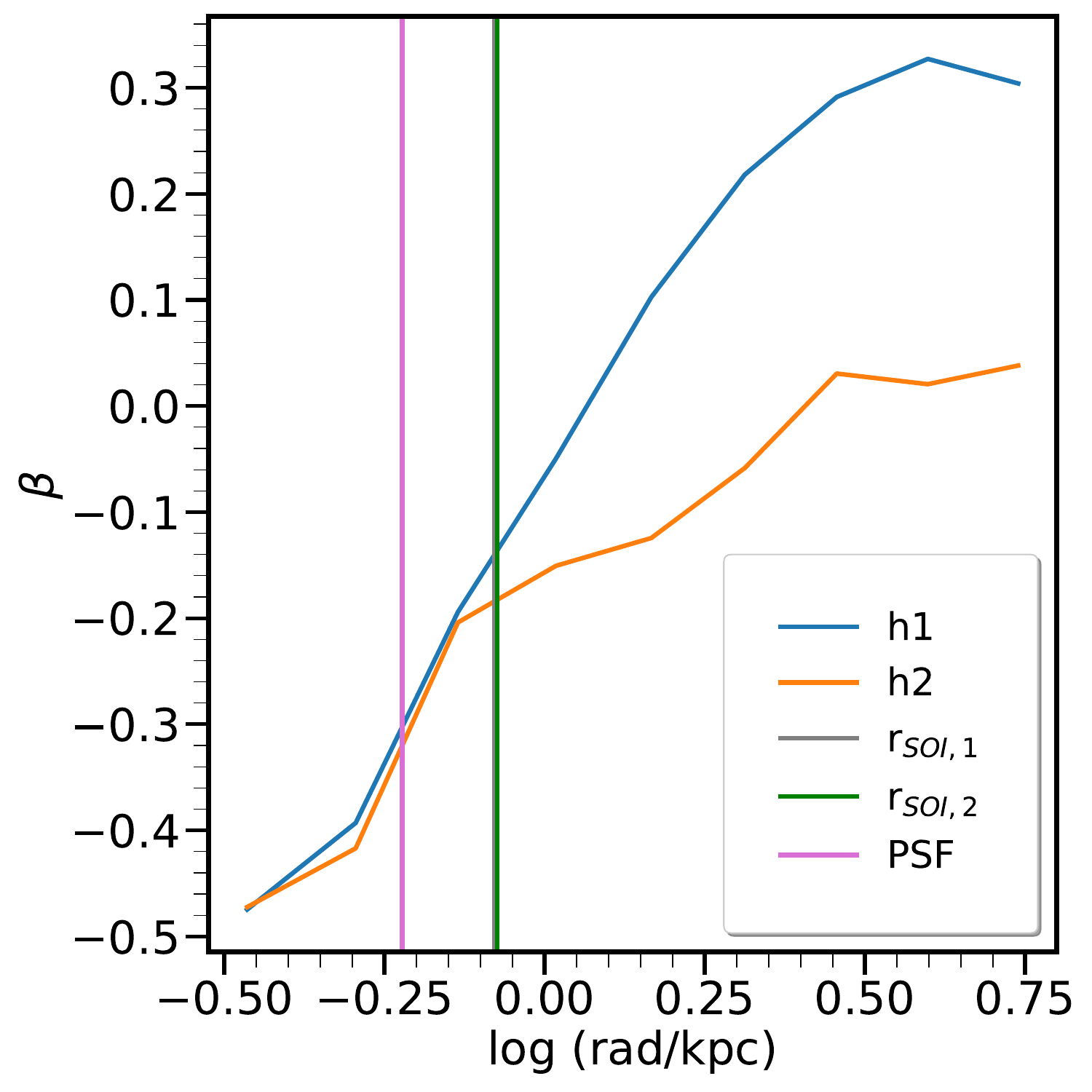}}
\subfloat[A2388, strong tangential anisotropy. \label{Fig.beta_strong_tang}]{\includegraphics[width=.3\linewidth]{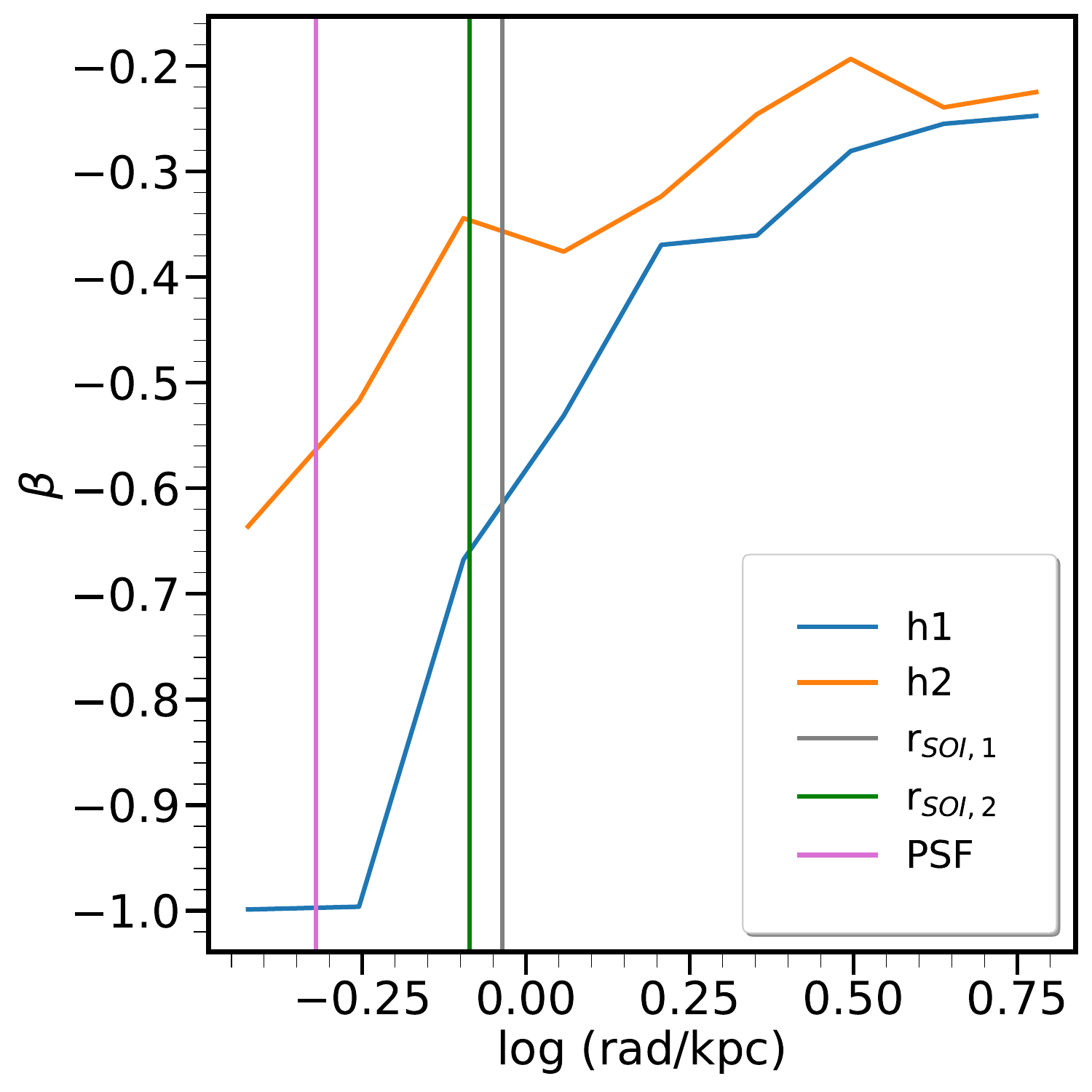}}

\subfloat[A399, weak tangential anisotropy.\label{Fig.beta_isotr}]{\includegraphics[width=.3\linewidth]{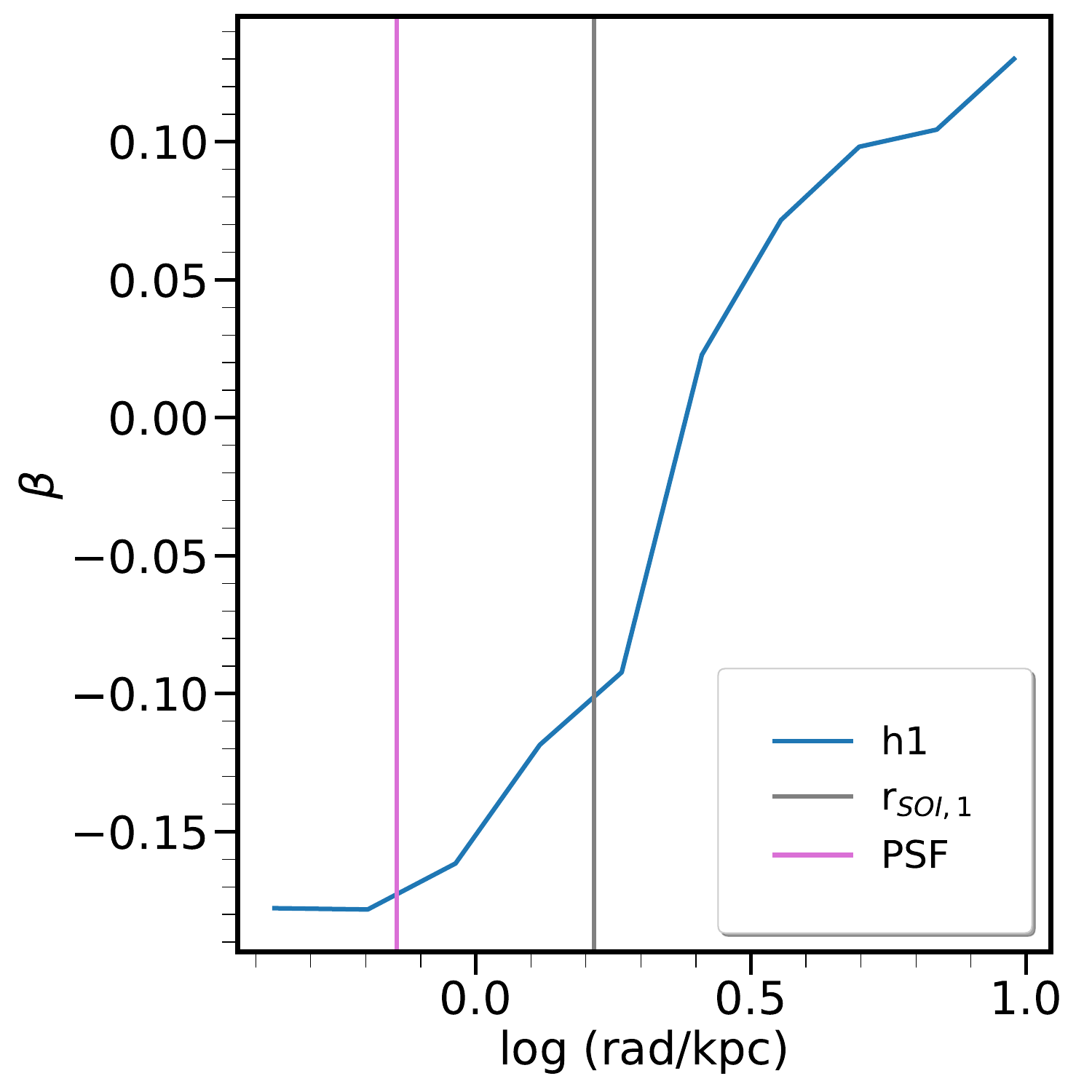}}
\subfloat[All BCGs with SINFONI galaxies. \label{Fig.beta_all}]{\includegraphics[width=.3\linewidth]{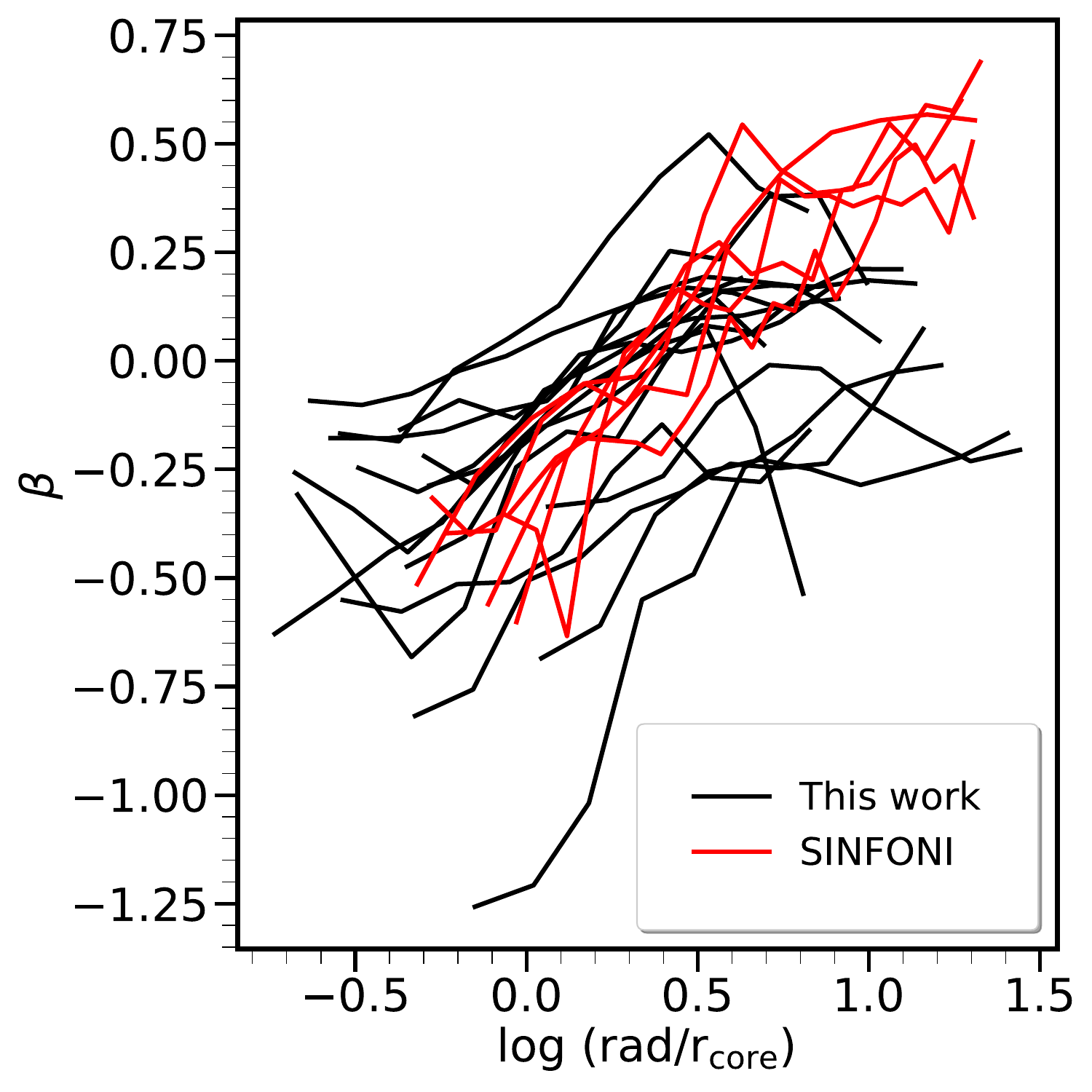}}

\caption{(a-b-c): Example of different anisotropy profiles derived from the dynamical models. Solid lines represent the anisotropy profile (one or two depending on whether both galaxy halves were modeled, see Tab.~\ref{Tab.dyn_results}), whereas vertical gray/green lines denote the sphere of influence, defined as \mbh\,= M$_*$ ($<$ r$_\mathrm{SOI}$). The purple line marks the FWHM of the kinematic PSF. (d): Anisotropy profiles for all BCGs (black lines) with superimposed SINFONI galaxies (red, see \citealt{Jens14}. Note that in this last figure the radius is scaled to the core sizes for better comparison.}
\label{Fig.anisotropy}
\end{figure*}

Among all our results, summarized in Tab.~\ref{Tab.dyn_results}, the highlight is the discovery of 8 new UMBHs in the galaxies A160, A292, A1185, A1749, A1775, A2107, A2147, and A2256, in addition to the already published case of A262. This discovery more than doubles the number of such systems known so far and helps fill the high-mass end of the \mbh-host scaling relations, as well as confirming that the presence of a large SB core is a strong indicator that the galaxy hosts a very massive BH. The \mbh-AIC$_\text{p}$ distributions for all BCGs are shown in App.~\ref{App.MBH_AICp}. \\
In Fig.~\ref{Fig.Schw_results} we show an example of AIC$_\text{p}$ plotted against the mass parameters of A2107 - the most massive BH of the sample - while the corresponding fit to the kinematic moments is presented in Fig.~\ref{Fig.kinfit}, showing that our models can accurately reproduce the observed kinematics. Furthermore, in Fig.~\ref{Fig.LOSVD_comparison} we show a model-to-data comparison for one of the LOSVDs belonging to A2107, measured along MN45. \\
From the best-fit mass parameters we derive the mass distribution of the galaxy and show it in Fig.~\ref{Fig.mass_profile}: in the central region (sphere of influence), the potential is Keplerian as the BH dominates, followed by a region where stellar mass contributes the most to the potential before DM takes over at large radii. This can be seen even better in Fig.~\ref{Fig.tot_star}, where we show the \textit{total} mass-to-light ratio M$_\text{tot}$/L as a function of radius. For most galaxies, the region where stars contribute most to the potential extends from the BH SOI up to 10 kpc; as shown, e.g., in \citet{dN24}, the minimum value can be used to derive a more accurate estimate of the \textit{stellar} mass-to-light ratio in cases where the DM traces the light. \\
To ensure that our kinematical data robustly resolve the BH sphere of influence, we evaluate its extension by computing the radius r$_\mathrm{SOI}$ as $\mathrm{M}_* (\mathrm{r}_\mathrm{SOI}) = \mbhm$, i.e. as the radius at which the integrated stellar mass equals the black hole mass. The fact that the radii are well larger than $\sigma_\mathrm{PSF} = $ FWHM/2.35 for all galaxies (see Fig.~\ref{Fig.rSOI_PSF_hist}) confirms the robustness of our estimates. \\
Nearly all our galaxies host massive DM halos, with $\rho_0 > 10^{7.5}$ M$_\odot$/kpc$^3$ at 10 kpc. In particular, galaxies A1314, A2256, and A2319 exhibit very massive halos ($\rho_0 > 10^{8}$ M$_\odot$/kpc$^3$). Similar to A262 \citep{dN24}, it is possible that DM traces the stellar distribution in these systems; a notable example is A2319, which shows a large $\Gamma$ value despite its high $\rho_0$.
To verify that the DM parameters $\rho_0$, p$_\text{DM}$, q$_\text{DM}$ and $\gamma$ are robustly determined, we calculate the radius r$_\mathrm{DM}$ where the enclosed DM mass equals the sum of the stellar and the black hole mass, and compare it to the largest radius r$_\mathrm{kin,max}$ among all modeled kinematical bins (Tab.~\ref{Tab.dyn_details}). This ensures that our kinematics samples the region where DM contributes the most to the potential. As shown in Fig.~\ref{Fig.rmax_rDM_hist}, for 11 out of 16 galaxies modeled in this work we find $r_\mathrm{DM} > r_\mathrm{kin,max}$; the five galaxies for which this is not the case are A292, A399, A634, A2107 and A2388. In these cases, while the formal uncertainties on the DM parameters are generally small, the inferred DM profile should be interpreted with caution, as degeneracies between stellar mass, orbital structure, and DM distribution are not fully broken. Nevertheless, from Fig.~\ref{Fig.Schw_results}, we see that even if this occurs, the best-fit estimates are not degenerate. \\
Interestingly, we observe a variety of halo shapes (see also Sec.~\ref{Sec.comments}): spherical halos are the most common, but we also identify axisymmetric halos — both oblate (e.g., A160) and prolate (e.g., A1749)—as well as triaxial systems (e.g., A2107). \\
The models also provide insight into the orbital structure through the anisotropy ($\beta$) profiles. As shown by simulations \citep{Rantala18}, SMBH scouring tends to eject stars on radial orbits, leading to tangential anisotropy inside the BH sphere of influence (SOI). Indeed, this is what we find for all BCGs: a typical example is the BCG of A1749 (Fig.~\ref{Fig.beta_tang}), with $\beta_\mathrm{min} \sim$ -0.5. A BCG showing a strong tangential anisotropy is A1749 (Fig.~\ref{Fig.beta_strong_tang}), one of the BCGs hosting a UMBH. However, we note that there are cases such as A399 (Fig.~\ref{Fig.beta_isotr}) which are only mildly tangential in the core despite hosting a (nearly) UMBH. In brief, we find strong tangential anisotropy ($\beta < -0.6$) in the galaxies A2147, A1185, A2107 and A2388 (these last two show tangential anisotropy at all radii), all hosting UMBHs, and a nearly isotropic profile for A240, A399, A1314 and A1775. In Fig.~\ref{Fig.beta_all} we plot the anisotropy profiles of all BCGs along with those already published in \citet{Jens14} for SINFONI galaxies, showing that BCGs display a broader variety of profiles in comparison with ordinary massive ETGs. Finding different kinds of anisotropy profiles is important, since it allows us to understand what kind of progenitors led to the formation of these galaxies. In fact, simulations \citep{Rantala19} predict that, while a merger of two cuspy galaxies can generate large cores, if two core-ellipticals merge the result will be a smaller core compared to the BH mass and a more isotropic orbit distribution, as seen for example in Holm 15A \citep{Kianusch19}. \\
Finally, we comment on intrinsic shapes and orientations. The results of our models yield $< p(r) > = $ 0.81 and $< q(r) > = $ 0.67. With the exception of A634 (nearly oblate) and A2107 (nearly prolate), the other BCGs have 0.35 $ \leq \langle T \rangle \leq$ 0.80, implying that they are (strongly) triaxial systems. The presence of isophotal twists implies that a galaxy is not observed exactly along a principal axis. While this is the case for most BCGs, we find an example with nearly constant position angle oriented along the major axis (A2319), while other galaxies (A292, A1314, A2506) are oriented such that the equatorial plane lies close to the line-of-sight (LOS). For the roundest galaxies (e.g., A2256), estimating orientation is challenging. However, most systems are mildly inclined ($40^\circ < \theta < 70^\circ$), and thus ideally positioned to robustly recover their intrinsic shapes. All these result agree very well with \citet{dN22BCGs}, where the results were obtained using photometry only.

%North, South -> h1, h2

%r -> |r|

%\newpage

\section{Comments on individual galaxies}  \label{Sec.comments}

In the following, we provide object-by-object comments, addressing aspects of the photometry, kinematics, and/or dynamical modeling.
%Galaxies NOT in deNicola+22: A292, A399, A592, A1314, A1749, A1982, A2107, A2255, A2319, A2388, A2506
%Possibly comments on rDM
\begin{itemize}

    \item \textbf{A150:} \\
    We omit this galaxy from our analysis due to contamination from a nearby massive galaxy, which prevents accurate reconstruction of the LOSVD wings. The contamination introduces asymmetric velocity profiles along MJ - where asymmetry in $\sigma$ is also visible - and along MJ45, which shows a double peak in $\sigma$. The velocity profile along MN is similarly asymmetric. \\

\item \textbf{A160:} \\
    This galaxy has HST photometry (GO program 8683, P.I. van der Marel). The photometric fit is particularly good, with RMS$_\text{SB} \sim$ 0.005, RMS$_\varepsilon \sim$ 0.025, and RMS$_\text{PA} \sim$ 2.5. The best-fit solutions are triaxial in the center, become almost prolate out to 15 kpc, and then return to a triaxial shape, where a PA twist of $\sim$30$^\circ$ also appears. \\
    A kink in $\sigma$ is visible at 1" (symmetric on both sides) without a corresponding velocity change. The galaxy shows no significant rotation, has a low $\sigma_0 \sim$ 220 km/s, an outward-increasing $\sigma$ beyond 3", and a slightly decreasing h$_4$ in the inner $\sim$2". The anisotropy remains negative out to at least 5 kpc. The DM halo is oblate. \\

\item \textbf{A240:} \\
    This very round galaxy ($\varepsilon < 0.3$) shows little twist, except in the region where $\varepsilon < 0.1$. Most deprojections are triaxial. We find nearly identical values for r$_\gamma$ and r$_\text{b}$. Different wavelength ranges had to be adopted (see Tab.~\ref{Tab.kinematics}) due to template mismatch leading to LOSVD artifacts. The galaxy exhibits negligible rotation, low central velocity dispersion ($\sigma_0 \sim$ 210 km/s), a monotonically increasing $\sigma$ profile, and nearly flat h$_3$ and h$_4$ profiles. MN and MN45 slits contain external objects at r $>$ 15", but these do not impact the models. Due to large discrepancies in \mbh\, between the two halves, we model all bins together. Despite a $\lesssim 10^{10}$ M$_\odot$ black hole, the inner regions are nearly isotropic. \\

\item \textbf{A292:} \\
    This is one of the two galaxies with no AO/ESM or HST data. The ellipticity increases significantly at large radii, while the PA remains roughly constant. The SB profile cannot be fitted by a CS profile, though the cusp radius of 1" ($\sim$1.25 kpc) matches \mbh\, well. The $q(r)$ profile is well-constrained, decreasing with radius to values $\sim$0.3 in the outskirts, while $p(r)$ is more variable. The orientation from dynamics aligns well with the photometric estimate. \\
    The galaxy shows no significant rotation. The velocity dispersion increases slightly toward the center ($\sigma_0 \sim$ 310 km/s), and more significantly beyond r $>$ 5". h$_4$ decreases slightly inward but remains low ($\lesssim$ 0.05). The DM halo is prolate. \\

\item \textbf{A399:} \\
     A $\sim$30$^\circ$ twist is present, and nearly all deprojections are triaxial. No rotation is detected. In contrast to most galaxies, $\sigma$ does not increase in the outskirts, and h$_4$ decreases at large radii. The MJ45 slit is omitted due to suspected incorrect orientation. \citet{Oegerle91} report $\sigma_0 = 230 \pm 30$ km/s, while we find $\sigma_0 = 265 \pm 10$ km/s. The galaxy is an outlier in the \mbh-r$_\text{core}$ relation (with a smaller BH), has a very large $\Gamma$ = 14.2, and a flat, nearly prolate triaxial DM halo. Finally, the galaxy is only mildly tangential ($\beta_\mathrm{min} = -0.15$) inside the core. \\

    \item \textbf{A592:} \\
    The noisy profiles suggest that this galaxy is not fully relaxed, and the core size estimate should be treated with caution. It exhibits a $\sim$30$^\circ$ twist and a significant ellipticity gradient, which complicates the deprojection and leads us to truncate it at r = 40 kpc. The best-fit model is triaxial with $p(r) \sim 0.9$, while $q(r) \sim 0.5$ in the centre, then rises up to 0.8 at 3 kpc before dropping to 0.3 at large radii. \\
    The central velocity dispersion is very low ($\sigma_0 \sim$ 210 km/s) and begins increasing from $\sim$2". The LOSVDs become asymmetric (h$_3 \lesssim$ –0.1) beyond 4". h$_4$ remains slightly negative at all radii, and no significant rotation is detected. We mask contaminants along MN and MJ45 and model kinematics only out to 5". This galaxy is an outlier in the \mbh–r$_\mathrm{b}$ relation, with the core radius overpredicting \mbh\,by a factor of 2. The DM halo is prolate and among the flattest in the sample (p$_\mathrm{DM}$ = q$_\mathrm{DM}$ = 0.4). \\ % The M/L from SSP models is $<$6 in the central 10", consistent with a Salpeter-like IMF. \\

\item \textbf{A634:} \\
    The galaxy is quite round in the center ($\varepsilon \sim$ 0.1), flattens to $\varepsilon \sim$ 0.3 at 10", and becomes round again at larger radii. The PA is constant until 30", beyond which a significant 25$^\circ$ twist is seen. It has HST photometry (GO program 8683, P.I. van der Marel). \\
    The photometry is well recovered, with the ellipticity bump explainable via $p(r)$, $q(r)$, or both, depending on orientation. Most models are mildly triaxial, though the best-fit becomes oblate beyond 10 kpc. \\
    The galaxy was re-observed in 2025 with MODS due to corrupted acquisition files, preventing accurate PSF measurement. $\sigma$ is nearly constant at all radii ($\sigma_0 \sim$ 270 km/s). h$_4$ fluctuates around 0, and no rotation is detected in any slit. \\
    Both halves show tangential anisotropy at all radii, with $\beta_\mathrm{min}$ = [–0.8, –0.35] despite the small core and the lack of a particularly massive BH. The DM halo is triaxial. \\

\item \textbf{A688:} \\
    At 551 Mpc, this is the farthest BCG in our sample. The deprojections are compatible with a range of geometries. However, the low S/N in the kinematics and the limited spatial resolution of MODS prevent us from placing strong constraints on the mass parameters, leading us to omit this galaxy. We could not obtain a Core-S{\'e}rsic fit with $\alpha_\text{CS} = 5$ as for the other galaxies in Tab.~\ref{Tab.alphaCS_fit}. \\

%Tonry: his Fig. 1i
\item \textbf{A1185:} \\
    For this galaxy we only took H-band data at LBT. The galaxy is a site of ongoing galactic cannibalism \citep{Lauer88}. A bump in ellipticity and SB corresponds to a 40$^\circ$ twist, which is due to a companion galaxy (see e.g. Fig. 14 of \citealt{Matthias20}); this is only visible from $\text{r} > 5$ arcseconds, thus likely not impacting the estimate of the core size using the cusp radius. It is difficult to find a 3D density model that fits the photometry well; orientations yielding RMS$_\text{grid} <$ 0.09 are rare. \\
    The galaxy was re-observed in 2025 due to poor seeing during the initial run. Kinematics were also published by \citet{Tonry85}. No significant rotation is observed. This is another case of low $\sigma_0 \sim$ 210 km/s increasing outward, with low h$_4$ across all radii. The bins along MJ are omitted due to contamination by the companion galaxy. Despite the challenges, we robustly detect an UMBH, which correlates well with the cusp radius and agrees with the tangential core with $\beta_\mathrm{min} \sim -0.6$. 
    \\ %The galaxy has a Kroupa IMF, but the massive DM halo suggests $\Gamma$ may underestimate M/L, potentially indicating a Salpeter-like IMF (\citealt{dN24}). \\

\item \textbf{A1314:} \\
    The galaxy has HST photometry (GO program 8683, P.I. van der Marel). A $\sim$20$^\circ$ twist rules out axisymmetry. Some models indicate prolate geometry within the central kpc, transitioning to strong triaxiality at larger radii. The inclination is high ($\theta = 81^\circ$). \\
    The BCG exhibits strong rotation along three slits; MAJOR is the only one without significant rotation, indicating minor-axis rotation and ruling out oblate geometry. The velocity dispersion increases towards the center ($\sigma_0 \sim$ 350 km/s) but not outward. h$_4$ decreases slightly towards the center. \citet{Smith00} report $\sigma_0 = 332 \pm 15$ km/s. The DM halo is massive and oblate, and the anisotropy within the core is nearly isotropic ($\beta_\mathrm{min} \sim$ -0.2). \\

\item \textbf{A1749:} \\
    The ellipticity stays between 0.2 and 0.3 across all radii. A $\sim$20$^\circ$ twist is present. The core radius of $\sim$0.8 kpc suggests a $10^{10}$ M$_\odot$ BH, which is confirmed by our modeling. The best-fit deprojection is nearly oblate ($p(r) \sim 0.93$) out to 10 kpc, becomes triaxial, and then prolate beyond 50 kpc. \\
    This is one of the few galaxies in our sample showing rotation, mostly along the major axis. The velocity dispersion increases strongly outward, reaching 500 km/s in the outskirts. The DM halo is massive, flat, and prolate. \\

\item \textbf{A1775:} \\
    This galaxy has HST photometry (GO program 7281, P.I. Fanti). Most deprojections yield a constant $p(r)$, and the best-fit is mildly triaxial. \\
    A secondary object along MN45 has been masked. The galaxy was re-observed in 2025 due to poor initial seeing. It has $\sigma_0 > 300$ km/s, and $\sigma$ does not increase in the outskirts. h$_4$ shows a similar trend. No significant rotation is observed. The core is the second largest in the sample (the only one $>$2 kpc), yet the detected UMBH is still $\sim$3 times smaller than expected. Despite the massive BH, the galaxy has only mild tangential anisotropy ($\beta_\mathrm{min} \sim$ -0.15). The DM halo is massive and strongly triaxial.  \\

\item \textbf{A1982:} \\
    This is a flat galaxy ($\varepsilon > 0.3$ at all radii) with no twist. Several deprojections show flat, nearly oblate geometries. The galaxy was re-observed in 2025 (under poor seeing conditions) due to two missing slits during the first run (see Tab.~\ref{Tab.kinematics}) leading to a low number of bins. We could not observe this galaxy at LBT and, hence, relied on Wendelstein photometry for the deprojection. All these factors do not allow to resolve the BH SOI, and no modeling was performed. \\

    \item \textbf{A2107:} \\
    The photometry shows an interesting ellipticity profile: $\varepsilon \sim 0.35$ in the centre, then the galaxy becomes round at 3" and then $\varepsilon$ increases up to $\sim$0.3 at large radii. $p(r), q(r)$ intersect due to the large twist. \\
    The galaxy possibly has a KDC in the central 2" with rotation up to 50 km/s, while we do not find significant rotation at larger radii. The dispersion profile constantly increases up to 500 km/s. We observe v–h$_3$ anti-correlation. This is the second most massive BH of our sample. The tangential anisotropy is particularly pronounced ($\beta_\mathrm{min}$ = [–1.75, –0.75]). However, the SB profile is cuspy, thus not showing a core at all. The galaxy has a triaxial DM halo. \\
    %cross p,q

\item \textbf{A2147:} \\
    The galaxy has HST photometry (GO program 8683, P.I. van der Marel). The $\sim$40$^\circ$ twist is well reproduced. Nearly all solutions are triaxial, in good agreement with the strong twist. \\
    The galaxy does not show significant rotation, whereas the dispersion increases towards the centre with $\sigma_0 \sim$ 300 km/s and h$_4$ decreases towards the centre with some bumps. At r $>$ 10" a second object along MN45 is visible in the spectrum. Our measurements are in good agreement with the findings of \citet{Loubser08, Loubser09}. \\
    The large core size hints at $\mbhm \sim 2 \times 10^{10}$ M$_\odot$, in line with our result. \\

\item \textbf{A2255:} \\
    The galaxy is quite round ($\varepsilon < 0.3$) and has a 25$^\circ$ twist in the outskirts. There are several geometries compatible with the photometry (the best-fit one is triaxial until 40 kpc and then prolate). \\
    The most prominent absorption lines are masked for some bins because they cannot be properly fitted (they are too deep), even using the full MILES library. $\sigma$ steadily rises towards the outskirts and does not increase towards the centre; the galaxy does not rotate. We measure $\sigma_0 = 257 \pm 4$ km/s; \citet{Oegerle91} report $285 \pm 30$ km/s. \\
    We attempt to model the two galaxy halves separately obtaining mass parameters differing by one order of magnitude and omit it because of no BH detection, due to unresolved BH SOI.  \\
    %triaxial, prolate from ~40 kpc

\item \textbf{A2256:} \\
    The photometry has also been published by \citet{Blakeslee92}, who report PA = 300$^\circ$. We assume PA = 140$^\circ$. This is one of the roundest galaxies of our sample ($\varepsilon < 0.2$). Therefore, it was analyzed following \citet{Rob24}. We only use the photometry up to 50" when deprojecting because the $\varepsilon$ and PA could not be estimated beyond that point. \\
    The galaxy was re-observed in 2025 due to poor seeing conditions during the first run. This is one of the few examples with $\sigma_0 > 300$ km/s, which still significantly underestimates \mbh. The profile increases outwards from 5", h$_4$ is negative, and we do not observe rotation. This is the most massive UMBH of our sample and the same time the BCG with the largest core size. It has a spherical and very massive DM halo. \\%potentially tracing stars as for NGC 708 \citep{dN24}. \\

\item \textbf{A2319:} \\
    The galaxy is round ($\varepsilon < 0.25$), with viewing angles along the major axis. Therefore, we test more deprojections to properly sample the degeneracy. The best-fit one has $p(r) \sim 0.8$. \\
    The velocity dispersion starts increasing outwards from 3", reaching $\sim$500 km/s at $r \sim 15$", while h$_4$ is negative at all radii. Several kinematic bins need extended masking beyond 4500 \AA. \\
    The DM halo is massive and prolate. The anisotropy is mildly tangential everywhere. It has one of the smallest cores of our sample (r$_\mathrm{b} \sim$ 0.3 kpc), which should be treated with caution. \\

\item \textbf{A2388:} \\
    For this galaxy we only took H-band data at LBT. The galaxy has several deprojections compatible with the photometry, showing mild triaxiality in the innermost regions and almost prolate in the outermost regions. There is little or no rotation visible in the kinematics, while $\sigma$ rises both towards the centre and outward from 4". \\
    We get nearly identical values for \mbh, $\Gamma$, and $\rho_0$ from both halves. Interestingly, the anisotropy inside the core is lower on one half ($\beta_\mathrm{min} \sim$ [-1.0, -0.6]). It has one of the most lightweight (slightly triaxial) DM halos of the sample. \\

\item \textbf{A2506:} \\
    The galaxy has a very small core. While the ellipticity smoothly increases from 0.1 in the centre to 0.5 in the outskirts, the PA is noisier but the twist is very mild ($\sim$10$^\circ$). \\
    Interestingly, we find intrinsic densities with changing geometry at different radii, e.g., triaxial solutions in the centre becoming prolate at large radii as well as oblate shapes. The best-fit one is triaxial and flat. \\
    We observe rotation up to 80 km/s along all slits except for MINOR, with MAJOR having the strongest signal. The dispersion remains roughly constant at all radii with $\sigma_0 \sim$ 220 km/s. A star along MAJOR+45 has been masked. \\
    This is one of the smallest BHs of our sample and has a slightly oblate DM halo. \\  %Salpeter IMF

\item \textbf{A2665:} \\
    The galaxy photometric profiles show several bumps, which reflect into noisy intrinsic shape profiles. We observed slits at 8 different PAs due to an error during the first observing run. The dispersion does not increase at large radii, while h$_4$ is slightly positive and decreases towards the centre. The poor quality of the kinematics does not allow for a robust modeling; we thus do not model the galaxy. \\

\end{itemize}

%Phototable:Depro/untitled3.py

%Gamma err 2 sig.digits, rho0 1
%Omitted (should be red in kintable): 688,2255,1982,2665(?),150
%if need to rotate the table uncomment landscape and table* -> table
\begin{landscape}
\begin{table}
    \centering
    \renewcommand{\arraystretch}{1.5}
    \setlength{\tabcolsep}{2pt}
    \begin{tabular}{c c c c c c c c c c c c c c}
       Cluster & Distance (Mpc) & halves (y/n) & \mbh\,(10$^9$ M$_\odot$) & $\Gamma$ & $\log (\rho_0$ / M$_\odot$\,kpc$^{-3}$) & p$_\text{DM}$  & q$_\text{DM}$ & $\gamma$ & \va$^\circ$ & $\langle p \rangle$ & $\langle q \rangle$ & r$_\mathrm{SOI}$ (arcsec) & r$_\mathrm{DM}$ (arcsec) \\ \hline \hline
       A160 & 179.24 & n & 13.2 $\pm$ 3.7 & 5.62 $\pm$ 0.84 & 7.7 $\pm$ 0.3 & 1.0$^{+0.0}_{-0.1}$ & 0.9 $\pm$ 0.1 & 1.0 $\pm$ 0.3 & (70,30,100) & 0.876 & 0.787 & 1.72 $\pm$ 0.15 &  5.91 \\  \hline
       A240 & 250.00 & n & 9.2 $\pm$ 2.6 & 2.76 $\pm$ 0.41 & 7.9 $\pm$ 0.3 & 1.0$^{+0.0}_{-0.1}$ & 1.0$^{+0.0}_{-0.1}$ & 0.0$^{+0.4}_{-0.0}$ & (62,157,151) & 0.762 & 0.691 & 1.04 $\pm$ 0.11 & 4.89 \\  \hline
       A292 & 257.83 & n & 13.8 $\pm$ 3.9 & 6.7 $\pm$ 1.0 & 6.9 $\pm$ 0.3 & 0.9 $\pm$ 0.1 & 0.9 $\pm$ 0.1 & 0.0$^{+0.4}_{-0.0}$ &  (85,60,86) & 0.635 & 0.516 & 0.85 $\pm$ 0.10 &  24.9 \\  \hline
       A399 & 281.96 & n & 7.2 $\pm$ 2.0 & 14.2 $\pm$ 2.1 & 7.5 $\pm$ 0.3 & 0.6 $\pm$ 0.1 & 0.5 $\pm$ 0.1 & 0.5 $\pm$ 0.3 & (71,36,144) & 0.707 & 0.531 & 1.20 $\pm$ 0.10 & 12.8 \\  \hline
       A592 & 261.54 & n & 8.6 $\pm$ 2.4 & 8.5 $\pm$ 1.3 & 7.8 $\pm$ 0.3 & 0.4 $\pm$ 0.1 & 0.4 $\pm$ 0.1 & 0.0$^{+0.4}_{-0.0}$ &  (71,144,66) & 0.891 & 0.651 & 1.00 $\pm$ 0.10 & 4.91 \\  \hline
       A634 & 113.03 & y & 5.6 $\pm$ 1.3 & 8.38 $\pm$ 0.86 & 7.4 $\pm$ 0.1 & 0.8 $\pm$ 0.1 & 0.7 $\pm$ 0.1 & 0.3 $\pm$ 0.3 & (60,130,80) & 0.976 & 0.744 & 0.67 $\pm$ 0.11 & 31.8 \\  \hline
       A1185 & 145.00 & n & 11.2 $\pm$ 3.1 & 5.21 $\pm$ 0.78 & 7.9 $\pm$ 0.3 & 1.0$^{+0.0}_{-0.1}$ & 1.0$^{+0.0}_{-0.1}$ & 0.4 $\pm$ 0.3 & (48,42,164) & 0.863 & 0.785 & 2.14 $\pm$ 0.18 & 6.95 \\  \hline
       A1314 & 137.99 & y & 5.3 $\pm$ 3.0 & 5.35 $\pm$ 0.43 & 8.1 $\pm$ 0.1 & 1.0$^{+0.0}_{-0.1}$ & 0.9 $\pm$ 0.1 & 1.0 $\pm$ 0.2 & (81,150,85) & 0.725 & 0.585 & 0.80 $\pm$ 0.23 & 4.56 \\  \hline
       A1749 & 224.83 & y & 10.5 $\pm$ 2.9 & 3.81 $\pm$ 0.81 & 8.17 $\pm$ 0.08 & 0.5 $\pm$ 0.1 & 0.5 $\pm$ 0.1 & 0.3 $\pm$ 0.3 & (61,62,88) & 0.932 & 0.676 & 0.771 $\pm$ 0.012 & 5.22 \\  \hline
       A1775 & 297.23 & y & 15.1 $\pm$ 3.3 & 4.578 $\pm$ 0.045 & 8.03 $\pm$ 0.05 & 0.7 $\pm$ 0.3 & 0.3 $\pm$ 0.1 & 0.7 $\pm$ 0.3 & (66,147,172) & 0.768 & 0.702 & 1.48 $\pm$ 0.14 & 3.37 \\  \hline
       A2107 & 172.23 & y & 22.4 $\pm$ 3.3 & 10.5 $\pm$ 1.6 & 7.7 $\pm$ 0.2 & 0.75 $\pm$ 0.05 & 0.65 $\pm$ 0.05 & 0.4 $\pm$ 0.4 & (48,42,124) & 0.902 & 0.865 & 1.24 $\pm$ 0.23 & 17.2 \\  \hline
       A2147 & 146.04 & y & 16.21 $\pm$ 0.90 & 4.60 $\pm$ 0.79 & 7.72 $\pm$ 0.02 & 1.0$^{+0.0}_{-0.1}$ & 1.0$^{+0.0}_{-0.1}$ & 1.0 $\pm$ 0.2 & (60,40,5) & 0.755 & 0.578 & 2.05 $\pm$ 0.10 & 8.36 \\  \hline
       A2256 & 237.62 & n & 24.7 $\pm$ 6.9 & 10.5 $\pm$ 1.6 & 7.8 $\pm$ 0.3 & 1.0$^{+0.0}_{-0.1}$ & 1.0$^{+0.0}_{-0.1}$ & 0.2$^{+0.3}_{-0.2}$ & (56,127,37) & 0.934 & 0.815 & 1.61 $\pm$ 0.11 & 3.96 \\  \hline
       A2319 & 220.29 & n & 8.0 $\pm$ 2.3 & 10.7 $\pm$ 1.6 & 8.0 $\pm$ 0.3 & 0.8 $\pm$ 0.1 & 0.8 $\pm$ 0.1 & 1.2 $\pm$ 0.3 & (90,0,175) & 0.799 & 0.663 & 0.93 $\pm$ 0.12 & 1.72 \\  \hline
       A2388 & 246.28 & y & 9.6 $\pm$ 1.0 & 9.46 $\pm$ 0.14 & 6.9 $\pm$ 0.4 & 0.9 $\pm$ 0.1 & 0.85 $\pm$ 0.05 & 0.3 $\pm$ 0.3 & (50,40,110) & 0.762 & 0.665 & 0.734 $\pm$ 0.042 & 28.2 \\  \hline
       A2506 & 106.43 & y & 2.8 $\pm$ 1.1 & 7.22 $\pm$ 0.66 & 7.7 $\pm$ 0.2 & 1.0$^{+0.0}_{-0.1}$ & 0.95 $\pm$ 0.05 & 0.1 $\pm$ 0.1 & (87,160,136) & 0.665 & 0.522 & 0.628 $\pm$ 0.082 & 17.8 \\  \hline \hline
    \end{tabular}

    \caption{Results from our dynamical models. \textit{Col. 1}: Cluster; \textit{Col. 2}: Adopted distance (see \citealt{Matthias20}); \textit{Col. 3}: Whether we model halves or not; \textit{Col. 34}: Black Hole mass; \textit{Col. 5}: Mass-to-light ratio; \textit{Col. 6}: Logarithmic Dark Matter density at 10 kpc; \textit{Cols. 7, 8}: Flattenings of the DM halo; \textit{Col. 9}: Inner logarithmic slope of the DM halo. \textit{Col. 10-12}: Best-fit orientation estimated from dynamics and corresponding $\langle p \rangle$ and $\langle q \rangle$.  \textit{Col. 13}: Radius of the BH sphere of influence, defined as $\mbhm = \mathrm{M}_*$ (r$_\mathrm{SOI}$). \textit{Col. 14}: Radius r$_\mathrm{DM}$ where the mass of the DM halo equals the sum of stellar and black hole mass. In cases where we manage to model a galaxy using two datasets — one for each half — we take the semi-width of the interval as the uncertainty estimate; otherwise, we scale the relative uncertainties found using the N-body simulations of \citet{Rantala18, Rantala19} to the respective values.}
    \label{Tab.dyn_results}
\end{table}
\end{landscape}

%Bottom-heavy -> more small stars than expected (e.g. Salpeter)
%lightweight -> the opposite (e.g. Kroupa, MW)

%M/L large --> small stars (Salpeter) or maybe DM with same trend as stars
%M/L small --> larger stars (Kroupa) or halo too massive & therefore M/L underestimated

\section{Summary and conclusions} \label{Sec.conclusions}
Following our recent study on the BCG of A262 \citep{dN24}, we present here the full dataset comprising 22 BCGs with long-slit kinematics obtained at the LBT. We combine high-resolution imaging—either from HST or LBT—with wide-field data from Wendelstein, as well as kinematics from LBT, to derive non-parametric deprojections and line-of-sight velocity distributions (LOSVDs) required for our Schwarzschild modelling. We successfully apply these models to 16 BCGs, in addition to A262. Our main findings are:

\begin{itemize}
    \item We identify 8 new UMBHs already discussed in a companion paper (de Nicola, submitted), more than doubling the number of such known systems. These rare objects help to populate the high-mass end of the BH–host galaxy scaling relations. 
    \item The dark matter halos span a wide range of intrinsic geometries, including spherical, axisymmetric, and triaxial configurations.
    \item The kinematical profiles reveal low central velocity dispersions, rising $\sigma$ profiles from 2–5" in most galaxies, the presence of one kinematically decoupled core (KDC), and three systems showing clear rotation aligned with - or close to - their projected principal axes.
    \item For BCGs not included in \citet{dN22BCGs}, we confirm the trend of strong triaxiality.

\end{itemize}

The large BH masses and the saturation of the central $\sigma$ values cause the  \mbh-$\sigma$ relation to break down at high masses, highlighting the need for novel correlations to predict \mbh. This topic is further discussed in a companion paper.  \\
These results could be further refined through the use of integral field unit (IFU) spectroscopy, which would enhance the spatial coverage, particularly in the outer regions, and enable more robust constraints on dark matter halo parameters that remain unconstrained in this work—such as the DM scale radius and the outer logarithmic slope. Additionally, stellar population analyses for the full sample would allow a comparison between dynamical and population-based mass-to-light ratios. \\
Finally, we note that the exceptional spatial resolution of the Euclid survey enables the resolution of cores as small as $\sim$1 kpc out to $z \sim 1$. Combined with upcoming spectroscopic facilities such as MICADO, this will open the door to Schwarzschild modelling of high-redshift systems, and ultimately to probing the evolution of the black hole mass function across cosmic time.

%We thank the anonymous referee for carefully reading the manuscript and providing us with useful comments which helped us improving the paper. \\
\section*{Acknowledgements}
We thank the referee for the insightful report, which helped us improve the paper. \\ 
Computations were performed on the HPC systems Raven and Cobra at the Max Planck Computing and Data Facility. \\
The LBT is an international collaboration among institutions in the United States, Italy and Germany. LBT Corporation partners are:  LBT Beteiligungsgesellschaft, Germany, representing the Max-Planck Society, the Astrophysical Institute Potsdam, and Heidelberg University; The University of Arizona on behalf of the Arizona university system; Istituto Nazionale di Astrofisica, Italy; The Ohio State University, and The Research Corporation, on behalf of The University of Notre Dame, University of Minnesota and University of Virginia. \\
modsCCDRed was developed for the MODS1 and MODS2 instruments at the Large Binocular Telescope Observatory, which were built with major support provided by grants from the U.S. National Science Foundation's Division of Astronomical Sciences
Advanced Technologies and Instrumentation (AST-9987045), the NSF/NOAO TSIP Program, and matching funds provided by the Ohio State University Office of Research and the Ohio Board of Regents. Additional support for modsCCDRed was provided by NSF Grant AST-1108693.
This paper made use of the modsIDL spectral data reduction reduction pipeline developed in part with funds provided by NSF Grant AST-1108693 and a generous gift from OSU Astronomy alumnus David G. Price through the Price Fellowship in Astronomical Instrumentation. \\
This work makes use of the data products from the HST
image archive.

\section*{Data Availability}
The data underlying this article will be shared on reasonable request to the corresponding author.

\bibliographystyle{mnras}
\bibliography{bibl}

\appendix

\section{ICL contamination} \label{App.ICL}
\citet{Matthias21} discuss several methods to separate BCGs from the intracluster light (ICL). Contamination from the ICL can be significant in the outermost regions of the galaxy. In this appendix, we evaluate the amount of ICL light that is carried over when performing the deprojections and constructing the dynamical models. \\
We follow the approaches illustrated by \citet{Matthias21} in their Fig. 1 (panels (c) and (d)):
\begin{itemize}
    \item If, following \citet{Matthias20}, a single S{\'e}rsic component is sufficient to fit the full surface brightness (SB) profile, we estimate the ICL contamination by fitting a \citet{deVaucouleurs48} profile in the region where SB $<$ 23 mag/arcsec$^2$, excluding the core. The excess light across the \textit{entire} SB profile relative to this fit is taken as the ICL contribution.
    \item If a double S{\'e}rsic component is required, we assume that the innermost component represents the BCG, while the outermost component corresponds to the ICL.
\end{itemize}

As reported in Tab.~\ref{Tab.BCG_ICL}, the fraction of ICL in the region used for deprojection never exceeds 30\% and is negligible in the region with kinematic data, which is fitted by our triaxial Schwarzschild models. This confirms that our results are robust against ICL contamination.

\begin{table*}
    \begin{tabular}{c c c}
    Galaxy & $\left(\mathrm{L}_{\mathrm{ICL}} / \mathrm{L}_{\mathrm{TOT}}\right)_\mathrm{depro}$ & $\left(\mathrm{L}_{\mathrm{ICL}} / \mathrm{L}_{\mathrm{TOT}}\right)_\mathrm{kinematics}$ \\
    \hline \hline
       A160 & 0.145 & 0.0298  \\
       A240 & 0.0537 & 0.00623 \\
       A292 & 0.0382 & 0.00571 \\
       A399 & 0.00202 & 0.000451 \\
       A592 & 0.190 & 0.0124 \\
       A634 & 0.0123 & 0.00209 \\
       A1185 & 0.0633 & 0.00890 \\
       A1314 & 0.292 & 0.0565 \\
       A1749 & 0.238 & 0.0477 \\
       A1775 & 0.235 & 0.0244 \\
       A2107 & 0.0375 & 0.00599 \\
       A2147 & 0.282 & 0.0486 \\
       A2256 & 0.0773 & 0.00845 \\
       A2319 & 0.178 & 0.0322 \\
       A2388 & 0.284 & 0.0401 \\
       A2506 & 0.104 & 0.0255 \\
       \hline
    \end{tabular}
    \caption{\textit{Col. 1}: Galaxy name. \textit{Col. 2-3}: Percentage of light belonging to the ICL in the photometry region we use for the deprojection and kinematics, respectively.}
\label{Tab.BCG_ICL}
\end{table*}

\section{Estimates of the core size}   \label{App.core_size_estimate}

\subsection{PSF Convolution}  \label{App.core_size_PSF}

The break radii used throughout this work are estimated by fitting PSF-convolved Core-S{\'e}rsic profiles to the observed 1D surface brightness profiles. To ensure a robust determination of the 6 parameters needed to specify the Core-S{\'e}rsic function, we derive an initial guess for the effective surface brightness, S{\'e}rsic index and effective radius by fitting a simple \citet{Sersic63} law to the SB profile, as well as using r$_\gamma$ as starting value for the break radius. \\
In principle, it would not be necessary to go through the 1D profile: we could do this directly from the images using the program IMFIT \citep{Erwin15IMFIT}. However, this presents two main problems:
\begin{itemize}
    \item IMFIT is not set up to fit more than one image at the time. Given that (except for A292 and A1982) the photometry has been derived using more than one image - high-resolution HST or LBT images combined with WWFI acquisitions - one should limit itself to the innermost regions, which would not allow for a robust estimation of the S{\'e}rsic index, the effective radius and the effective surface brightness.
    \item IMFIT assumes constant ellipticity and PA values, which is almost never the case for our objects.
\end{itemize}

\begin{figure}
    \centering
    \includegraphics[width=\linewidth]{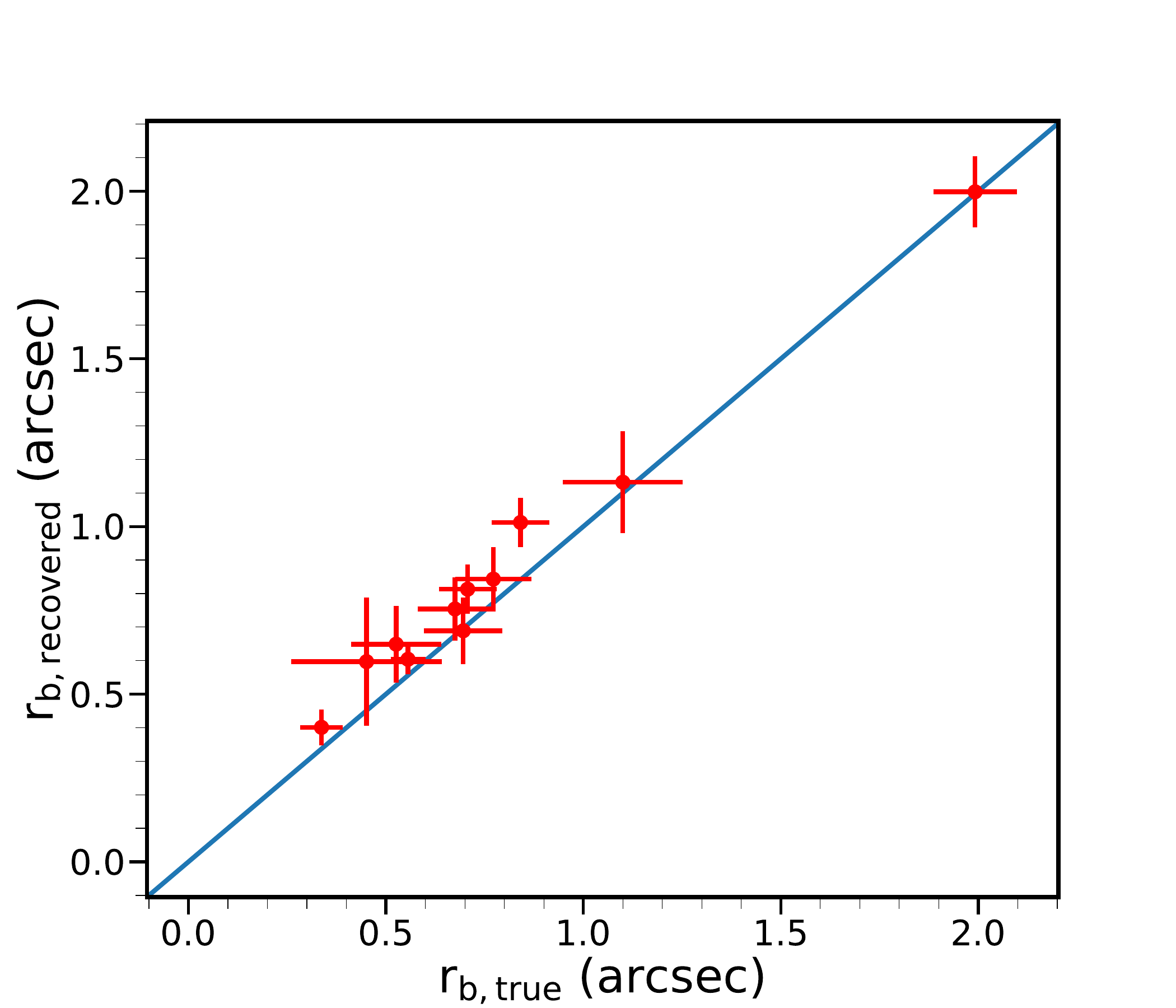}
    \caption{Comparison between r$_\mathrm{b}$ used to generate mock IMFIT images (see text) and those recovered the images themselves using our procedure. The recovered r$_\mathrm{b}$ values tend to be overestimated, albeit never exceeding 15\%.}
    \label{Fig.rb_PSF_test}
\end{figure}

We then model the PSFs for each galaxy using circularized Moffat profiles:
\begin{equation}
   \mathrm{PSF} (R) = \frac{\beta-1}{\pi \alpha^2}\left[1+\left(\frac{R}{\alpha}\right)^2\right]^{-\beta}
    \label{eq.Moffat}
\end{equation}

\noindent where we adopt $\beta = 2.5$ and $\alpha$ comes from Tab.~\ref{Tab.photometry}. The FWMH is then $2\alpha\sqrt{2^{1/\beta}-1}$ \citep{Rob93}. The convolved Core-S{\'e}rsic profile I$_c (R)$ reads

\begin{equation}
  \mathrm{I}_c (R) = \int_0^{+\infty} I(R')\,\mathrm{PSF}(R-R')\,dR'
    \label{eq.psfconvolution}
\end{equation}

\noindent with SB = -2.5$\log_{10} \mathrm{I}$.
This is an approximation: the exact way of performing the PSF convolution - described in App. A of \citet{Rob93} - would require performing the convolution on the 2D image directly. %or using a circular PSF and an angular average of the SB profile, 
To test the robustness of this approach, we construct for each BCG a mock IMFIT image using two components, an innermost PSF-convolved Core-S{\'e}rsic and an outermost S{\'e}rsic profile, and see how well we can recover r$_\mathrm{b}$ using our procedure. We take the necessary parameters to construct the mock images from Tabs.~\ref{Tab.photometry} and~\ref{Tab.core_fit}. \\
In Fig.~\ref{Fig.rb_PSF_test} we show the comparison between the original break radii r$_\mathrm{b,original}$ and those we recover (r$_\mathrm{b,recovered}$). The comparison shows that with our approach we might be overestimating the core sizes by 15\% at most, well within the intrinsic scatter of the \mbh-r$_\mathrm{c}$ relation (de Nicola et al., submitted). \\
Finally, for the two galaxies A292 and A1185 we could not fit any Core-S{\'e}rsic profile to the SB profile. For A292 this might be due to the WWFI image, whose larger PSF of 1.2 arcseconds does not allow for a clear resolution of the core. Instead, the profile of A1185 points to a not fully relaxed galaxy. Thus, in this case we follow our study of A85 \citep{Kianusch19} and use the cusp radius r$_\gamma$ \citep{Carollo97} to estimate the core size.

%We show in Fig.~\ref{Fig.rb_rgamma} the comparison between r$_\mathrm{b}$ and r$_\gamma$ for all the BCGs for which we could also derive an estimate of r$_\mathrm{b}$. \R{conclusions?}

%\begin{figure}
%    \centering
%    \includegraphics[width=\linewidth]{Figures/Data/rb_rgamma.pdf}
%    \caption{For all the BCGs with r$_\mathrm{b}$ measurements, we show the comparison between core size estimated used the cusp radius r$_\gamma$ or r$_\mathrm{b}$ itself. No clear trend or systematics is evident.}
%    \label{Fig.rb_rgamma}
%\end{figure}

\subsection{Testing double S{\'e}rsic components}  \label{App.CSS_fit}

Tab.~\ref{Tab.CSS_fit} reports the results of fitting a Core-S{\'e}rsic + S{\'e}rsic (CSS) model to those BCGs for which \citet{Matthias20} found that the SB profile is better reproduced by a double S{\'e}rsic component rather than a single one. The only exception is A2388, for which we could not obtain a reasonable CSS fit to the SB profile. All core size estimates are consistent within 2$\sigma$.

%Errors with 2 sign. digits
\begin{table*}
  \centering
    \setlength{\tabcolsep}{2pt}
  \begin{tabular}{cccccccccc}
    \hline
    Galaxy & r$_\mathrm{e,1}$ (arcsec) & $\mu_\mathrm{e,1}$ (mag/arcsec$^2$) & $n_1$ & r$_\mathrm{e,2}$ (arcsec) & $\mu_\mathrm{e,2}$ (mag/arcsec$^2$) & $n_2$ & r$_\mathrm{b}$ (arcsec) & $\gamma_\mathrm{CS}$ & $\alpha_\mathrm{CS}$ \\ \hline
    A160 & $44.9 \pm 4.4$ & $24.77 \pm 0.19$ & $4.53 \pm 0.27$ & $250 \pm 12$ & $26.587 \pm 0.038$ & $0.214 \pm 0.027$ & $0.840 \pm 0.048$ & $0.058 \pm 0.017$ & $4.5 \pm 1.3$ \\
    A1314 & $37 \pm 10$ & $24.72 \pm 0.55$ & $18.0 \pm 1.9$ & $110.85 \pm 0.97$ & $25.589 \pm 0.013$ & $1.153 \pm 0.037$ & $1.37 \pm 0.13$ & $0.279 \pm 0.016$ & $3.112 \pm 0.082$ \\
    A2147 & $12.6\pm 3.5$ & $22.21 \pm 0.48$ & $2.9 \pm 1.9$ & $128.4 \pm 8.6$ & $25.39 \pm 0.19$ & $1.38 \pm 0.16$ & $1.3 \pm 1.3$ & $0.138 \pm 0.055$ & $1.50 \pm 0.87$ \\
    A2319 & $20.56\pm 0.86$ & $24.16 \pm 0.12$ & $0.48 \pm 0.10$ & $188 \pm 26$ & $27.43 \pm 0.28$ & $5.56 \pm 0.29$ & $0.375 \pm 0.036$ & $0.083 \pm 0.049$ & $7.9 \pm 7.1$ \\
    A2665 & $63.0\pm 4.4$ & $25.57 \pm 0.13$ & $4.59 \pm 0.16$ & $163 \pm 14$ & $27.857 \pm 0.078$ & $0.198 \pm 0.034$ & $0.517 \pm 0.039$ & $0.042 \pm 0.035$ & $2.69 \pm 0.52$ \\
    \hline
  \end{tabular}
  \caption{Best-fit Core-S{\'ersic}-S{\'ersic} parameters. The columns are the same as in Tab.~\ref{Tab.core_fit} with the subscripts 1 and 2 denoting the two S{\'ersic} components.}
  \label{Tab.CSS_fit}
\end{table*}

\subsection{Testing low $\alpha_\text{CS}$ values}  \label{App.alphaCS_fit}

In Tab.~\ref{Tab.alphaCS_fit} we show the CS fits for all BCGs in our sample that originally exhibited high values of $\alpha_\text{CS} > 10$, after fixing $\alpha_\text{CS} = 5$. No significant differences are found.

\begin{table*}
  \centering
  \begin{tabular}{cccccc}
    \hline
    Galaxy & r$_\mathrm{e}$ (arcsec) & $\mu_\mathrm{e}$ (mag/arcsec$^2$) & $n$ & r$_\mathrm{b}$ (arcsec) & $\gamma_\mathrm{CS}$ \\ \hline
    A399 & $49.8 \pm 1.1$ & $25.202 \pm 0.051$ & $2.919 \pm 0.078$ & $1.07 \pm 0.18$ & $0.131 \pm 0.025$ \\
    A592 & $187 \pm 28$ & $27.06 \pm 0.29$ & $4.72 \pm 0.31$ & $0.39 \pm 0.21$ & $0.10^{+0.13}_{-0.10}$ \\
    A1982 & $39.6 \pm 2.8$ & $24.56 \pm 0.17$ & $4.75 \pm 0.37$ & $0.94 \pm 0.31$ & $0.00^{+0.52}_{-0.00}$ \\
    A2255 & $75 \pm 10$ & $26.34 \pm 0.27$ & $6.68 \pm 0.43$ & $0.60 \pm 0.14$ & $0.14^{+0.26}_{-0.14}$ \\
    A2319 & $101.9 \pm 6.2$ & $26.08 \pm 0.12$ & $4.29 \pm 0.14$ & $0.310 \pm 0.072$ & $0.04^{+0.10}_{-0.04}$ \\
    A2506 & $53.4 \pm 1.9$ & $24.540 \pm 0.078$ & $5.41 \pm 0.15$ & $0.56 \pm 0.25$ & $0.24^{+0.43}_{-0.24}$ \\
    \hline
  \end{tabular}
  \caption{Same as Tab.~\ref{Tab.core_fit} for galaxies with $\alpha_\text{CS} > 10$ setting $\alpha_\text{CS} = 5$.}
  \label{Tab.alphaCS_fit}
\end{table*}

\section{Deprojection details}  \label{App.depro}

In Tab.~\ref{Tab.depro_parameters} we report the details about deprojections for those galaxies who have been observed at LBT after \citet{dN22BCGs} and, hence, are not included in that paper.

\begin{table*}
%\begin{center}
%\resizebox{1.5\columnwidth}{.4\paperheight}{%
\begin{tabular}{ccccccccc}
Galaxy & r$_{\text{min}}$ (kpc) & r$_{\text{max}}$ (kpc) & I$_{\text{obs}}$ grid & $\rho$\,grid & $10^2\times$ RMS$_{\text{MIN}}$ (\%) & Good Deprojections \\
\hline \hline
A292 & 0.573 & 108.2 & 50 $\times$ 15 & 60 $\times$ 15 $\times$ 15 & 1.7 & 51 \\
A399 & 0.601 & 109.6 & 40 $\times$ 10 & 60 $\times$ 11 $\times$ 11 & 2.7 & 55 \\
A592 & 0.490 & 109.0 & 50 $\times$ 15 & 70 $\times$ 13 $\times$ 13 & 2.1 & 51 \\
A1314 & 0.231 & 101.8 & 50 $\times$ 15 & 60 $\times$ 15 $\times$ 15 & 1.6 & 51 \\
A1749 & 0.329 & 105.9 & 40 $\times$ 10 & 60 $\times$ 11 $\times$ 11 & 1.9 & 72 \\
A1982 & 0.457 & 102.7 & 40 $\times$ 10 & 60 $\times$ 11 $\times$ 11 & 2.6 & 60 \\
A2107 & 0.229 & 106.5 & 60 $\times$ 10 & 70 $\times$ 11 $\times$ 11 & 2.1 & 42 \\
A2255 & 0.409 & 106.6 & 50 $\times$ 10 & 60 $\times$ 11 $\times$ 11 & 1.9 & 96 \\
A2319 & 0.898 & 109.5 & 40 $\times$ 10 & 60 $\times$ 11 $\times$ 11 & 2.0 & 68 \\
A2388 & 0.346 & 106.5 & 40 $\times$ 10 & 60 $\times$ 11 $\times$ 11 & 2.7 & 61 \\
A2506 & 0.164 & 60.11 & 50 $\times$ 12 & 65 $\times$ 13 $\times$ 13 & 2.9 & 49 \\
\hline
\end{tabular}

%\end{center}
%\vspace{.5cm}
\caption{Details about the deprojection parameters as in Table 2 of \citet{dN22BCGs} for BCGs not considered in that paper. \textit{Col. 1}: Galaxy. \textit{Cols. 2-3}: Smallest and largest isophotal radii. \textit{Cols. 4-5}: $I_{\text{obs}}$ and $\rho$-grid dimensions. \textit{Col. 6}: Percentage RMS for the best-fit model. \textit{Col. 7}: Number of deprojection for which RMS $\leq 1.2\times\text{RMS}_\mathrm{MIN}$ holds.}
\label{Tab.depro_parameters}
\end{table*}

%Avg v.a.
%A292 (70,120,75); A399 (56,143,103); A592 (70,30,80); A1314 (73,148,80); A1749 (70,160,60); A1982 (54,127,112); A2107 (64,164,59); A2255 (64,34,139); A2319 (84,8,175); A2388 (71,36,94); A2506 (48,138,106)

\section{Kinematics details}   \label{App.kin}

Tab. \ref{Tab.kinematics} reports specifications about the spectroscopic observations.

%Does rad. extent match with radial profiles of the kinematics?
%A1314 MN45 Bin12: template mismatch
%A1185MJ: corrupted absorption, impossible to derive reliable LOSVDs. Two galaxies along the MJ, A is the BCG
%A240: 4900-5300 in general has template mismatch
%A399: MJ45 uncertain orientation
%A1749: MN45 sigma balorda

%Missing: exp hours for 2 glxs, 2025 observations
\begin{table*}

    \begin{tabular}{c c c c c c c c c}
       \hline
       Cluster & Configuration & Date & Slits & $\lambda_\mathrm{int}$ & r (arcsec) & Seeing (arcsec) & PA & Exp. hours \\
       \hline
       A150 & b & 20231005 & MJ & [4000-5400] & [-11.6,16.1] & 1.3 & 170 & 1.5 \\
        & b & 20231005 & MN & [4000-5400] & [-13.3,15.9] & 1.3 & 260 & 1.5 \\
        & b & 20231007 & MJ & [4000-5400] & [-10.0,15.6] & 1.4 & 170 & 1.8 \\
        & b & 20231007 & MN & [4000-5400] & [-14.1,15.2] & 1.5 & 260 & 1.8 \\
        & b & 20231008 & MJ & [4000-5400] & [-12.9,10.3] & 1.1 & 170 & 1.0 \\
        & b & 20231008 & MN & [4000-5400] & [-9.2,14.9] & 0.8 & 260 & 1.0 \\
       & b & 20231008 & MJ45 & [4000-5400] & [-14.0,17.3] & 1.0 & 215 & 4.0 \\
        & b & 20231008 & MN45 & [4000-5400] & [-12.2,17.5] & 0.80 & 305 & 4.0 \\ \hline
        A160 & b & 20211204 & MJ & [4000-5400] & [-13.9,17.4] & 0.65 & 90 & 2.0 \\
        & b & 20211204 & MN & [4000-5400] & [-12.4,19.7] & 0.76 & 180 & 2.0 \\
        & b & 20211204 & MJ45 & [4000-5400] & [-10.8,26.3] & 1.1 & 135 & 2.0 \\
        & b & 20211204 & MN45 & [4000-5400] & [-10.3,20.1] & 1.2 & 225 & 2.0 \\
        & b & 20211205 & MJ & [4000-5400] & [-11.1,14.3] & 0.75 & 90 & 2.0 \\
        & b & 20211205 & MN & [4000-5400] & [-14.5,8.90] & 0.8 & 180 & 2.0 \\
        & b & 20211205 & MJ45 & [4000-5400] & [-12.2,10.6] & 0.9 & 135 & 2.0 \\ 
        & b & 20211205 & MN45 & [4000-5400] & [-10.5,15.6] & 1.0 & 225 & 2.0 \\ \hline
        A240 & b & 20231009 & MJ & [4100-4900] & [-12.2,14.6] & 1.1 & 0 & 4.0 \\
        & b & 20231009 & MN & [4000-5400] & [-13.9,14.9]  & 1.1 & 90 & 4.0 \\
        & b & 20231010 & MJ45 & [4300-5300] & [-12.2,13.3] & 0.90 & 225 & 3.0 \\
        & b & 20231010 & MN45 & [4000-5400] & [-13.0,15.0]  & 0.80 & 315 & 3.0 \\
        & m & 20231010 & MJ45 & [4300-5300] & [-11.5,11.3] & 1.2 & 225 & 1.0 \\
        & m & 20231011 & MN45 & [4000-5400] & [-10.9,14.1]  & 0.6 & 315 & 1.0 \\ \hline
        A292 & b & 20211206 & MJ & [4100-5300] & [-13.6,24.5] & 0.80 & 95 & 2.0 \\
        & b & 20211206 & MN & [4100-5300] & [-9.50,12.9] & 0.90 & 185 & 2.0 \\
        & b & 20211206 & MJ45 & [4100-5300] & [-15.5,8.15] & 0.80 & 140 & 2.0 \\
        & b & 20211206 & MN45 & [4100-5300] & [-12.6,21.4] & 0.70 & 230 & 2.0 \\
        & b & 20211208 & MJ & [4100-5300] & [-11.9,12.3] & 1.0 & 95 & 2.0 \\
        & b & 20211208 & -MJ & [4100-5300] & [-11.9,12.3] & 0.80 & 275 & 2.0 \\
        & b &  20211208 & MJ45 & [4100-5300] & [-18.5,13.1] & 1.2 & 140 & 2.0 \\
        & b &  20211208 & MN45 & [4100-5300] & [-19.8,15.0] & 1.0 & 230 & 2.0 \\ \hline
        A399 & b & 20231008 & MJ & [4100-5300] & [-13.1,17.3] & 0.90 & 50 & 0.5 \\
        & b & 20231008 & MN & [4100-5300] & [-13.1,4.51] & 0.90 & 140 & 0.5 \\
        
        & b & 20231009 & MJ & [4100-5300] & [-11.3,20.6] & 1.0 & 50 & 0.8 \\
        & b & 20231009 & MN & [4100-5300] & [-20.0,7.94] & 1.1 & 140 & 0.8 \\
        & b & 20231010 & MJ & [4100-5300] & [-11.0,24.7] & 1.0 & 50 & 1.0 \\
        & b & 20231010 & MN & [4100-5300] & [-19.1,6.54] & 1.0 & 140 & 1.0 \\
        & b & 20240912 & MJ & [4100-5300] & [-18.6,22.5] & 1.2 & 50 & 2.0 \\
        & b & 20240912 & MN & [4100-5300] & [-23.4,9.32] & 0.7 & 140 & 1.4 \\
        & m & 20240913 & MN & [4100-5300] & [-15.5,9.94]  & 1.9 & 140 & 0.5 \\
        & m & 20240913 & MN45 & [4100-5300] & [-14.3,18.9] & 1.6 & 5 & 2.5 \\
        & b & 20240914 & MN45 & [4100-5300] & [-13.1,4.51] & 0.80 & 5 & 1.4 \\
        & b & 20240914 & MJ45 & [4100-5300] & [-13.6,11.6] & 0.60 & \R{275} & 1.0 \\
        & m & 20240930 & MJ45 & [4100-5300] & [-12.2,15.9] & 0.80 & \R{275} & 3.0 \\ \hline
        A592 & b & 20211204 & MJ & [4400-5300] & [-12.1,8.40] & 0.60 & 110 & 3.5 \\
        & b & 20211204 & MN & [4400-5300] & [-13.3,16.0] & 0.70 & 200 & 3.5 \\
        & b & 20211208 & MJ45 & [4400-5300] & [-12.8,18.6] & 0.70 & 155 & 2.5 \\
        & b & 20211208 & MN45 & [4400-5300] & [-12.8,16.0] & 0.80 & 245 & 3.5 \\
        & b & 20221001 & MJ & [4400-5300] & [-10.0,6.54] & 1.0 & 110 & 1.0 \\
        & b & 20221001 & MN & [4400-5300] & [-18.7,17.7] & 1.1 & 200 & 1.0 \\
        & b & 20221001 & MJ45 & [4400-5300] & [-13.3,21.2] & 1.1 & 155 & 1.5 \\ 
        & b & 20221001 & MN45 & [4400-5300] & [-13.0,25.5] & 1.2 & 245 & 1.5 \\ \hline
        A634 & b & 20220401 & MJ & [3950-5250] & [-11.2,9.54] & 0.90 & 100 & 2.0 \\
        & b & 20220401 & MN & [3950-5250] & [-14.5,15.0] & 0.90 & 190 & 2.0 \\
        & b & 20220401 & MJ45 & [3950-5250] & [-10.1,13.6] & 0.85 & 145 & 2.0 \\
        & b & 20220401 & MN45 & [3950-5250] & [-12.3,10.2] & 0.89 & 235 & 2.0 \\
        & b & 20250319 & MJ & [3950-5250] & [-13.8,8.01] & 1.8 & 100 & 0.5 \\
        & b & 20250319 & MN & [3950-5250] & [-17.8,11.2] & 1.7 & 190 & 0.5 \\
        & b & 20250320 & MJ & [3950-5250] & [-19.9,7.88] & 0.80 & 100 & 2.5 \\
        & b & 20250320 & MN & [3950-5250] & [-13.1,15.5] & 0.80 & 190 & 2.5 \\
        & b & 20250321 & MJ45 & [3950-5250] & [-14.4,18.1] & 1.3 & 145 & 2.0 \\
        & b & 20250321 & MN45 & [3950-5250] & [-11.4,11.6] & 1.4 & 235 & 2.0 \\
        & b & 20250327 & MJ45 & [3950-5250] & [-13.2,16.9] & 1.3 & 145 & 2.0 \\
        & b & 20250327 & MN45 & [3950-5250] & [-12.9,10.0] & 1.3 & 235 & 2.0
        \\ \hline
    \end{tabular}

\caption{\textit{Col. 1}: Cluster; \textit{Col. 2}: Whether the observations were executed in monocular or binocular mode; \textit{Col. 3}: Acquisition date (YYYYMMDD); \textit{Col. 4}: Slit configuration (see text); \textit{Col. 5} Fitted wavelength interval (observed values) to derive the kinematics, if not given see above rows; \textit{Col. 6} Radial extent of the kinematics, if not given see above rows; \textit{Col. 7}: PSF, quantified using the FWHM in arcsec; \textit{Col. 8}: Position angle of the slits, canonically measured counterclockwise from the North; \textit{Col. 9}: Total exposure time. (Continues on next page)} %Objects/slit configurations in red are omitted from the dynamical modeling.
\label{Tab.kinematics}

\end{table*}

\begin{table*}
\begin{tabular}{c c c c c c c c c}
       \hline
       Cluster & Configuration & Date & Slits & $\lambda_\mathrm{int}$ & r (arcsec) & Seeing (arcsec) & PA & Exp. hours \\
    \hline
    A688 & b & 20220225 & MJ & [4300-5500] & [-6.50,6.35]  & 1.2 & 40 & 4.0 \\
        & b & 20220225 & MN & [4300-5500] & [-7.53,18.5] & 0.90 & 310 & 4.0 \\
        & b & 20220226 & MJ45 & [4300-5500] & [-8.78,5.65] & 1.0 & 85 & 4.0 \\
        & b & 20220226 & MN45 & [4300-5500] & [-12.4,5.66] & 0.90 & 175 & 4.0
        \\ \hline
A1185 & b & 20210222 & MJ & [4000-5400] & [-3.23,13.6] & 1.0 & 30 & 1.5 \\
        & b & 20210222 & MN & [4000-5400] & [-16.8,19.2] & 1.3 & 120 & 1.5 \\
        & b & 20210223 & MJ45 & [4000-5400] & [-10.9,13.5]  & 1.1 & 75 & 1.5 \\
        & b & 20210223 & MN45 & [4000-5400] & [-7.49,12.5] & 1.0 & 165 & 1.5 \\
        & b & 20250322 & MJ & [4000-5400] & [-5.41,11.7] & 2.3 & 30 & 1.5 \\
        & b & 20250322 & MN & [4000-5400] & [-13.2,19.9] & 1.8 & 120 & 1.5 \\
        & b & 20250323 & MJ & [4000-5400] & [-3.40,10.9] & 1.2 & 30 & 1.5 \\
        & b & 20250323 & MN & [4000-5400] & [-16.0,16.0] & 1.1 & 120 & 1.5 \\
        & b & 20250324 & MJ45 & [4000-5400] & [-9.98,16.4] & 1.2 & 75 & 1.5 \\
        & b & 20250324 & MN45 & [4000-5400] & [-6.66,11.4] & 0.90 & 165 & 1.5 \\
        & b & 20250325 & MJ45 & [4000-5400] & [-7.50,12.0] & 1.3 & 75 & 1.0 \\
        & b & 20250325 & MN45 & [4000-5400] & [-5.24,17.9] & 1.4 & 165 & 1.0 
 bb        \\ \hline
        A1314 & b & 20230411 & MJ & [4000-5200] & [-12.1,12.5] & 1.2 & 90 & 2.0 \\
        & b & 20230411 & MN & [4000-5200] &  [-12.0,13.0] & 1.2 & 0 & 2.0 \\
        & b & 20230412 & MJ45 & [4000-5200] & [-12.3,8.46]  & 0.90 & 135 & 2.0 \\
        & b & 20230412 & MN45 & [4000-5200] &  [-12.1,13.3] & 0.80 & 45 & 2.0 \\ \hline
A1749 & b & 20240407 & MJ & [4000-5400] & [-12.2,14.1] & 1.4 & 275 & 2.5 \\
        & b & 20240407 & MN & [4000-5400] & [-14.1,13.4] & 1.3 & 185 & 2.5 \\
        & b & 20240408 & MJ & [4000-5400] & [-13.5,13.6] & 0.86 & 185 & 1.5 \\
        & b & 20240408 & MN & [4000-5400] & [-12.9,15.1] & 0.68 & 275 & 1.5 \\
        & b & 20240408 & MJ45 & [4000-5400] & [-10.8,11.0] & 0.86 & 140 & 2.0 \\
        & b & 20240408 & MN45 & [4000-5400] &  [-14.3,11.1] & 0.72 & 50 & 2.0 \\
        & b & 20240409 & MJ45 & [4000-5400] & [-14.1,14.0] & 1.1 & 140 & 3.0 \\
        & b & 20240409 & MN45 & [4000-5400] & [-15.6,12.9] & 1.1 & 50 & 3.0
        \\ \hline
A1775 & b & 20230413 & MJ & [4000-5400] & [-11.7,13.0] & 2.5 & 340 & 2.0 \\
       & b & 20230413 & MN & [4000-5400] & [-19.6,11.1] & 2.2 & 70 & 2.0 \\
       & b & 20230416 & MJ45 & [4000-5400] & [-11.1,11.5] & 0.60 & 205 & 1.0 \\
       & b & 20230416 & MN45 & [4000-5400] & [-11.8,19.7] & 0.60 & 295 & 1.0 \\
       & b & 20230417  & MJ45 & [4000-5400] & [-15.0,14.8] & 0.60 & 205 & 1.0 \\
       & b & 20230417  & MN45 & [4000-5400] & [-15.5,10.2] & 0.60 & 295 & 1.0 \\
       & b & 20250321  & MJ & [4000-5400] & [-18.5,20.2] & 1.2 & 340 & 4.0 \\
       & b & 20250321  & MN & [4000-5400] & [-13.4,15.9] & 1.3 & 70 & 4.0 \\
       & b & 20250322  & MJ & [4000-5400] & [-10.1,15.5] & 1.4 & 340 & 1.0 \\
       & b & 20250322  & MN & [4000-5400] & [-17.9,13.0] & 1.4 & 70 & 1.0 \\
       & b & 20250323  & MJ & [4000-5400] & [-16.6,13.1] & 1.6 & 340 & 1.5 \\
       & b & 20250323  & MN & [4000-5400] & [-14.5,23.9] & 1.2 & 70 & 1.5 \\
       & b & 20250324  & MJ45 & [4000-5400] & [-11.2,16.6] & 0.90 & 205 & 3.5 \\
       & b & 20250324  & MN45 & [4000-5400] & [-13.2,13.2] & 1.2 & 295 & 3.5 \\
       & b & 20250325 & MJ45 & [4000-5400] & [-20.1,11.1] & 1.1 & 205 & 1.0 \\
       & b & 20250325 & MN45 & [4000-5400] & [-21.8,13.9] & 1.0 & 295 & 1.0 \\
       \hline
A1982 & b & 20240409 & MJ & [4000-5400] &  [-9.49,11.1] & 0.86 & 47 & 2.0 \\
        & b & 20240409 & MN & [4000-5400] &  [-9.02,6.15] & 0.97 & 317 & 2.0 \\
        & b & 20250325 & MJ45 & [4000-5400] &  [-11.6,16.1] & 1.9 & 272 & 3.5 \\
        & b & 20250325 & MN45 & [4000-5400] &  [-13.3,10.9]  & 1.9 & 182 & 3.5 \\
        & b & 20250327 & MJ45 & [4000-5400] & [-18.9,17.7] & 1.3 & 272 & 1.0 \\
        & b & 20250327 & MN45 & [4000-5400] & [-11.9,17.0] & 1.3 & 182 & 1.0 \\
\hline
A2107 & b & 20230417 & MJ & [4000-5200] & [-11.3,12.7] & 0.70 & 290 & 2.0 \\
      & b & 20230417 & MN & [4000-5200] &  [-10.5,10.0] & 0.60 & 200 & 2.0 \\
      & b & 20240403  & MJ45 & [4000-5200] & [-12.9,14.5] & 0.86 & 335 & 2.0 \\
      & b & 20240403  & MN45 & [4000-5200] &  [-13.6,11.3]  & 0.76 & 245 & 2.0 \\
\hline
A2147 & b & 20220401 & MJ & [4000-5400] & [-11.3,14.9] & 0.80 & 13 & 2.0 \\
        & b & 20220401 & MN & [4000-5400] &  [-13.9,17.4]  & 0.90 & 103 & 2.0 \\
    & b & 20220409 & MJ45 & [4000-5400] &  [-10.4,10.5] & 0.60 & 58 & 2.0 \\
        & b & 20220409 & MN45 & [4000-5400] & [-13.8,13.4]  & 0.50 & 148 & 2.0 \\ \hline
A2255 & b &  20231005 & MJ & [4100-5300] & [-12.4,10.1] & 1.0 & 50 & 2.0 \\
      & b &  20231005 & MN & [4100-5300] & [-13.0,16.2] & 1.0 & 320 & 2.0 \\
       & b &  20231007 & MJ & [4100-5300] & [-10.0,10.1] & 1.1 & 50 & 0.3 \\
       & b &  20231007 & MN & [4100-5300] & [-19.9,21.5] & 1.1 & 320 & 0.3 \\
      & b &  20231008 & MJ & [4100-5300] & [-13.1,16.9] & 0.80 & 50 & 1.0 \\
      & b &  20231008 & MN & [4100-5300] & [-15.5,15.5] & 0.80 & 320 & 1.0 \\
      & b &  20231010 & MJ45 & [4100-5300] & [-13.2,13.0] & 0.90 & 95 & 2.0 \\
       & b &  20231010 & MN45 & [4100-5300] & [-10.6,17.8] & 0.70 & 5 & 2.0 \\
    & b &  20231012 & MJ & [4100-5300] & [-23.8,12.0] & 0.90 & 50 & 0.6 \\
    & b &  20231012 & MN & [4100-5300] & [-12.8,11.4] & 0.80 & 320 & 0.6 \\
        & m &  20240910 & MJ45 & [4100-5300] & [-12.3,14.1] & 1.2 & 275 & 2.0 \\
       & m &  20240911 & MN45 & [4100-5300] & [-11.3,15.9] & 0.60 & 5 & 1.5 \\
\hline

\end{tabular}

\caption*{Continued.}

\end{table*}

\begin{table*}
\begin{tabular}{c c c c c c c c c}
       \hline
       Cluster & Configuration & Date & Slits & $\lambda_\mathrm{int}$ & r (arcsec) & Seeing (arcsec) & PA & Exp. hours \\
    \hline
   A2256 & b & 20210223 & MJ & [4000-5400] & [-9.05,12.4]  & 0.80 & 140 & 0.5 \\
    & b & 20210223 & MN & [4000-5400] & [-13.3,13.4]  & 1.0 & 50 & 0.5 \\
    & b &  20210321 & MJ & [4000-5400] & [-7.92,10.5] & 1.6 & 140 & 2.0 \\
    & b &  20210321 & MN & [4000-5400] & [-15.6,15.3] & 1.4 & 50 & 2.0 \\
     & b & 20210323 & MJ45 & [4000-5400] & [-8.17,16.1]  & 1.5 & 5 & 1.0 \\
    & b & 20210323 & MN45 & [4000-5400] & [-10.1,11.1] & 1.5 & 95 & 1.0 \\
    & b & 20210420  & MJ45 & [4000-5400] & [-9.10,14.2] & 0.90 & 5 & 1.0 \\
    & b & 20210420  & MN45 & [4000-5400] & [-13.2,14.8] & 0.90 & 95 & 1.0 \\
    & b & 20250322 & MJ & [4000-5400] & [-8.11,19.8] & 0.90 & 140 & 1.5 \\
    & b & 20250322 & MN & [4000-5400] & [-11.4,18.1] & 1.1 & 50 & 1.5 \\
    & b & 20250323  & MJ45 & [4000-5400] & [-6.55,10.8] & 1.0 & 5 & 2.0 \\
    & b & 20250323  & MN45 & [4000-5400] & [-17.0,13.0] & 1.3 & 95 & 2.0 \\
\hline
A2319 & m & 20240930 & MN & [4100-5250] & [-14.2,17.6] & 0.70 & 53 & 4.0 \\
        & m & 20241003 & MJ & [4100-5250] & [-11.0,12.6] & 0.90 & 143 & 4.0 \\
      & m & 20241004 & MN45 & [4100-5250] & [-13.4,15.7] & 0.60 & 98 & 4.0 \\
      & m & 20241005 & MJ45 & [4100-5250] &  [-13.9,15.7] & 0.60 & 8 & 4.0 \\
\hline
A2388 & b &  20231007 & MJ & [4100-5400] & [-11.7,13.7]  & 0.90 & 105 & 4.0 \\
     & b &  20231007 & MN & [4100-5400] & [-14.5,7.48]   & 0.90 & 15 & 4.0 \\
      & b & 20231010 & MJ & [4100-5400] & [-18.9,12.0] & 1.0 & 105 & 4.0 \\
      & b & 20231010 & MN & [4100-5400] & [-10.0,9.91] & 1.0 & 15 & 4.0 \\
      & m &  20240910 & MJ45 & [4100-5400] & [-8.24,5.40]  & 0.70 & 150 & 2.0 \\
     & m &  20240911 & MN45 & [4100-5400] & [-7.04,6.80]   & 0.60 & 60 & 2.0 \\
\hline
A2506  & b & 20221001 & MJ & [4000-5400] & [-10.1,9.67]  & 1.0 & 40 & 1.0 \\
       & b & 20221001 & MN & [4000-5400] & [-8.28,14.2]  & 0.90 & 130 & 1.0 \\
       & b & 20221001 & MJ45 & [4000-5400] & [-7.63,6.27]  & 0.60 & 85 & 1.0 \\
       & b & 20221001 & MN45 & [4000-5400] & [-11.5,17.2]  & 0.65 & 175 & 1.0 \\
\hline
A2665 & b & 20191006 & MJP & [4000-5400] & [-11.4,11.2]  & 1.1 & 35 & 1.0 \\
        & b & 20191006 & MNP & [4000-5400] & [-14.9,18.2] & 1.3 & 125 & 1.0 \\
        & b & 20191006 & MJ45P & [4000-5400] & [-19.7,19.8]  & 0.80 & 80 & 1.0 \\
        & b & 20191006 & MN45P & [4000-5400] & [-19.2,12.3]  & 1.1 & 170 & 1.0 \\
        & b & 20201006 & MJ & [4000-5400] & [-8.65,11.3]  & 1.7 & 100 & 1.0 \\
        & b & 20201006 & MN & [4000-5400] & [-15.3,7.31]  & 1.7 & 190 & 1.0 \\
        & b & 20201006 & MJ45 & [4000-5400] & [-13.8,7.89]   & 1.0 & 145 & 1.0 \\
        & b & 20201006 & MN45 & [4100-4800] & [-15.1,6.52]  & 0.94 & 235 & 1.0 \\
        & b & 20201008 & MJ & [4000-5400] & [-9.00,14.5] & 2.0 & 100 & 1.0 \\
         & b & 20201008 & MN & [4000-5400] & [-11.8,4.55] & 1.7 & 190 & 1.0 \\
        & b & 20201008 & MJ45 & [4000-5400] & [-11.5,8.50] & 1.2 & 145 & 1.0 \\
        & b & 20201008 & MN45 & [4100-4800] & [-13.5,8.67] & 1.0 & 235 & 1.0
        \\
\hline
 
\end{tabular}

\caption*{Continued.}

\end{table*}

\section{Assessing the effect of negative LOSVDs}
\begin{figure*}

\subfloat{\includegraphics[width=.3\linewidth]{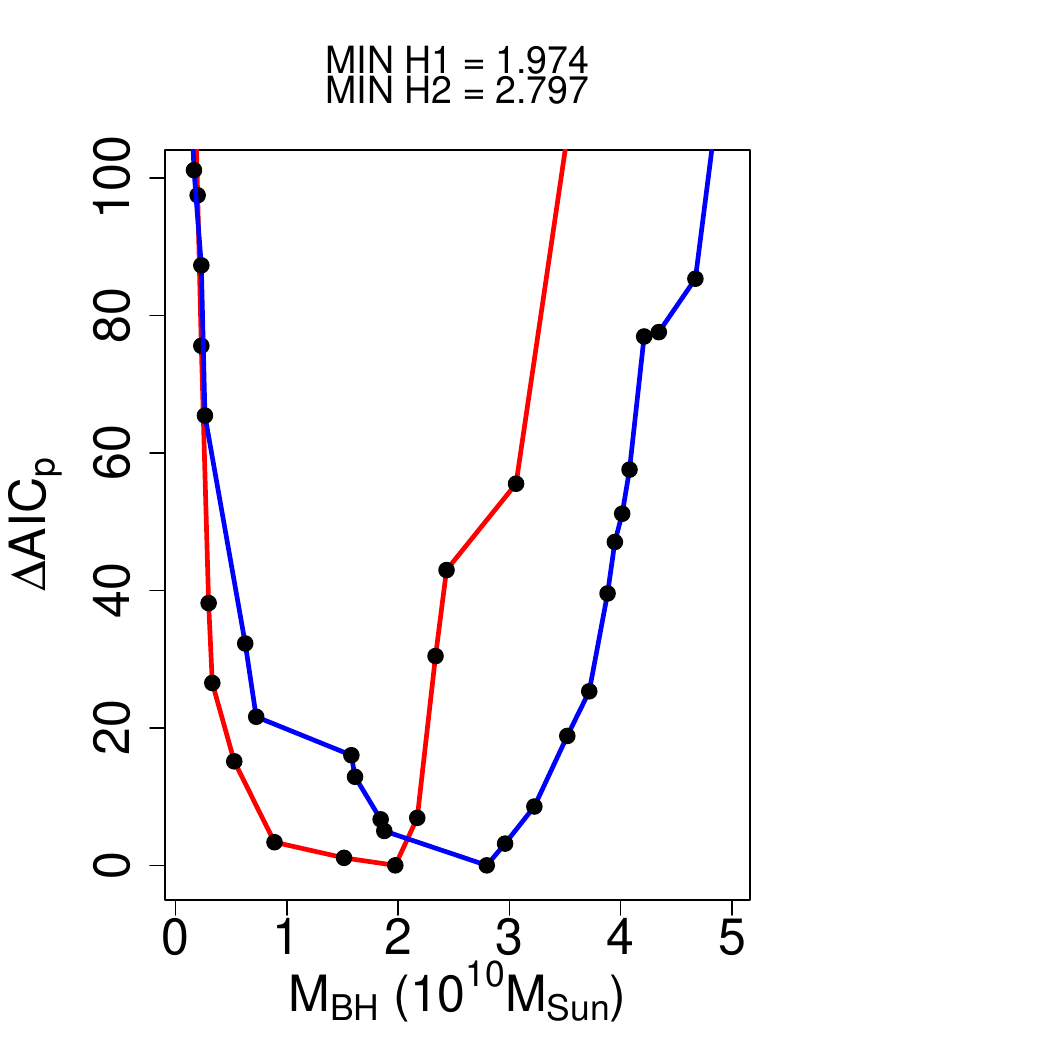}}
\subfloat{\includegraphics[width=.3\linewidth]{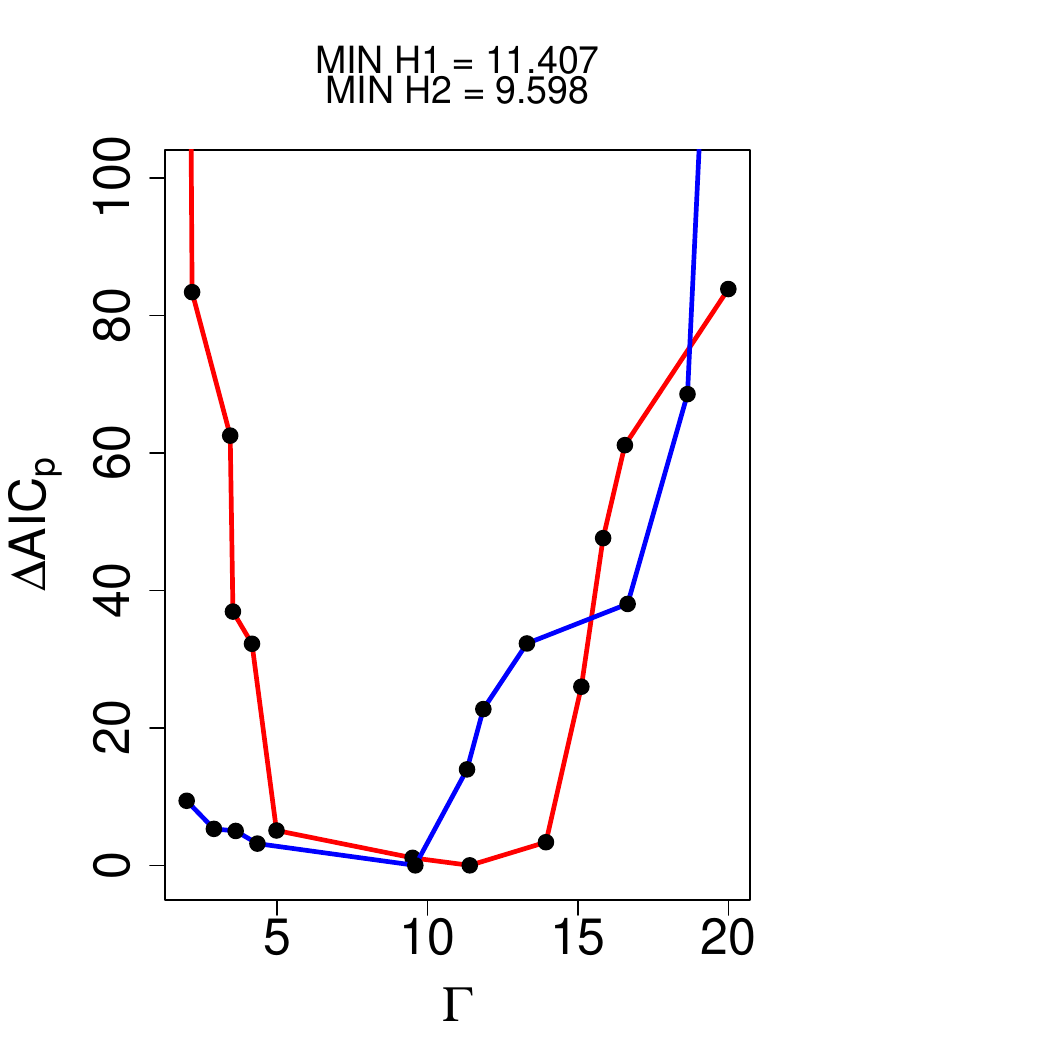}}
\subfloat{\includegraphics[width=.3\linewidth]{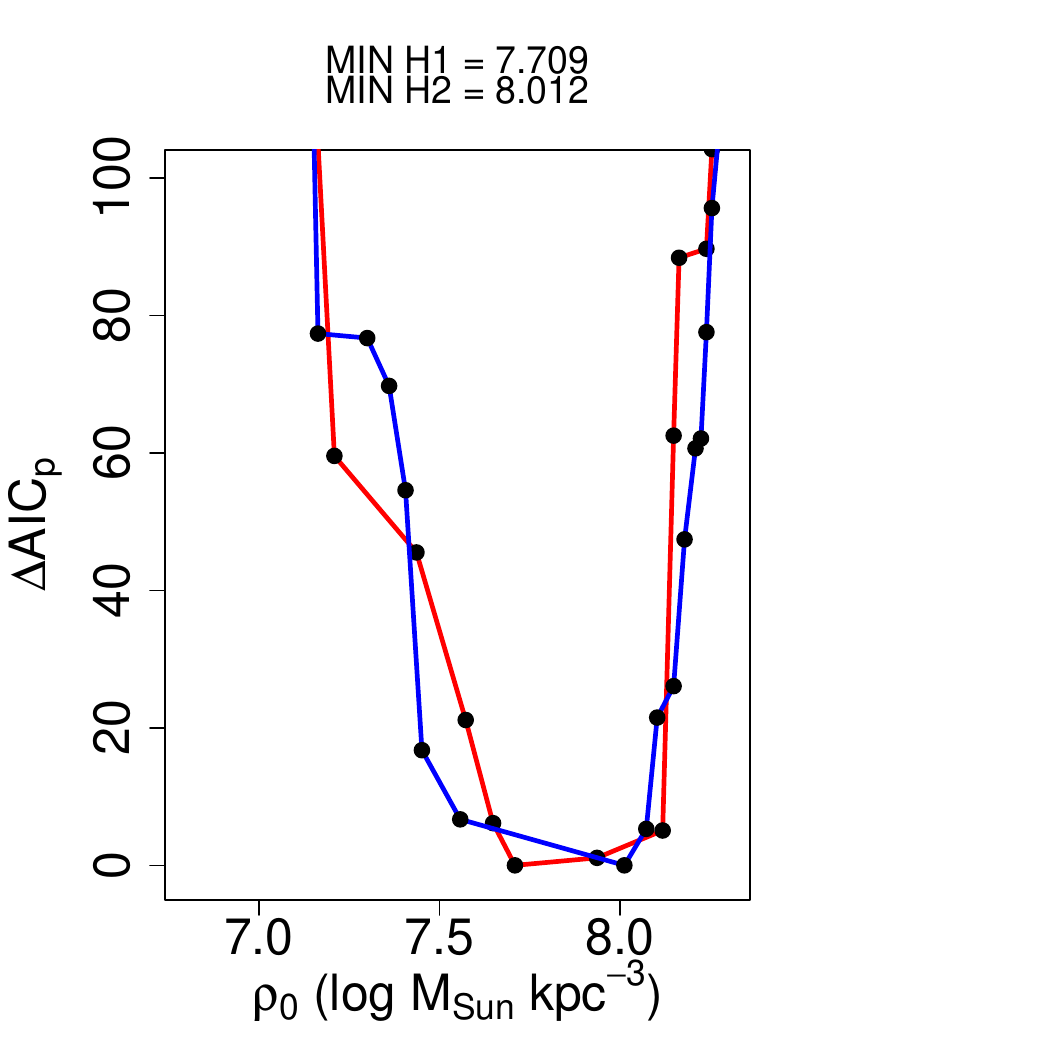}}

\subfloat{\includegraphics[width=.3\linewidth]{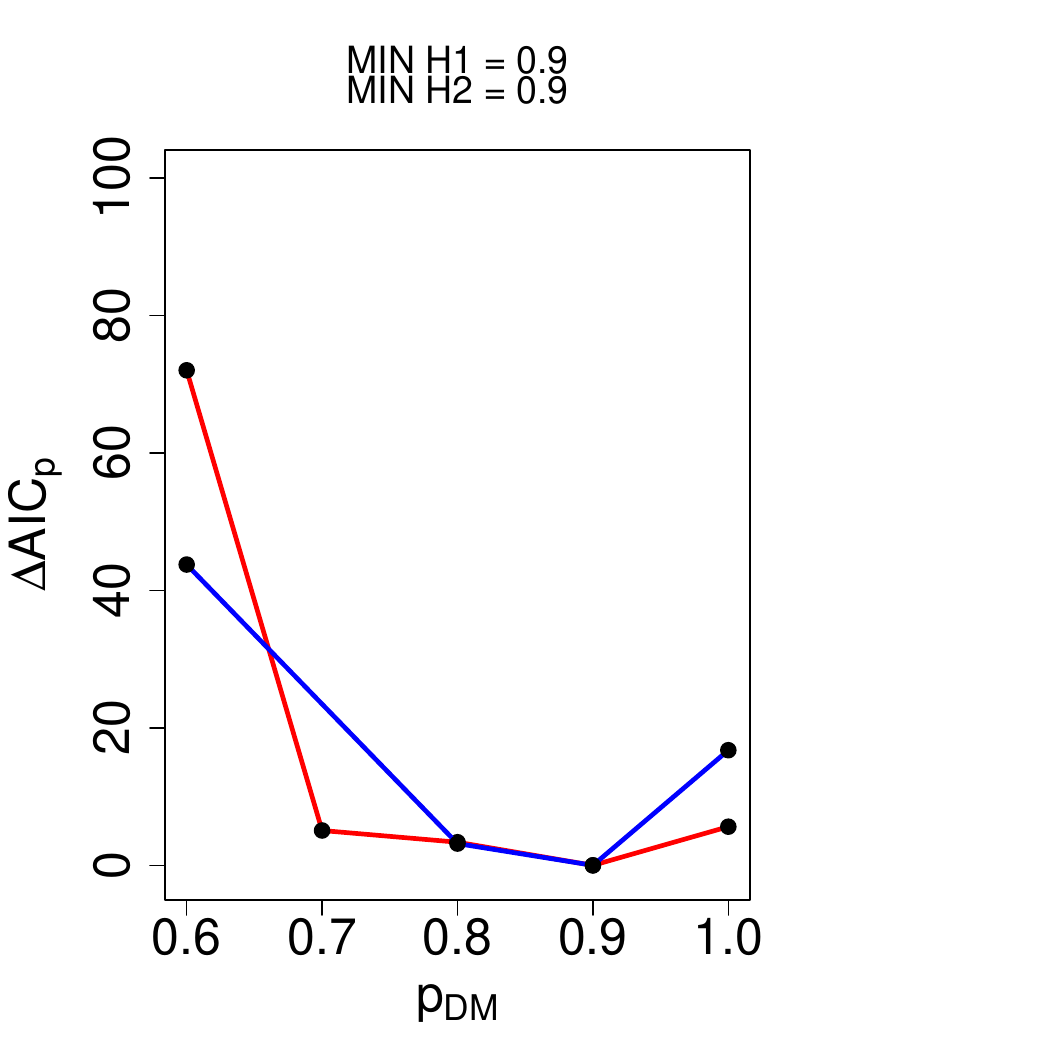}}
\subfloat{\includegraphics[width=.3\linewidth]{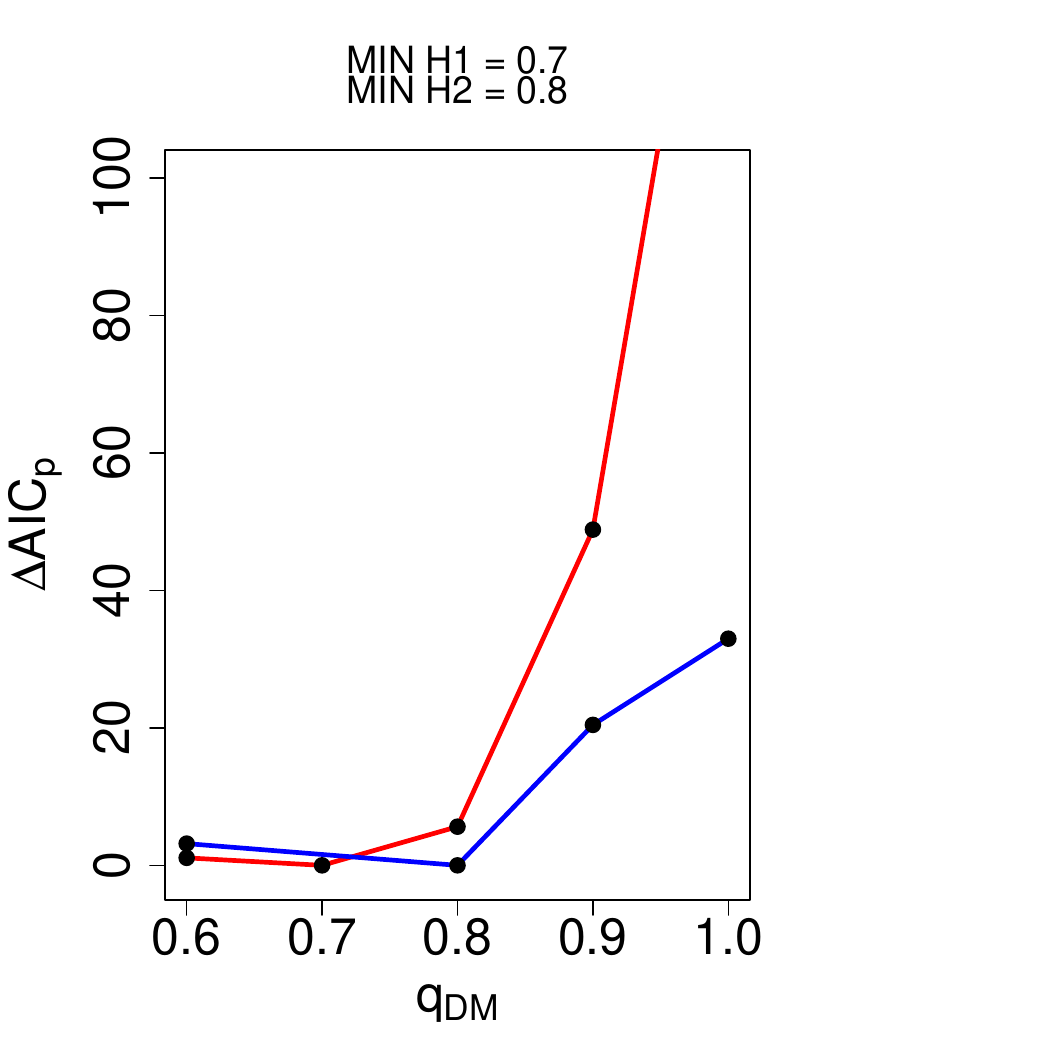}}
\subfloat{\includegraphics[width=.3\linewidth]{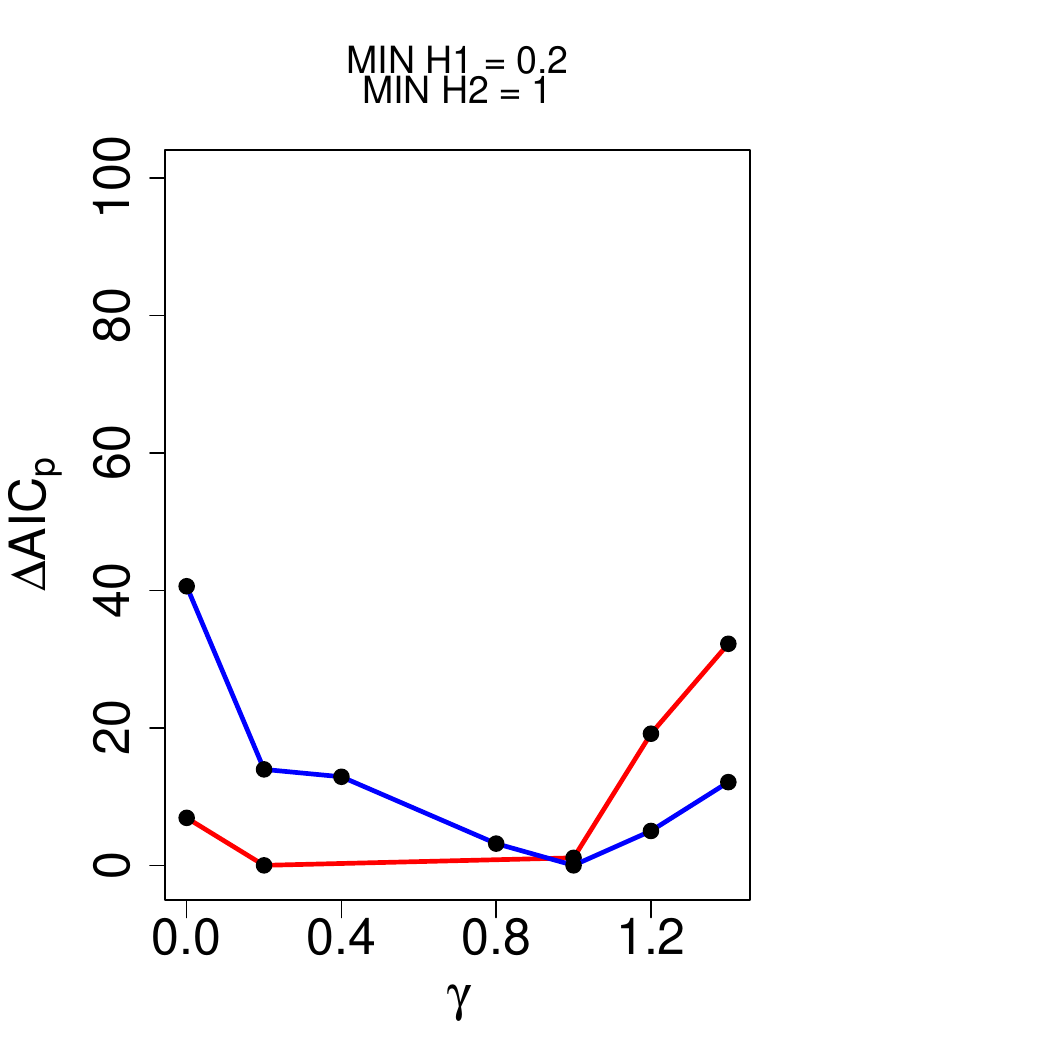}}

\caption{Same as Fig.~\ref{Fig.Schw_results}, with LOSVDs constrained to remain non-negative.}
\label{Fig.Schw_results_LOSVDs}
\end{figure*}

As stated in Sec.~\ref{Ssec.kinematics}, the model LOSVDs can take negative values, even when they are consistent with zero within the error bars. This appendix presents the results of the dynamical modeling of A2107 with positivity enforced in the LOSVDs (see Fig.~\ref{Fig.Schw_results}), showing in Tab.~\ref{Tab.Schw_results_LOSVDs} and Fig.~\ref{Fig.Schw_results_LOSVDs} that no significant differences are found.

\begin{table}
    \centering
    \begin{tabular}{c c}
     Variable & Result \\
    \hline \hline
       M$_\text{BH} \left(10^{9} \text{M}_\odot\right)$ & 23.9 $\pm$ 4.1  \\
       $\Gamma$ & 10.5 $\pm$ 0.9 \\
       $\log \left(\rho_0/\text{M}_\odot\,\text{kpc}^{-3} \right)$ & 7.86 $\pm$ 0.15\\
       p$_\text{DM}$ & 0.9 $\pm$ 0.1 \\
       q$_\text{DM}$ & 0.75 $\pm$ 0.05 \\
       $\gamma$ & 0.6 $\pm$ 0.2 \\
       \hline
       \end{tabular}
    \caption{Results of the dynamical modeling of A2107 with positive LOSVDs enforced.}
    \label{Tab.Schw_results_LOSVDs}
\end{table}

\section{M$_\text{BH}$ - AIC$_\text{p}$ plots}   \label{App.MBH_AICp}

\begin{figure*}
\subfloat{\includegraphics[scale=.23]{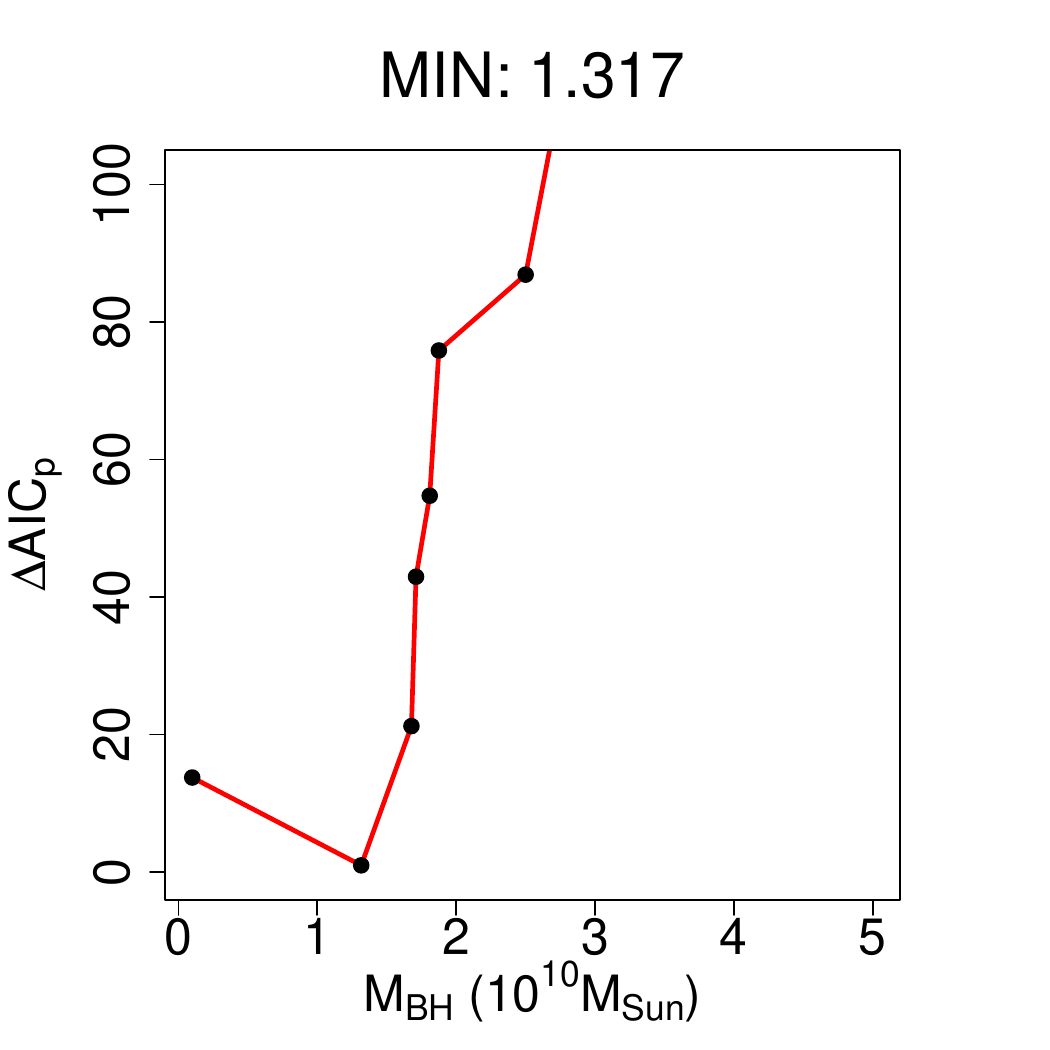}}
\subfloat{\includegraphics[scale=.23]{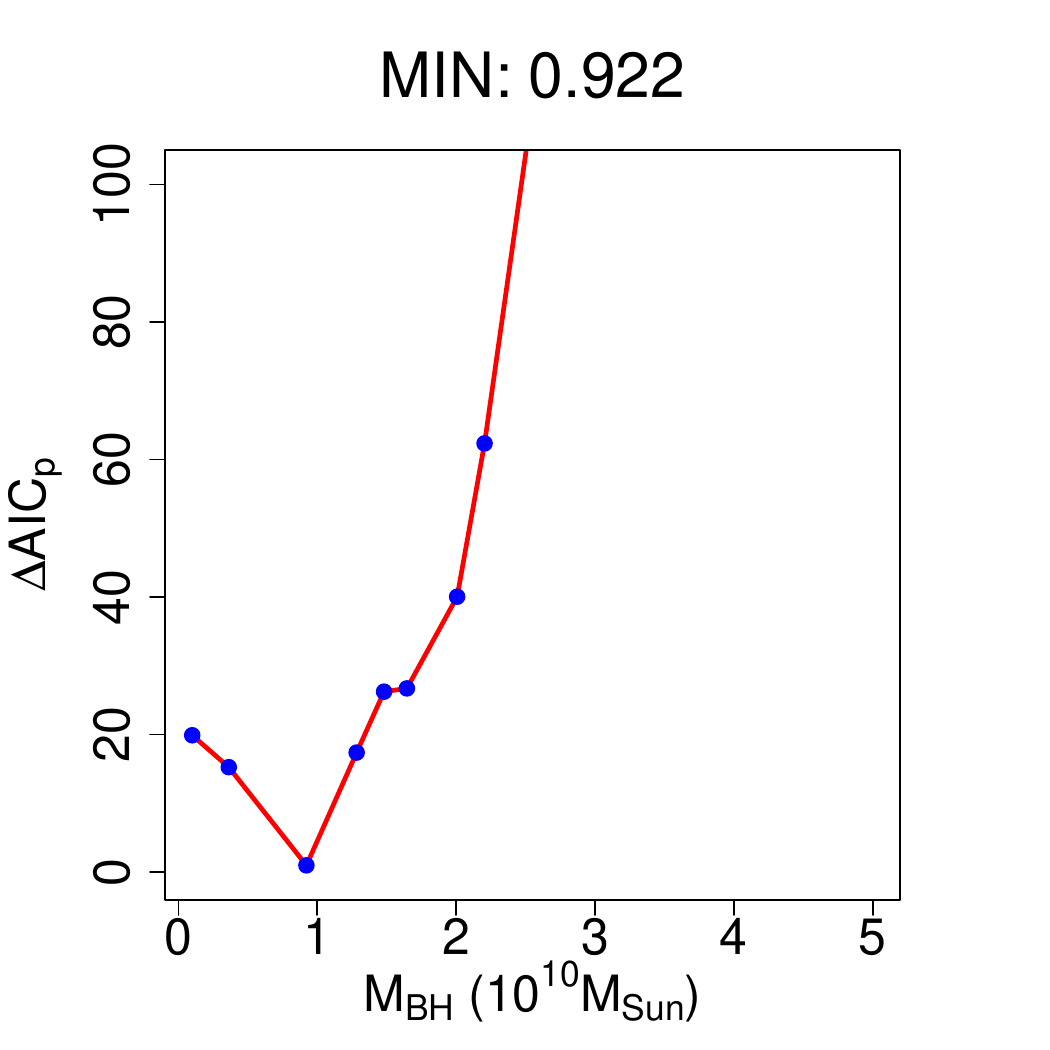}}
\subfloat{\includegraphics[scale=.23]{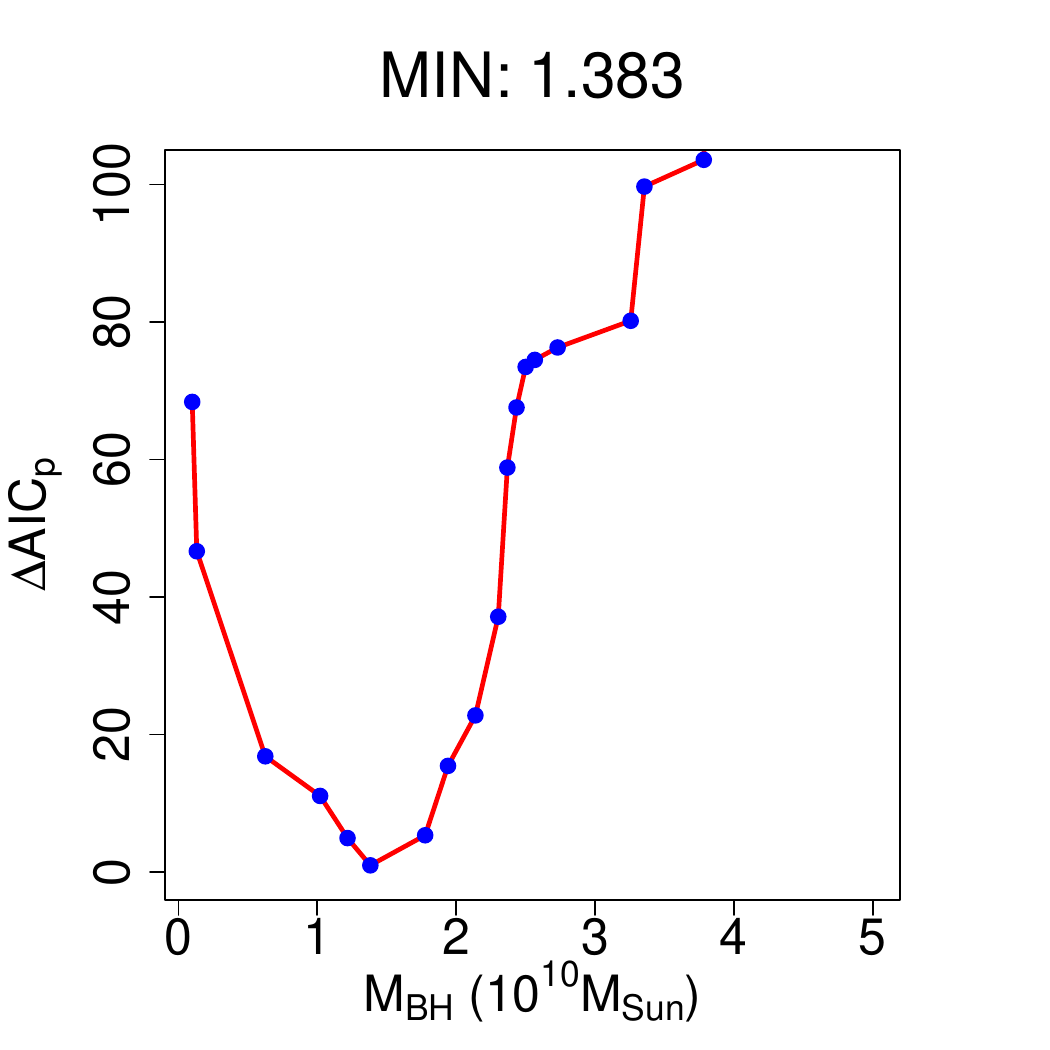}}
\subfloat{\includegraphics[scale=.23]{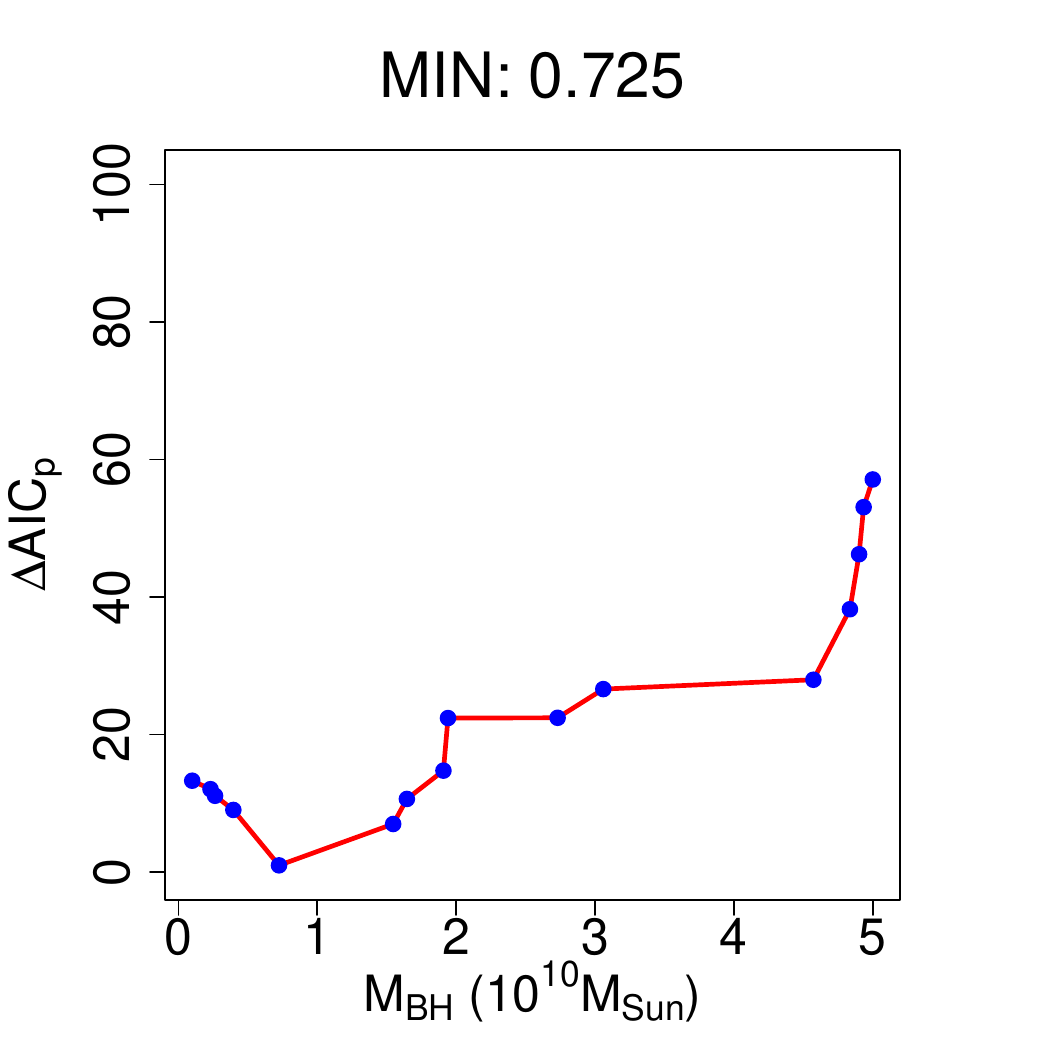}}

\subfloat{\includegraphics[scale=.23]{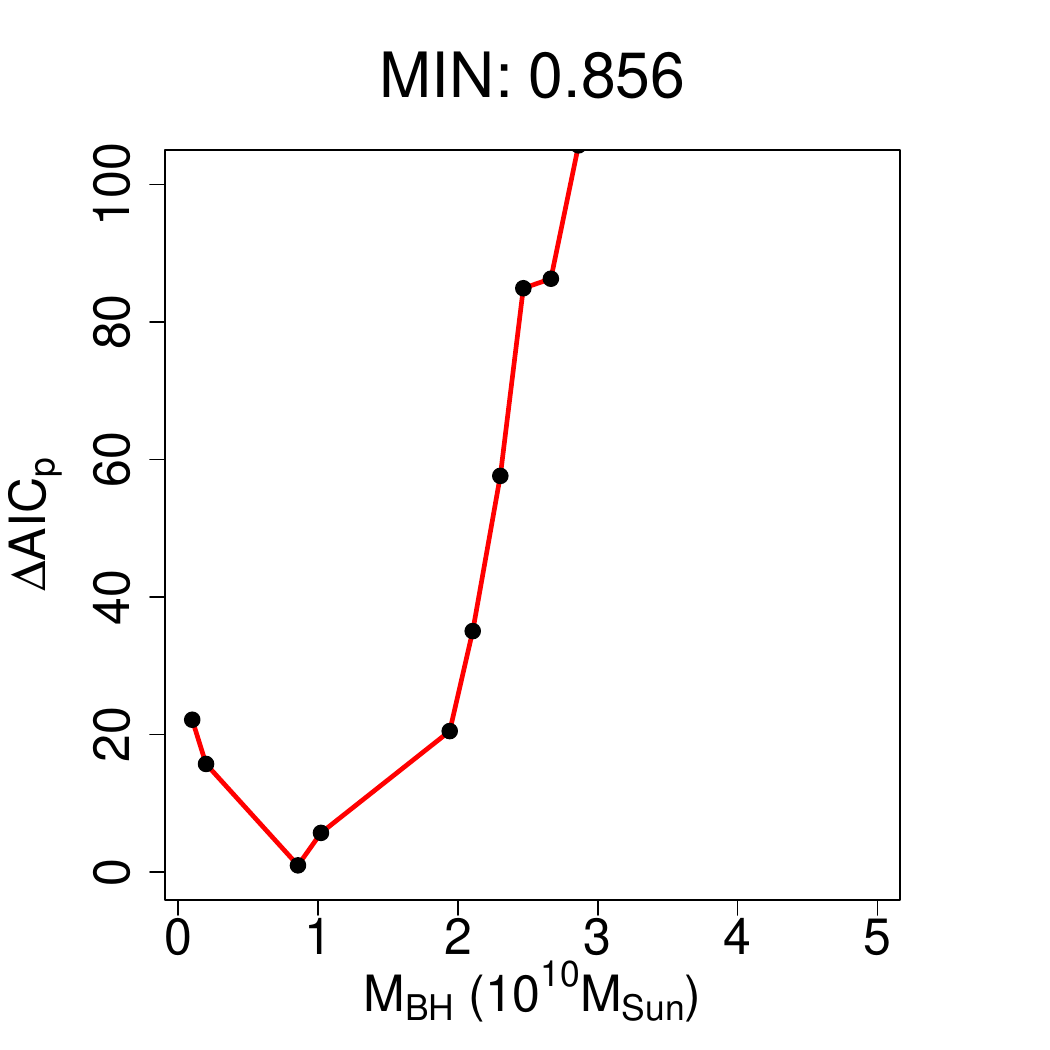}}
\subfloat{\includegraphics[scale=.23]{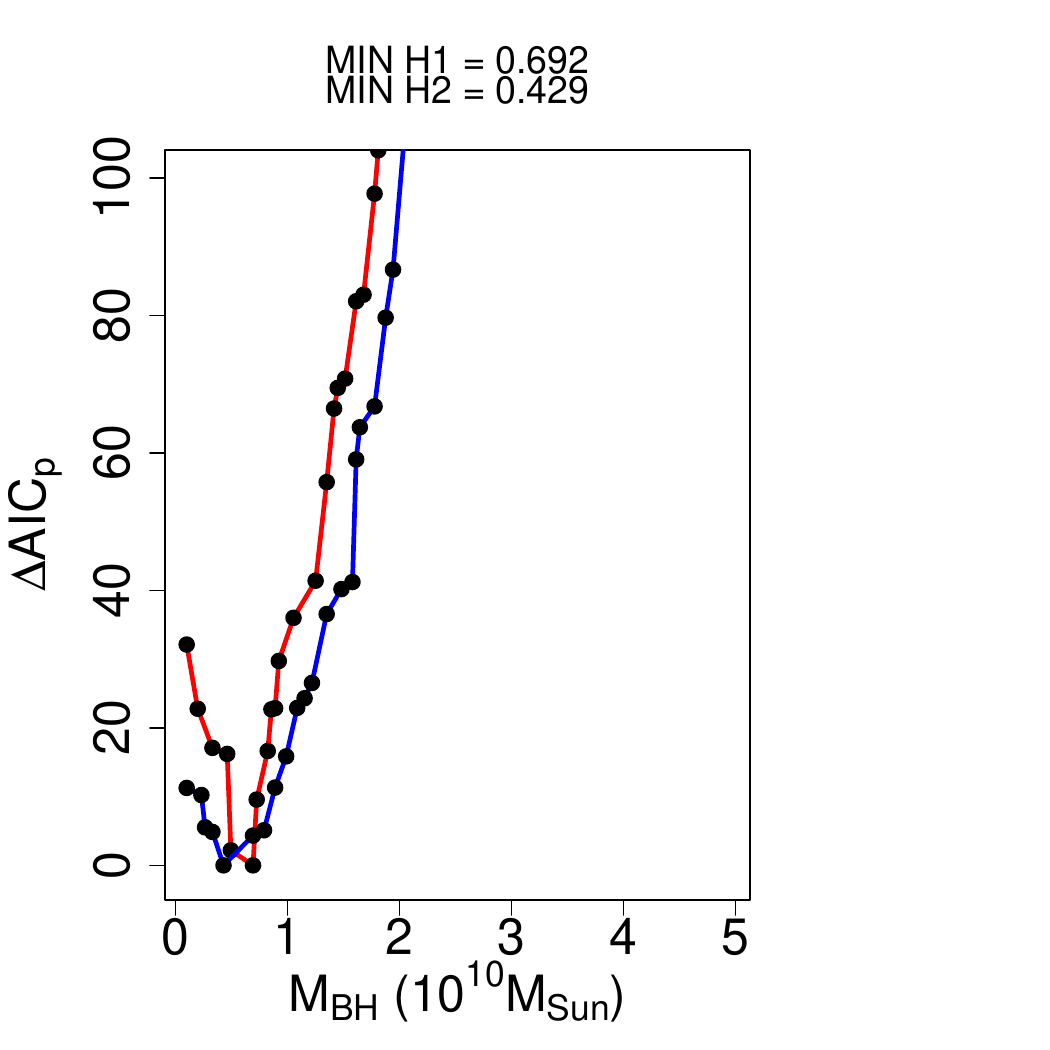}}
\subfloat{\includegraphics[scale=.23]{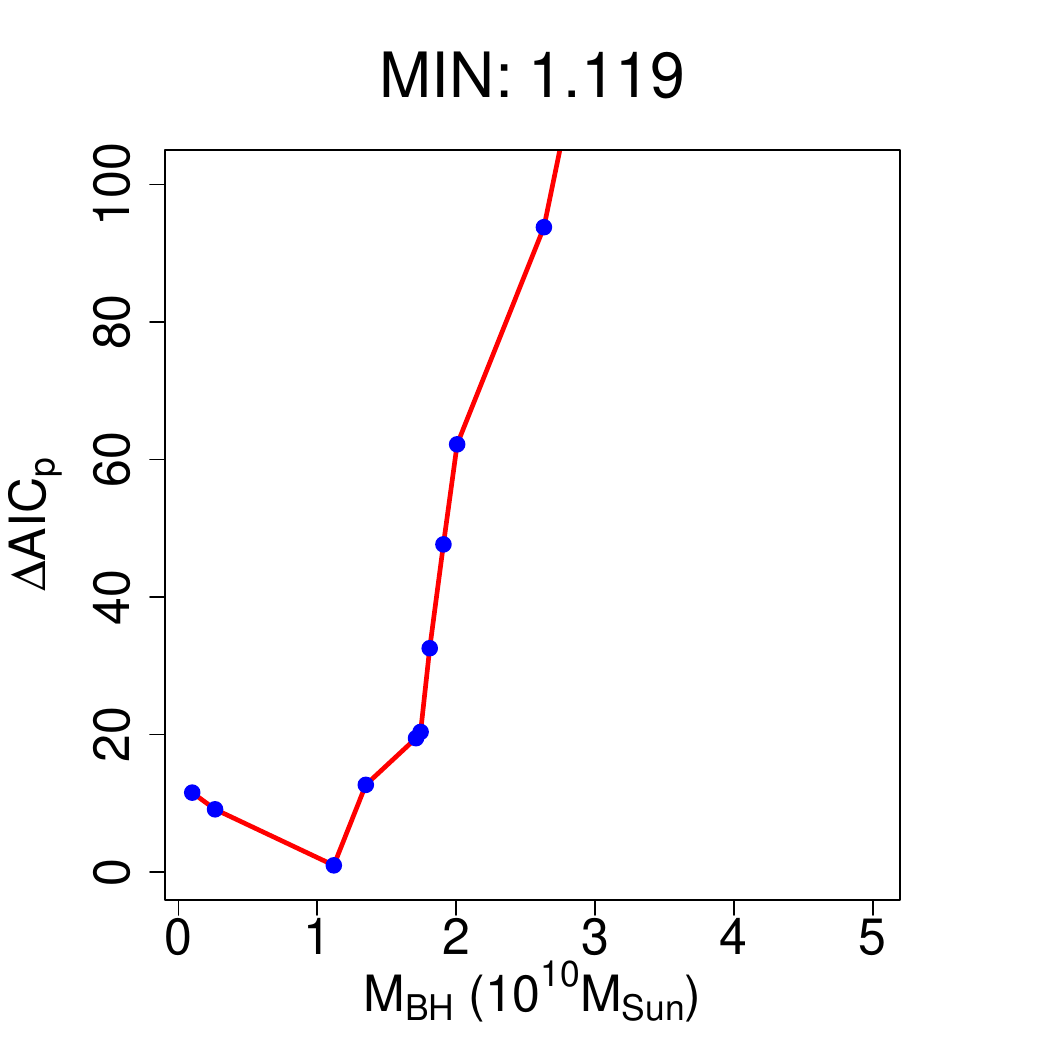}}
\subfloat{\includegraphics[scale=.23]{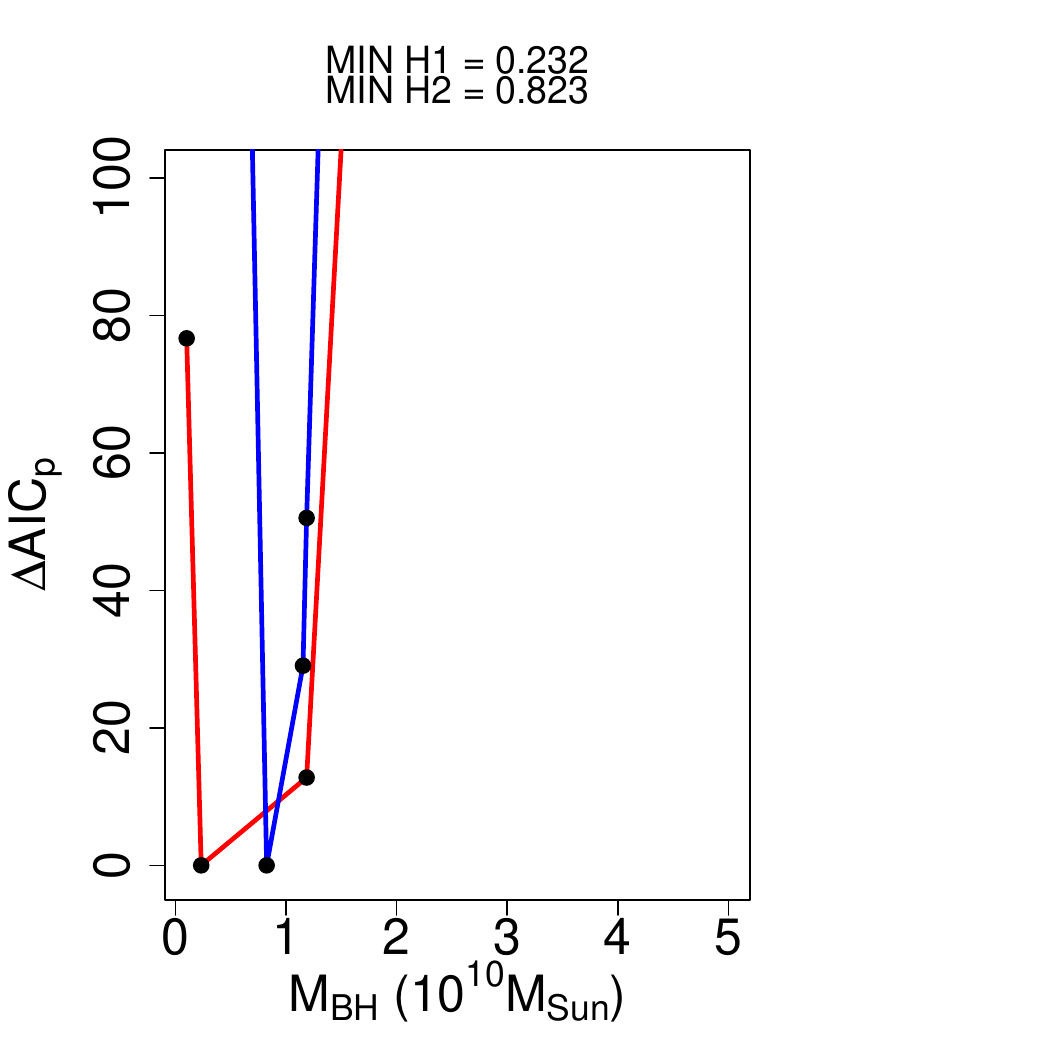}}

\subfloat{\includegraphics[scale=.23]{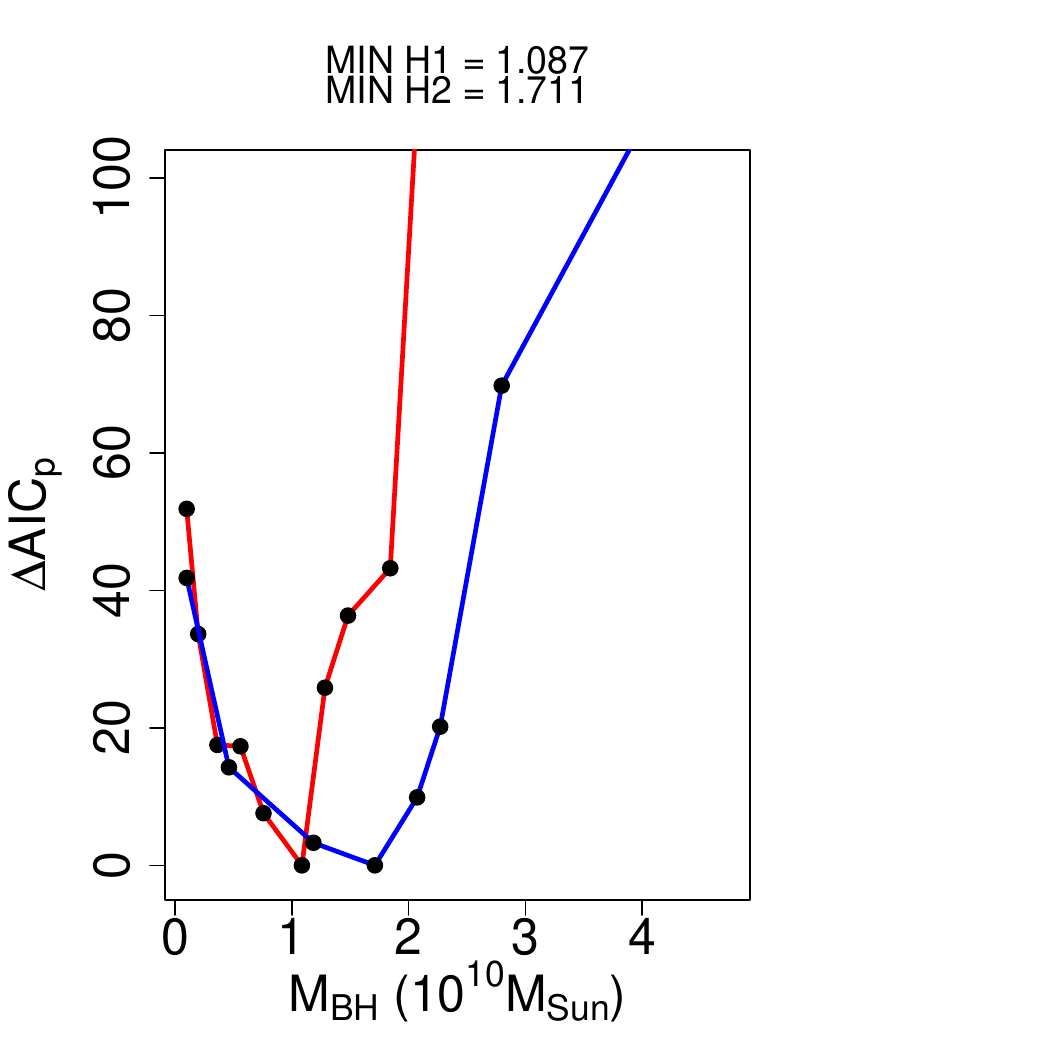}}
\subfloat{\includegraphics[scale=.23]{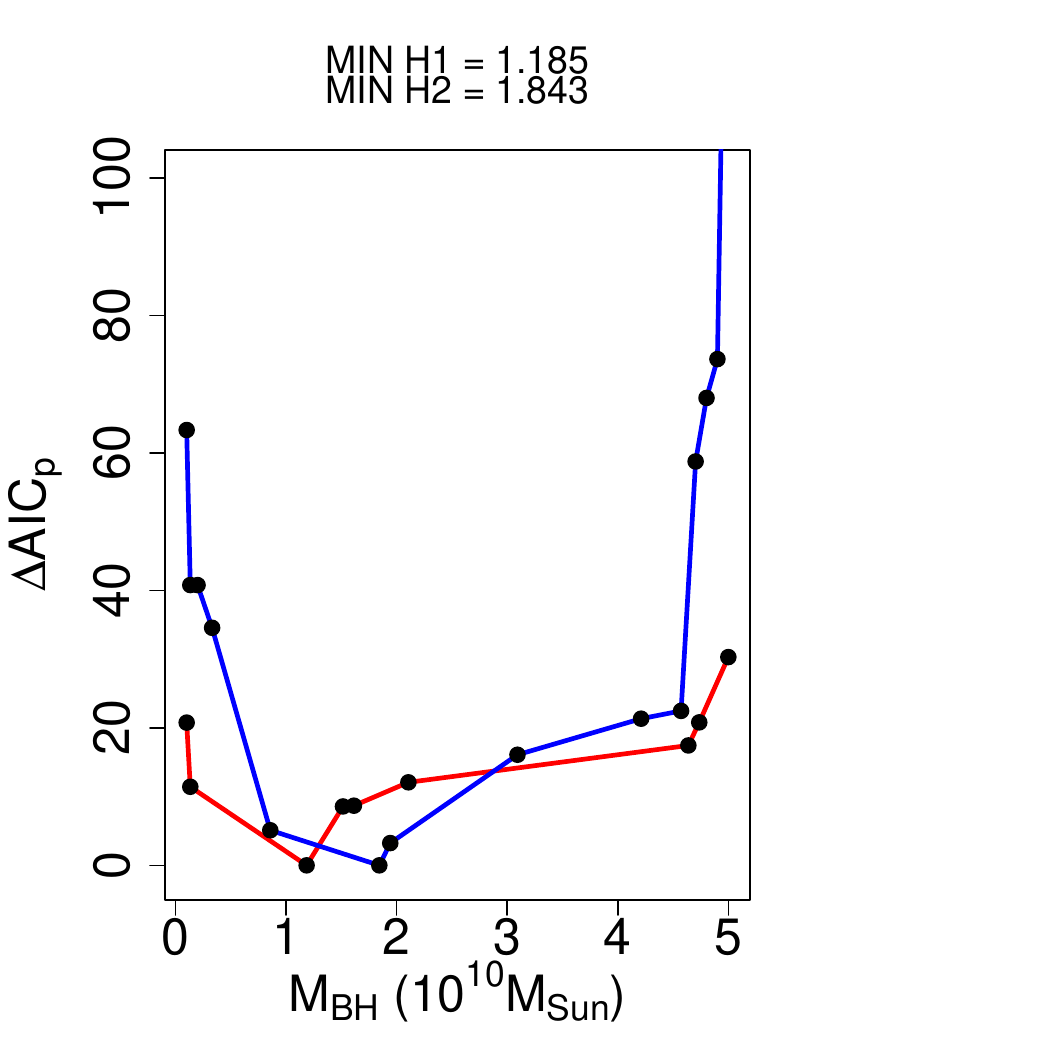}}
\subfloat{\includegraphics[scale=.23]{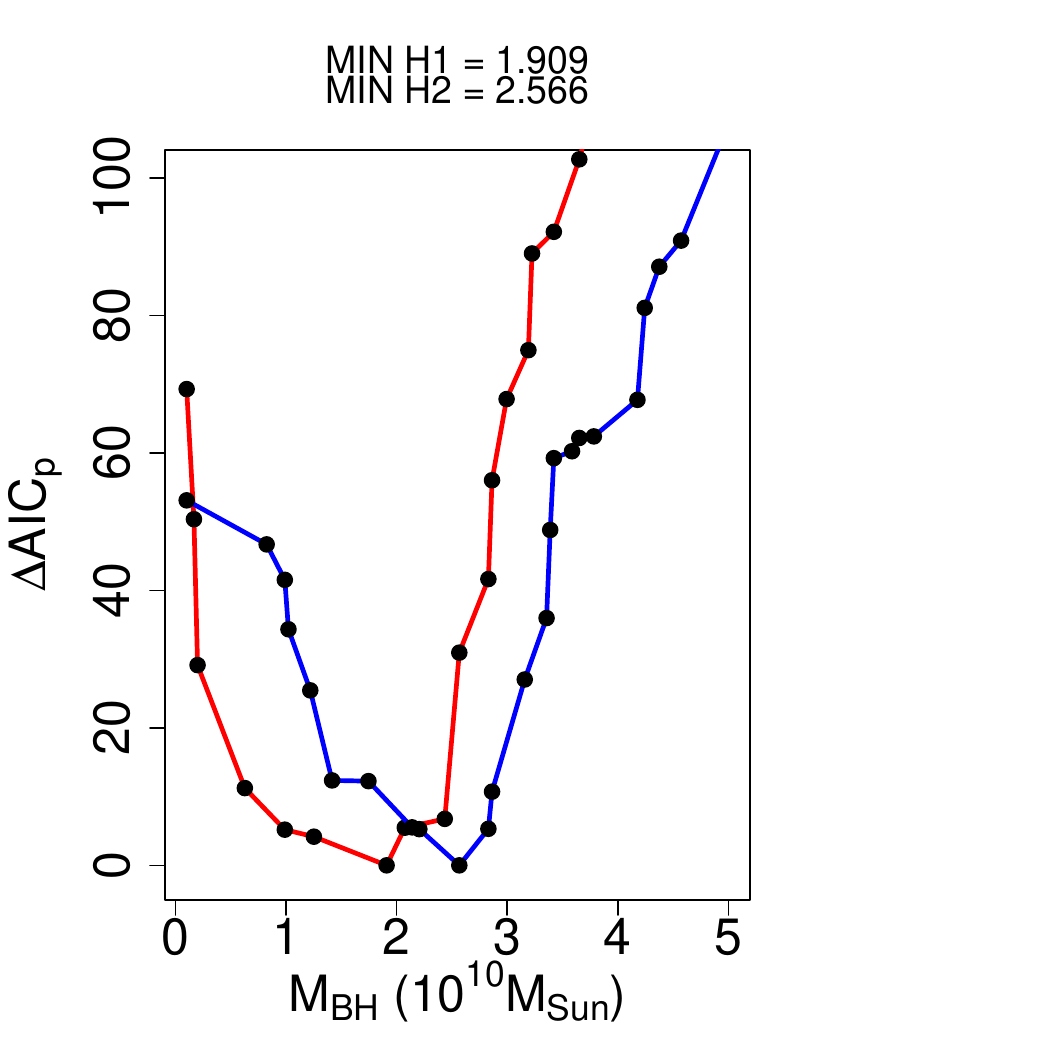}}
\subfloat{\includegraphics[scale=.23]{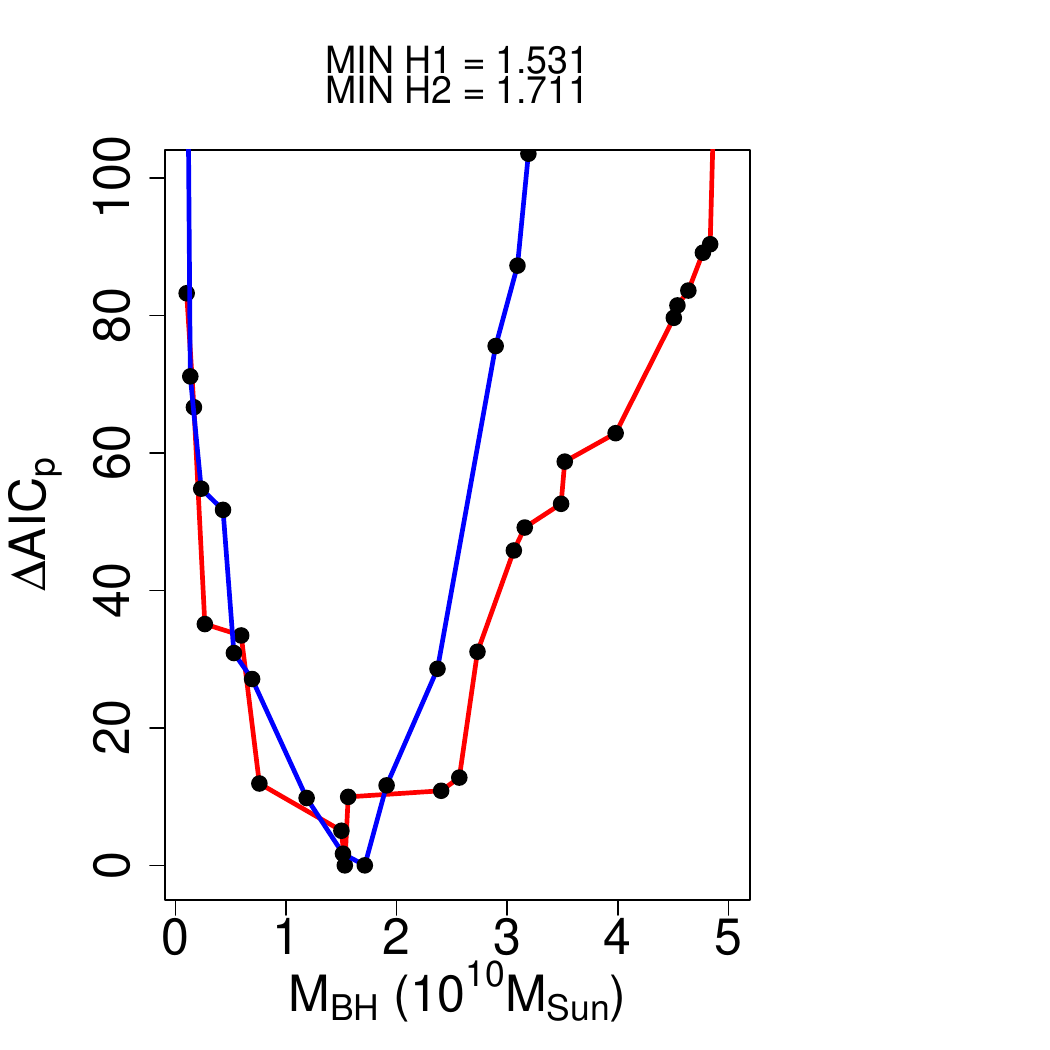}}

\subfloat{\includegraphics[scale=.23]{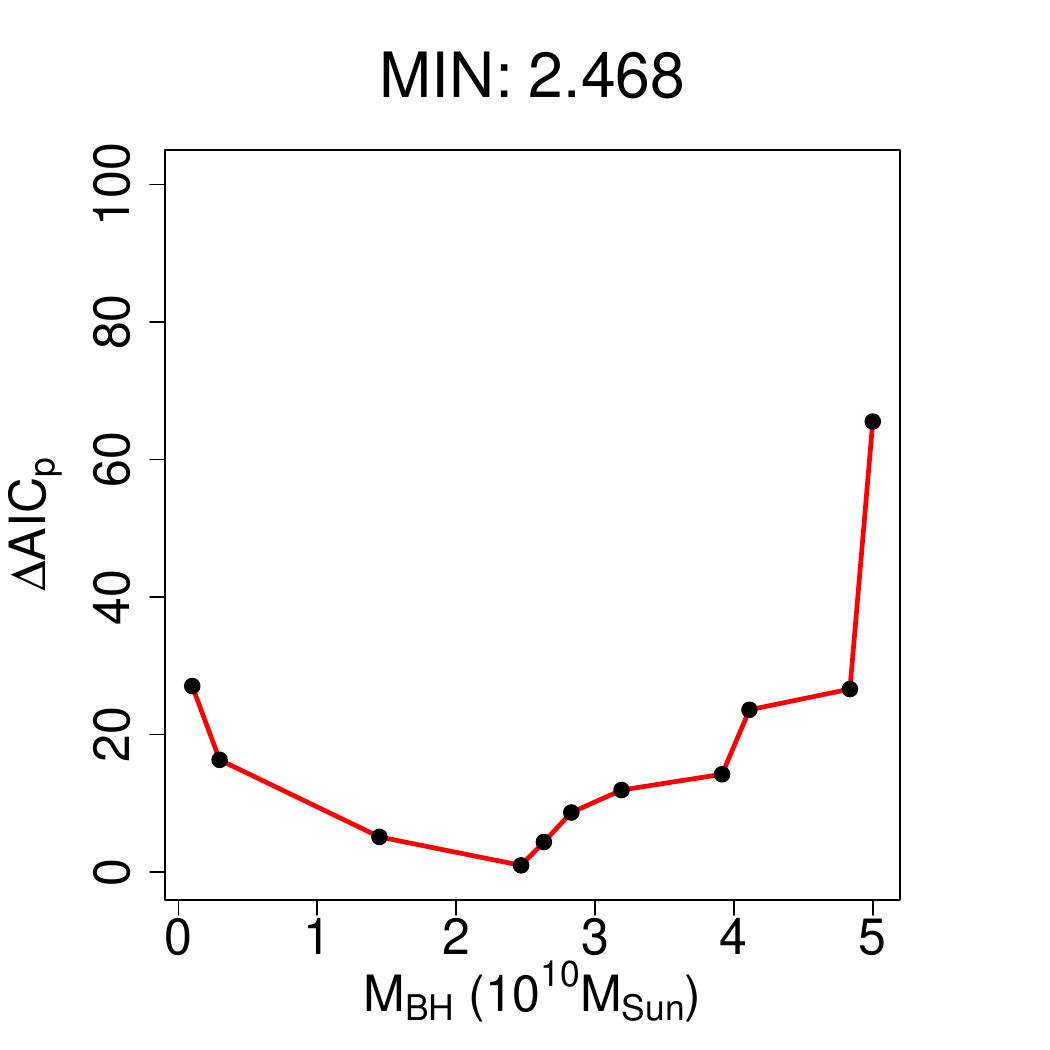}}
\subfloat{\includegraphics[scale=.23]{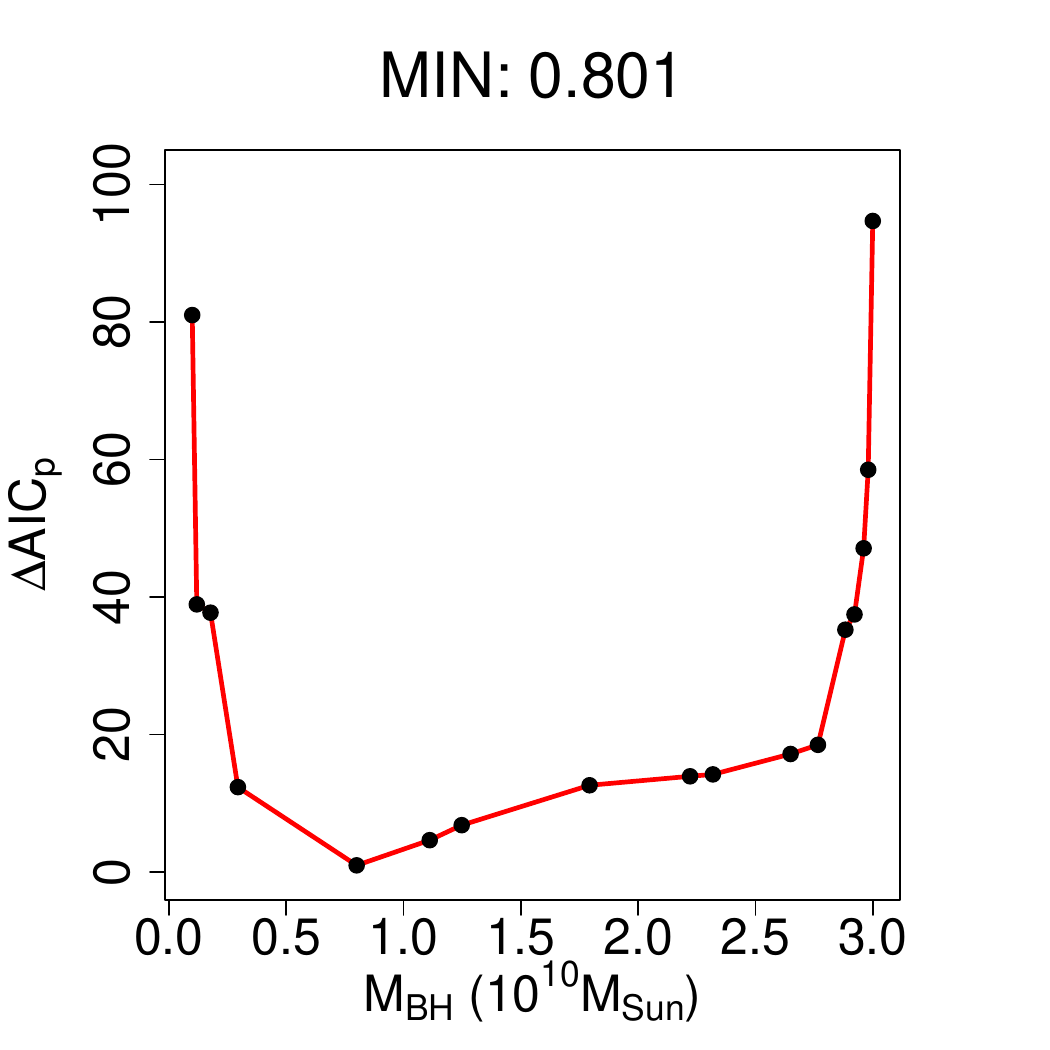}}
\subfloat{\includegraphics[scale=.23]{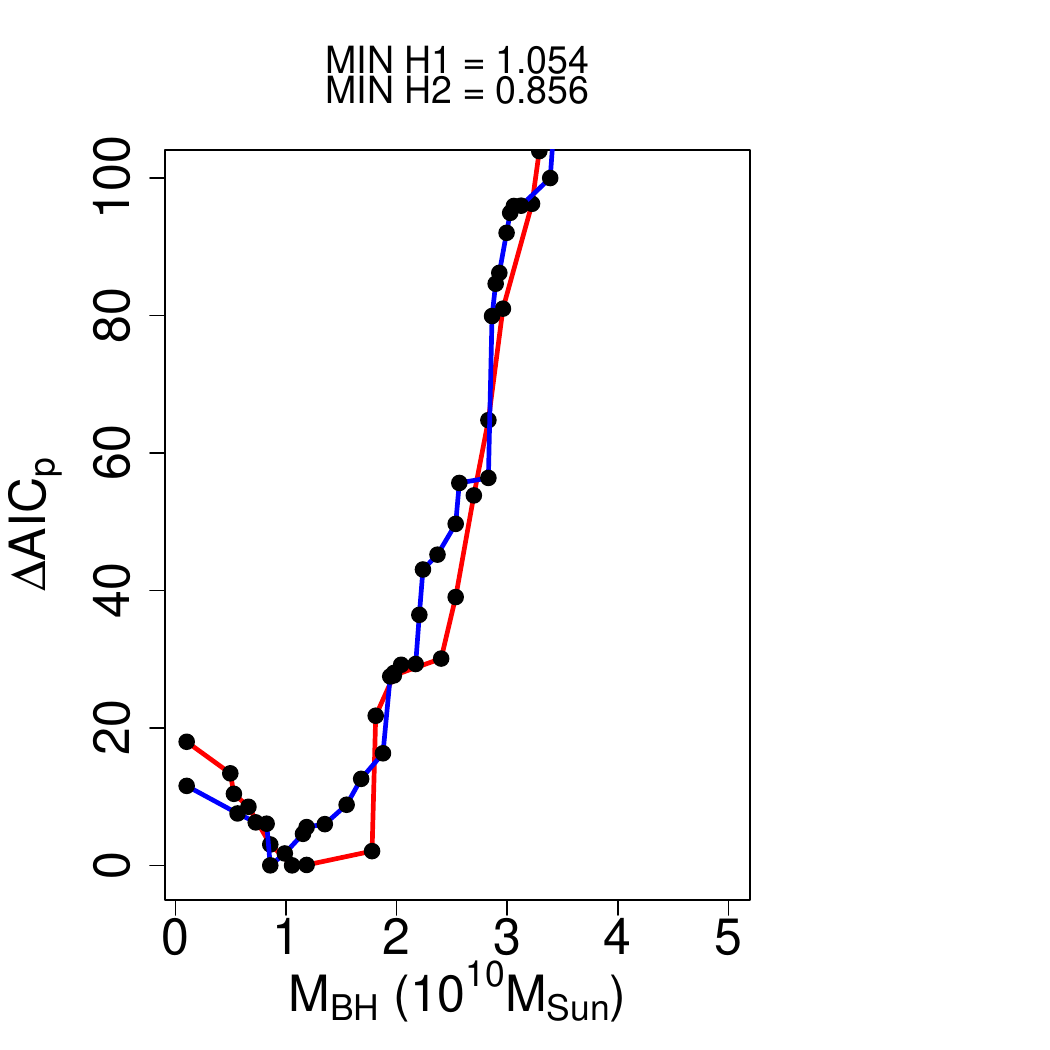}}
\subfloat{\includegraphics[scale=.23]{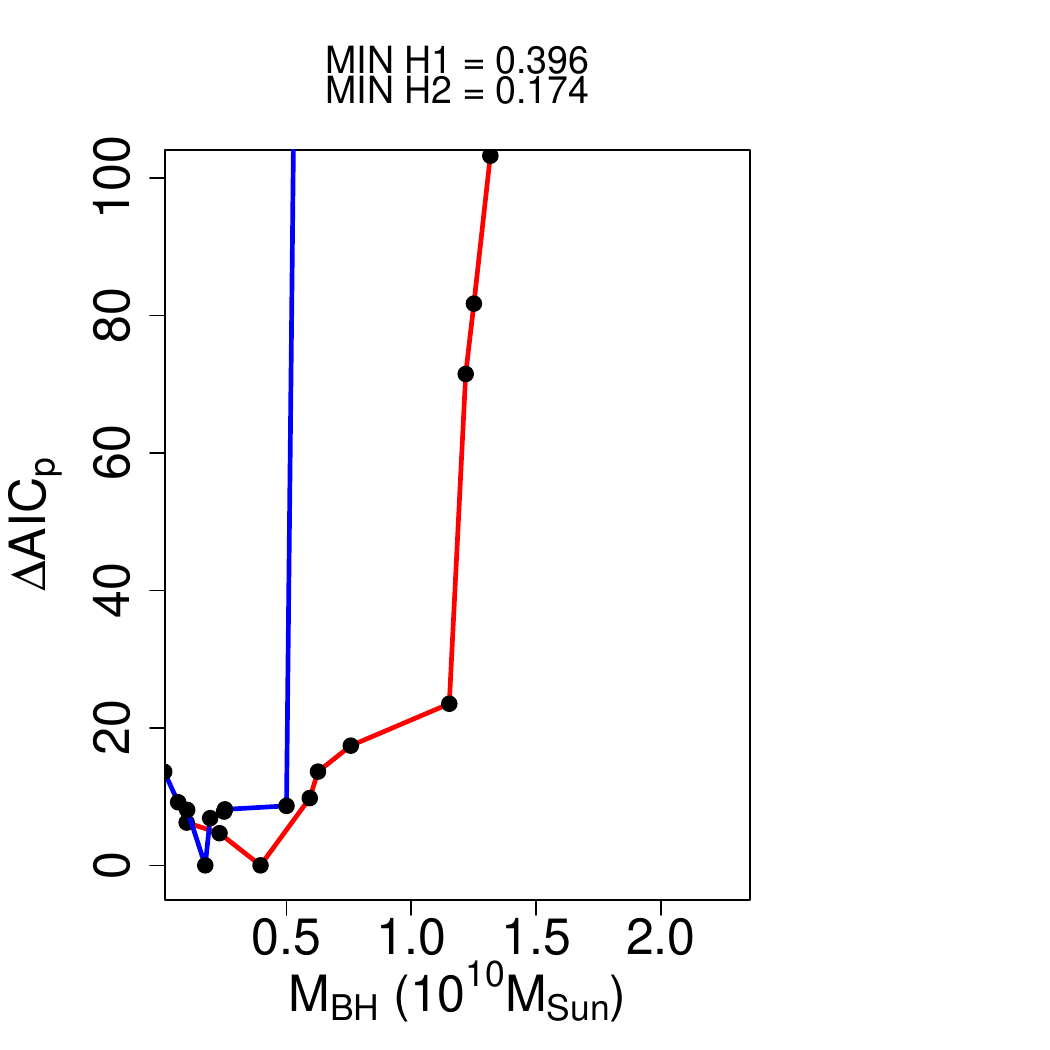}}

    \caption{\mbh–AIC$_\text{p}$ distributions for all 16 dynamically modeled BCGs. A single line is shown for those BCGs for which we modeled all bins together (see Tab.~\ref{Tab.dyn_results}). Note the different $y$-axis range with respect to the top-left panel of Fig.~\ref{Fig.Schw_results}.}
    \label{Fig.AICp_MBH}
\end{figure*}

Fig.~\ref{Fig.AICp_MBH} shows the M$_\text{BH}$ - AIC$_\text{p}$ distributions for the 16 modeled BCGs considered throughout this work.

%\section{Notes on individual galaxies}  \label{App.notes}

% DON'T TOUCH THESE LINES!
\bsp	% typesetting comment
\label{lastpage}
\end{document}